\documentclass[prd,preprint,tightenlines,preprintnumbers,aps,eqsecnum,nofootinbib,
superscriptaddress,notitlepage,showpacs,floatfix]{revtex4-1}

\usepackage{amsmath}    % need for subequations
\usepackage{amssymb}
\usepackage{graphicx}   % need for figures
\usepackage{verbatim}   % useful for program listings
\usepackage{color}      % use if color is used in text
\usepackage{subfigure}  % use for side-by-side figures
\usepackage[colorlinks=true,citecolor=blue,linkcolor=blue,urlcolor=blue]{hyperref}
\usepackage[geometry]{ifsym}

\usepackage{bm}         % need for better and more bold math symbols
\let\mathbf=\bm         % use the bm package's symbols instead

\definecolor{orange}{rgb}{1.0, 0.647059, 0.0}
\definecolor{newmagenta}{rgb}{0.8, 0.0, 0.8}
\definecolor{darkgreen}{rgb}{0.0, 0.545098, 0.0}

\newcommand{\sealight}{\hat{m}'}
\newcommand{\seaheavy}{m'_s}

\newcommand{\hs}{\ensuremath{\hphantom{5}}}
\newcommand{\NA}{\multicolumn{2}{c}{$-$}}

\newcommand{\rhoV}[1]{\ensuremath{\rho_{V^{#1}}}}
\newcommand{\ZV}[1]{\ensuremath{Z_{V^{#1}}}}

\newcommand{\ZVbb}{\ensuremath{Z_{V^4_{bb}}}}
\newcommand{\ZVcc}{\ensuremath{Z_{V^4_{cc}}}}
\newcommand{\ZVbc}[1]{\ensuremath{Z_{V^{#1}_{bc}}}}
\newcommand{\ZVcb}[1]{\ensuremath{Z_{V^{#1}_{cb}}}}

\renewcommand{\case}[2]{\ensuremath{{\textstyle\frac{#1}{#2}}}}
\newcommand{\half}{\case{1}{2}}

\newcommand{\Dslash}{\ensuremath{D\kern-0.65em/\kern0.15em}}
\newcommand{\Eslash}{\ensuremath{E\kern-0.65em/\kern0.15em}}
\newcommand{\dslash}{\ensuremath{\partial\kern-0.65em/\kern0.15em}}
\newcommand{\vslash}{\ensuremath{v\kern-0.55em/\kern0.05em}}
\newcommand{\eslash}{\ensuremath{\kern0.1em\epsilon\kern-0.4em/\kern-0.1em}}
\newcommand{\lDslsh}{\ensuremath{\loarrow{D}\kern-0.65em/\kern+0.15em}}

\newcommand{\ie}{\emph{i.e.}}

\begin{document}
\title{The $B \to D \ell \nu$ form factors at nonzero recoil and $|V_{cb}|$ from $2+1$-flavor lattice QCD }
\author{Jon A.~Bailey}
\affiliation{Department of Physics and Astronomy, Seoul National University, \\ Seoul, South Korea}
\author{A.~Bazavov}
\altaffiliation[Present address:~]%
{Department of Physics and Astronomy, University of Iowa, Iowa City, IA, USA}
\affiliation{Physics Department, Brookhaven National Laboratory, Upton, New York, USA}
\author{C.~Bernard}
\affiliation{Department of Physics, Washington University, St.~Louis, Missouri, USA}
\author{C.~M.~Bouchard}
\affiliation{Physics Department, College of William and Mary, Williamsburg, VA, USA}
\affiliation{Department of Physics, The Ohio State University, Columbus, Ohio, USA}
\author{C.~DeTar}\email{detar@physics.utah.edu}
\affiliation{Department of Physics and Astronomy, University of Utah, \\ Salt Lake City, Utah, USA}
\author{Daping~Du}
\affiliation{Physics Department, University of Illinois, Urbana, Illinois, USA}
\affiliation{Department of Physics, Syracuse University, Syracuse, New York, USA}
\author{A.~X.~El-Khadra}
\affiliation{Department of Physics, University of Illinois, Urbana, Illinois, USA}
\author{J.~Foley}
\affiliation{Department of Physics and Astronomy, University of Utah, \\ Salt Lake City, Utah, USA}
\author{E.~D.~Freeland}
\affiliation{Liberal Arts Department, School of the Art Institute of Chicago, Chicago, Illinois, USA}
\author{E.~G\'amiz}
\affiliation{CAFPE and Departamento de F\'{\i}sica Te\'orica y del Cosmos, Universidad de Granada, Granada, 
Spain}
\author{Steven~Gottlieb}
\affiliation{Department of Physics, Indiana University, Bloomington, Indiana, USA}
\author{U.~M.~Heller}
\affiliation{American Physical Society, Ridge, New York, USA}
\author{J.~Komijani}
\affiliation{Department of Physics, Washington University, St.~Louis, Missouri, USA}
\author{A.~S.~Kronfeld}
\affiliation{Fermi National Accelerator Laboratory, Batavia, Illinois, USA}
\affiliation{Institute for Advanced Study, Technische Universit\"at M\"unchen, Garching, Germany}
\author{J.~Laiho}
\affiliation{Department of Physics, Syracuse University, Syracuse, New York, USA}
\author{L.~Levkova}
\affiliation{Department of Physics and Astronomy, University of Utah, \\ Salt Lake City, Utah, USA}
\author{P.~B.~Mackenzie}
\affiliation{Fermi National Accelerator Laboratory, Batavia, Illinois, USA}
\author{E.~T.~Neil}
\affiliation{Department of Physics, University of Colorado, Boulder, CO 80309, USA}
\affiliation{RIKEN-BNL Research Center, Brookhaven National Laboratory, Upton, NY 11973, USA}
\author{Si-Wei Qiu} \email{qiu@physics.utah.edu}
\altaffiliation[Present address:~]%
{Laboratory of Biological Modeling, NIDDK, NIH, Bethesda, Maryland, USA}
\affiliation{Department of Physics and Astronomy, University of Utah, \\ Salt Lake City, Utah, USA}
\author{J.~Simone}
\affiliation{Fermi National Accelerator Laboratory, Batavia, Illinois, USA}
\author{R.~Sugar}
\affiliation{Department of Physics, University of California, Santa Barbara, California, USA}
\author{D.~Toussaint}
\affiliation{Department of Physics, University of Arizona, Tucson, Arizona, USA}
\author{R.~S.~Van~de~Water}
\affiliation{Fermi National Accelerator Laboratory, Batavia, Illinois, USA}
\author{Ran Zhou}
\affiliation{Department of Physics, Indiana University, Bloomington, Indiana, USA}
\affiliation{Fermi National Accelerator Laboratory, Batavia, Illinois, USA}
\collaboration{Fermilab Lattice and MILC Collaborations}
\noaffiliation
\date{\today}

\begin{abstract}
We present the first unquenched lattice-QCD calculation of the
hadronic form factors for the exclusive decay $\overline{B}
\rightarrow D \ell \overline{\nu}$ at nonzero recoil. We carry out numerical
simulations on fourteen ensembles of gauge-field configurations
generated with 2+1 flavors of asqtad-improved staggered sea quarks.
The ensembles encompass a wide range of lattice spacings
(approximately 0.045 to 0.12~fm) and ratios of light (up and down) to
strange sea-quark masses ranging from 0.05 to 0.4.  For the $b$ and
$c$ valence quarks we use improved Wilson fermions with the Fermilab
interpretation, while for the light valence quarks we use
asqtad-improved staggered fermions.  We extrapolate our results to the
physical point using rooted staggered heavy-light meson chiral
perturbation theory.  We then parameterize the form factors and extend
them to the full kinematic range using model-independent functions
based on analyticity and unitarity.  We present our final results for
$f_+(q^2)$ and $f_0(q^2)$, including statistical and systematic
errors, as coefficients of a series in the variable~$z$ and the
covariance matrix between these coefficients.  We then fit the lattice
form-factor data jointly with the experimentally measured differential
decay rate from BaBar to determine the CKM matrix element,
$|V_{cb}|=(39.6 \pm 1.7_{\rm QCD+exp} \pm 0.2_{\rm QED})\times
10^{-3}$. As a byproduct of the joint fit we obtain the form factors
with improved precision at large recoil. Finally, we use them to
update our calculation of the ratio $R(D)$ in the Standard Model,
which yields $R(D) = 0.299(11)$.
\end{abstract}

\maketitle

%%%%%%%%%%%%%%%%%%%%%%%%%%%%%%%
\section{Introduction}%
\label{sec:intro}
%%%%%%%%%%%%%%%%%%%%%%%%%%%%%%%

Precision tests of the Standard Model (SM) seek to find discrepancies that
may indicate the presence of new physics.  The SM requirement of a
unitary Cabibbo--Kobayashi--Maskawa (CKM) weak mixing matrix provides
a good opportunity for such a test.  The unitarity-triangle test
checks the orthogonality of the first and third rows of the CKM matrix.
It requires consistency between results extracted from the experimental
measurements and theoretical calculations of a wide variety of flavor-
and $CP$-violating observables.  Although recent results have been
roughly consistent with unitarity \cite{AidaLat13,Charles:2013aka},
some disagreements persist and require further attention. The CKM
parameter $|V_{cb}|$ plays an important role in the unitarity triangle
test, since it normalizes the lengths of sides of the triangle and
contributes to tension in the unitarity constraint.

The SM parameter $|V_{cb}|$ is determined through the combination of an
experimental measurement of a branching fraction and the theoretical
calculation of the underlying hadronic form factor(s).  There are two
common approaches \cite{Amhis:2012bh} using (1) the exclusive
processes $\overline{B} \rightarrow D \ell \nu$ and $\overline{B}
\rightarrow D^* \ell \overline{\nu}$ with lattice-QCD determinations of the
relevant hadronic form factors \cite{B2Dlat13,B2Dstar} or (2) the
inclusive decay $\overline{B} \rightarrow X_c \ell \overline{\nu}$ to any
charm-containing final state $X_c$ and the operator-product and
heavy-quark expansions to describe strong-interaction
effects~\cite{Gambino:2013rza}.  It is troublesome that the most
recent results for $|V_{cb}|$ from these exclusive and inclusive
determinations disagree at the $3 \sigma$
level~\cite{Amhis:2012bh,B2Dstar}.  It is unlikely that this
difference is due to new physics effects \cite{Crivellin:2014zpa}, and
further work is needed to refine the determinations.

Reducing the error in the determination of $|V_{cb}|$ requires both
experimental and theoretical effort.  Recent work by the BaBar
collaboration \cite{Aubert:2009ac} has provided better measurements of
the decay rate. The latest results from the Belle collaboration for
this process are still preliminary \cite{BelleICHEP14}.  Further
improvements will come from Belle II.  In this work we
improve the exclusive determination of $|V_{cb}|$ from the decay
process $\overline{B} \rightarrow D \ell \overline{\nu}$ by providing the first
unquenched lattice-QCD calculation of the relevant form factors with a
complete error budget and small statistical and systematic errors.

Traditionally, experimental measurements are first extrapolated to
zero recoil, where the comparison with theory to obtain $|V_{cb}|$
occurs, using a parameterization of the momentum dependence from
Caprini, Lellouch, and Neubert (CLN)~\cite{Caprini:1997mu}.  Indeed
lattice calculations at zero recoil momentum are simpler, and for the
exclusive process $\overline{B} \rightarrow D^* \ell \overline{\nu}$ this method
yields a very precise determination of $|V_{cb}|$, which is described
in our companion work \cite{B2Dstar}.  However, due to the more severe
phase space suppression of the $\overline{B} \rightarrow D \ell \overline{\nu}$
rate near zero recoil (compared with $\overline{B} \rightarrow D^*
\ell \overline{\nu}$) the extrapolation of the experimental data to zero recoil
is a source of significant uncertainty. This results in determinations
of $|V_{cb}|$ from $\overline{B} \rightarrow D \ell \overline{\nu}$ that are less
precise than they have to be.  Here we calculate the form factors for
$\overline{B} \rightarrow D \ell \overline{\nu}$ for a range of recoil momenta
and parameterize their dependence on momentum transfer using the
model-independent $z$-expansion of Boyd, Grinstein and Lebed
\cite{Boyd:1994tt}.  We fit the experimental and lattice data together
as a function of momentum transfer and determine $|V_{cb}|$ from the
relative normalization over the entire range of recoil momenta.

Where previous calculations of this process at nonzero recoil ignored
effects of sea quarks~\cite{deDivitiis:2007ui}, ours includes them.
The use of asqtad-improved staggered fermions and improved Wilson
(``clover") quarks reduces lattice discretization errors.  A
preliminary determination of $|V_{cb}|$ from $\overline{B} \rightarrow
D \ell \overline{\nu}$ with a very small subset of the present asqtad ensembles
was presented in Ref.~\cite{Okamoto:2004xg}.  That calculation was
done only at zero recoil and used only lattices with spacing
approximately 0.12~fm, so a continuum extrapolation was not
possible. The broad range of lattice spacings and sea-quark-mass
ratios in our current study gives confidence in the extrapolation to
physical quark masses and zero lattice spacing.  More recently, in a
related project of ours \cite{Bailey:2012rr}, the ratio of form
factors for $B_s \to D_s \ell\nu$ to $B \to D \ell\nu$ decays was
obtained using only four asqtad ensembles, \ie, with two different
light sea-quark masses at two lattice spacings.  This data set
was also used to obtain the first Standard-Model prediction for $R(D)
\equiv {\cal B}(B\to D \tau \nu )/{\cal B}(B \to D \ell\nu ) $ from
unquenched lattice QCD in Ref.~\cite{Bailey:2012jg}. The present work uses
all fourteen ensembles and uses a slightly different analysis. These
are the first such calculations to combine all of the ingredients
listed above.

Preliminary results for the $B\to D$ form factors from this project
were presented in~\cite{B2Dlat13}.  The final analysis presented here
includes a more sophisticated treatment of the matching factors as
well as more refined estimates for the renormalization and heavy-quark
discretization errors.

This article is organized as follows.  In Sec.~\ref{sec:formalism} we
review the formalism and our strategy for extracting the form factors
at nonzero recoil.  In Sec.~\ref{sec:analysis} we give details of the
ensembles and simulation and discuss our determination of the form
factors and the chiral-continuum extrapolation.  We discuss systematic
errors in Sec.~\ref{sec:err}. In Sec.~\ref{sec:Vcb} we present fits to
our lattice data for the two form factors $f_+$ and $f_0$ and a joint
fit that combines our lattice data with the 2009 BaBar measurements
\cite{Aubert:2009ac}, leading, finally, to our result for $|V_{cb}|$.
We discuss our results in Sec.~\ref{sec:discussion}.
Appendix~\ref{app:kappa} discusses technical details regarding the
tuning of the bare-quark masses.  Appendix~\ref{app:hqerror} derives
the pattern of heavy-quark discretization effects and discusses some
details of matching lattice gauge theory with heavy quarks to
continuum QCD.

%%%%%%%%%%%%%%%%%%%%%%%%%%%%%%%
\section{Form factors}
\label{sec:formalism}
%%%%%%%%%%%%%%%%%%%%%%%%%%%%%%%

\subsection{Continuum form factors}

The hadronic interaction in the process $\overline{B} \rightarrow D
\ell \overline{\nu}$ is determined by the transition matrix element of the vector
current $\mathcal{V}^\mu = \bar c \gamma^\mu b$, which is
conventionally decomposed in terms of the vector and scalar form
factors $f_+(q^2)$ and $f_0(q^2)$ as
\begin{equation}
    \langle D(p_D) | \mathcal{V}^\mu | B(p_B) \rangle =
        f_+(q^2) \left[ (p_B+p_D)^\mu - \frac{M_B^2-M_D^2}{q^2}q^\mu \right] +
        f_0(q^2) \frac{M_B^2-M_D^2}{q^2} q^\mu \, .
\end{equation}
Here $p_B$ and $p_D$ are the momenta of the $B$ and $D$ mesons, $M_B$
and $M_D$ are the respective masses, and $q = p_B - p_D$ is the
momentum transferred to the leptons.  In the approximation that the
masses of the leptons $\ell = e, \mu, \nu_e, \nu_\mu$ are much smaller
than the $B$ and $D$ mass difference $M_B - M_D$, the differential
decay rate is
\begin{eqnarray}
    \frac{d\Gamma}{dw}(\overline{B}\rightarrow D \ell \overline{\nu})=
    |\bar \eta_{\rm EW}|^2
    \frac{G_F^2|V_{cb}|^2 M_B^5}{48\pi^3}(w^2-1)^{3/2}r^3(1+r)^2\mathcal{G}(w)^2\,\,,
\label{eq:diffrate}
\end{eqnarray}
where $|\bar\eta_{\rm EW}|^2$ accounts for electroweak corrections
discussed below, $G_F$ is the Fermi weak decay constant, $|V_{cb}|$ is
the desired CKM matrix element, $w = v\cdot v^\prime$ is the recoil
parameter, $v = p_B/M_B$ and $v^\prime = p_D/M_D$ are the hadronic
velocities, and $\mathcal{G}$ is related to $f_+$ through
\begin{equation}
    f_+(w)^2=\frac{(1+r)^2}{4r}\mathcal{G}(w)^2\,\,.
\end{equation}
for $r = M_D/M_B = 0.354$.

The alternative parameterization in terms of the form factors $h_+$
and $h_-$ is convenient in heavy-quark effective theory (HQET) and
heavy-light meson chiral perturbation theory:
\begin{equation}
    \frac{\langle D(p_D) | \mathcal{V}^\mu | B(p_B) \rangle }{ \sqrt{M_B M_D}} =
        h_+(w)(v + v^\prime)^\mu  + h_-(w) (v-v^\prime)^\mu \, .
    \label{eq:h+-}
\end{equation}
These form factors are related to $f_+$ and $f_0$ through
\begin{eqnarray}
    f_+(q^2) & = & \frac{1}{2\sqrt{r}} \left[ (1+r) h_+(w) - (1-r) h_-(w) \right] , \label{eq:f+fromh} \\
    f_0(q^2) & = & \sqrt{r} \left [ \frac{w+1}{1+r} h_+(w) - \frac{w-1}{1-r} h_-(w) \right ],  \label{eq:f0fromh}
\end{eqnarray}
where $ q^2=M_B^2+M_D^2-2wM_BM_D $.  We note, also, the kinematic
constraint $f_+(0) = f_0(0)$ at $q^2 = 0$, which corresponds to $w =
(M_B^2+M_D^2)/(2M_BM_D) \approx 1.59$. We also have
\begin{equation}
    \mathcal{G}(w) = h_+(w) - \left(\frac{1-r}{1+r}\right)h_-(w)  \,.
    \label{eq:G=h}
\end{equation}

\subsection{Form factors from lattice matrix elements}
\label{sec:lattice_formalism}

We use the local Fermilab-improved vector current for the quark
transition $x\to y$
\begin{equation}
    V^\mu_{xy} = \bar{\Psi}_x\gamma^\mu\Psi_y,
    \label{eq:Vlat}
\end{equation}
where the subscripts denote flavor, $\Psi$ is the ``rotated'' field \cite{ElKhadra:1996mp} 
\begin{equation}
    \Psi = \left(1+d_1\bm{\gamma}\cdot\bm{D}_\text{lat}\right)\psi,
    \label{eq:Psi}
\end{equation}
and $\psi$ is the heavy-quark field in the action.
The lattice current $V^\mu$ is related to the continuum current $\mathcal{V}^\mu$ through
\begin{equation}
  Z^\mu_{xy} V^\mu_{xy} \doteq \mathcal{V}^\mu_{xy} \,,
\end{equation}
where ``$\doteq$" denotes the equality of matrix elements.  Following
\cite{ElKhadra:2001rv,Harada:2001fj}, we define the correction
matching factor as the double ratio of matching factors for flavor
off-diagonal currents to those for flavor-diagonal currents:
\begin{equation}
    \rhoV{\mu}^2 = \frac{\ZVcb{\mu}\ZVbc{\mu}}{\ZVcc\ZVbb},
    \label{eq:rhoV_ratio}
\end{equation}
where $\rhoV{\mu} = 1 +
4\pi\alpha_s(q^*)\rhoV{\mu}^{[1]} + {\cal O}(\alpha_s(q^*)^2)$ is
determined to one-loop order in lattice perturbation theory
\cite{Harada:2001fj}.  It is found to be quite close to 1 because of
cancellations in the ratio of similar quantities, including
cancellations of tadpole diagrams.  The truncation error is expected to
be small because $\alpha_s(q^*=2/a) \approx 0.2$.  

The matching factor $\rho_{V^\mu}(w)$ depends, in principle, upon the
velocity transfer $w$.  At present we have calculated only
$\rhoV{4}(1)$ for the quark masses and lattice spacings in our
project.  Calculation of the spatial correction $\rhoV{i}$ is more
difficult because, even for zero recoil, one must calculate it for
nonzero momentum.  Thus we have calculated $\rho_{V^i}(1)$ only for
the simpler case $m_c a = 0$, but our lack of knowledge of the $m_c$
dependence of the one-loop correction to $\rhoV{i}$ makes only a small
contribution to our final uncertainty. The $w$ dependence of
$\rho_{V^i}$ is also unavailable.  Below we note where these issues
arise.

To compute the form factors $h_+$ and $h_-$ at arbitrary recoil, we
need the lattice matrix elements of both the temporal and spatial
vector currents, $V^4$ and $\bm{V}$.  In practice, we use ratios of
lattice correlators in which the flavor-conserving renormalization
factors are automatically included, as discussed below.  These ratios
also suppress statistical fluctuations and systematic errors.  The
remaining correction factors $\rhoV{4}$ and $\rhoV{i}$ are applied
after fitting the ratios.  We apply this correction in
Sec.~\ref{sec:chiral}.

Our calculation is done in the $B$-meson rest frame for any recoil
$D$-meson momentum $\bm{p}$.  We compute the double ratio
\begin{equation}
    R_+ = \frac{ \langle D (\bm{0})| V_{cb}^4 |B (\bm{0}) \rangle 
                \langle B(\bm{0}) | V_{bc}^4 |D(\bm{0}) \rangle }
    { \langle D(\bm{0}) | V_{cc}^4 |D(\bm{0}) \rangle 
      \langle B(\bm{0}) | V_{bb}^4 |B(\bm{0}) \rangle } \,\, 
\label{eq:Rplus}
\end{equation}
and the single ratios
\begin{eqnarray}
    Q_+(\bm{p}) & \equiv &
      \frac{\langle D(\bm{p})|V^4|B(\bm{0})\rangle}
           {\langle D(\bm{0})|V^4|B(\bm{0})\rangle} \,,
    \label{eq:Qplus} \\
    \bm{R}_-(\bm{p}) &\equiv& \frac{\langle D(\bm{p})|\bm{V}|B(\bm{0})\rangle}
    {\langle D(\bm{p})|V^4|B(\bm{0})\rangle} \, ,
    \label{eq:Rminus} \\
        \bm{x}_f(\bm{p}) &\equiv& \frac{\langle D(\bm{p})|\bm{V}|D(\bm{0})\rangle}
	{\langle D(\bm{p})|V^4|D(\bm{0})\rangle} \, .
    \label{eq:xf} 
\end{eqnarray}
Note that $Q_+(\bm{p})$ is the ratio of $B \to D$ matrix elements
at nonzero and zero recoil, and that $\bm{x}_f(\bm{p})$ is computed
only from the flavor-diagonal transition $D\to D$.  As spelled out
below, we use $R_+(\bm{p})$, $Q_+(\bm{p})$, and
$\bm{R}_-(\bm{p})$ to obtain $h_+(w)$ and $h_-(w)$, and
$\bm{x}_f(\bm{p})$ to obtain the recoil $w$. The flavor-conserving
renormalization factors $\ZVbb$ and $\ZVcc$ cancel exactly
in the double ratio $R_+$, which was introduced by Hashimoto {\it et
  al.} and used to obtain the $B\to D \ell\nu$ form factor at zero
recoil in quenched lattice QCD~\cite{Hashimoto:2001nb}.

%More
%generally, working with such ratios of similar quantities throughout
%reduces statistical and systematic errors significantly.

From Eq.~(\ref{eq:x=v}), the 3-vector $\bm{x}_f$ yields the velocity
without any matching ambiguities:
\begin{equation}
    \bm{x}_f = \frac{\bm{v}'}{w+1}\,.
\end{equation}
Because $w^2=1+{\bm{v}'}^2$ (when the initial meson is at rest), one finds
\begin{equation}
    w(\bm{p}) = \frac{1+\bm{x}_f^2(\bm{p})}{1-\bm{x}_f^2(\bm{p})}\,.
    \label{eq:w}
\end{equation}
Thus, even the kinematic variable $w$ is determined dynamically from a ratio of matrix elements.

The other ratios require matching factors.
It is convenient to define
\begin{eqnarray}
    \mathcal{R}_+          & = & \rhoV{4}^2(1) R_+ \,,
    \label{eq:renormR+} \\
    \mathcal{Q}_+      (\bm{p}) & = & \frac{\rhoV{4}(w)}{\rhoV{4}(1)} Q_+(\bm{p}) \,,
    \label{eq:renormQ+} \\
    \bm{\mathcal{R}}_- (\bm{p}) & = & \frac{\rhoV{i}(w)}{\rhoV{4}(w)} \bm{R}_-(\bm{p}) \,.
    \label{eq:renormR-} 
\end{eqnarray}
We derive these factors and discuss how we handle them in Appendix~\ref{app:hqerror}.
Note that $\mathcal{R}_+$ reduces to
\begin{equation}
    \sqrt{\mathcal{R}_+} = h_+(1) + \text{matching \& discretization errors}\,.
    \label{eq:h+(1)}
\end{equation}
Also, $\mathcal{Q}_+(\bm{0})=Q_+(\bm{0})=1$ by construction.
We then can obtain $h_+$ and $h_-$ from
\begin{eqnarray}
    h_+\left(w(\bm{p})\right) & = & 
        \sqrt{\mathcal{R}_+} \mathcal{Q}_+(\bm{p}) \left[ 1 - \rule{0pt}{1.2em}
        \bm{\mathcal{R}}_-(\bm{p})\cdot\bm{x}_f(\bm{p})\right] \,, 
    \label{eq:hplusQ} \\
    h_-\left(w(\bm{p})\right) & = & 
        \sqrt{\mathcal{R}_+} \mathcal{Q}_+(\bm{p}) \left[ 1 -
        \frac{\bm{\mathcal{R}}_-(\bm{p})\cdot\bm{x}_f(\bm{p})}{\bm{x}^2_f(\bm{p})}\right] \,,
    \label{eq:hminusQ}
\end{eqnarray}
as in Eq.~(\ref{eq:h+(1)}) up to matching and discretization errors.

%%%%%%%%%%%%%%%%%%%%%%%%%%%%%%%
\section{Analysis}
\label{sec:analysis}
%%%%%%%%%%%%%%%%%%%%%%%%%%%%%%%

\subsection{Lattice action and parameters}%

\begin{figure}
    \includegraphics[width=0.5\textwidth]{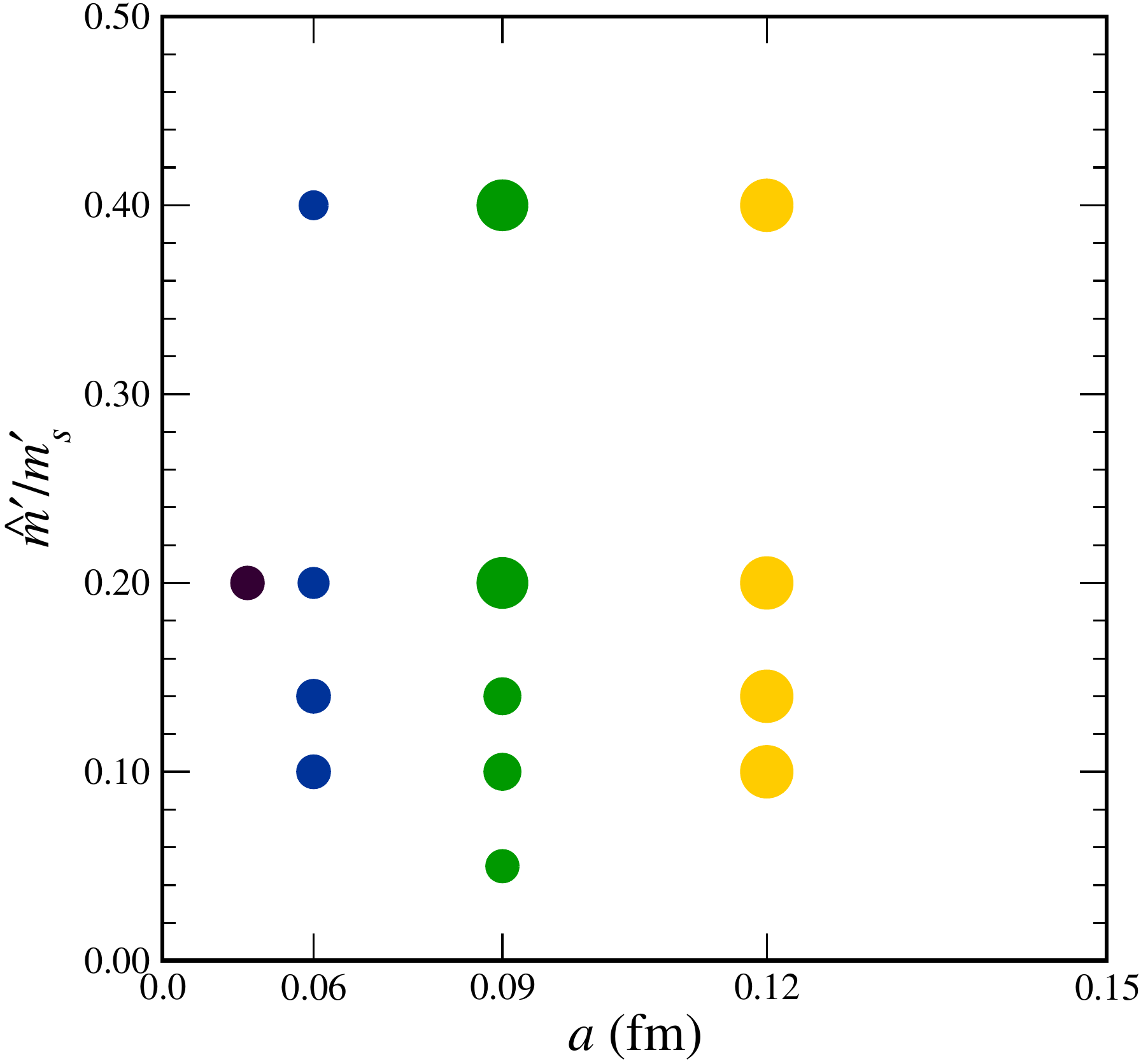}
    \caption[fig:a-vs-m]{(color online) Range of lattice spacings and
      light-quark masses used here.  The area of each disk is
      proportional to the number of configurations in the ensemble.}
    \label{fig:a-vs-m}
\end{figure}

Our calculation uses fourteen ensembles of gauge-field configurations
generated in the presence of 2+1 flavors of asqtad-improved staggered
sea quarks by the MILC collaboration \cite{Bazavov:2009bb}.  Ensembles
are indicated graphically in Fig.~\ref{fig:a-vs-m}, and they are
tabulated in Table~\ref{tab:params}. There are four lattice spacings,
$a \approx 0.12$ fm, 0.09 fm, 0.06 fm, and 0.045 fm, and light
sea-quark to strange sea-quark mass ratios $\hat{m}^\prime/m^\prime_s$
ranging from $0.4$ to $0.05$.  The strange sea-quark mass is set
approximately to its physical value.  For the light valence quarks we
use the asqtad action.  Light-quark propagators are converted to
improved ``naive'' propagators as in Ref.~\cite{Wingate:2002fh} to
implement the standard Dirac spin algebra.  In this study, masses of
the light valence quarks are always equal to the sea-quark masses.
For the heavy valence quarks we use the Fermilab interpretation of the
clover action with the parameters listed in Table
  \ref{tab:params2}.

Two-point and three-point correlators are computed from four
equally-spaced source times per configuration, but with random offsets
in time and space to reduce correlations between successive
gauge-field configurations within an ensemble.  We performed a
blocking study to look for residual autocorrelations, and found that
the statistical errors did not change significantly with block size.
Thus we do not block the data in this work. The masses of the heavy
valence quarks were tuned so that the kinetic masses of the $D_s$ and
$B_s$ mesons were equal to their physical values.  A detailed
discussion of tuning is given in the appendix of Ref.~\cite{B2Dstar},
where we show that we get good agreement between the lattice values of
the $D_s$ and $B_s$ hyperfine splittings and their experimental
values. The simulation values of the heavy-quark masses are not quite
the same as our best-tuned values, which were determined {\it a
  posteriori}. Post-simulation adjustment for heavy-quark-mass tuning
is described in Sec.~\ref{sec:kappa}.

After fixing the lattices to Coulomb gauge, two types of interpolating
operators for the $D$ meson are used, namely, a local operator and a
smeared operator based on a Richardson 1S wave
function~\cite{Menscher:2005kj}.  For the $B$ meson we use only the
$1S$ operator. These two operators have different overlap with excited
states, so computing both helps us remove excited-state contributions.
We generate three-point functions in a standard way by fixing the
position of the $D$ and $B$ mesons to a separation $T$ in imaginary
time and then varying the time $t$ of the vector current.
Calculations at two adjacent time separations $T$ are carried out in
each case to control the effects of oscillating staggered-fermion
propagators.  We rotate the heavy-quark fields as in
Eq.~(\ref{eq:Psi}) using the tadpole-improved tree-level values for
$d_1$ listed in Table~\ref{tab:params2}, so that the vector
current is tree-level improved.  Calculations are made at several
choices of three-momentum.  In units of $2\pi /L$, for this study
we use five momenta (0,0,0), (0,0,1), (0,1,1), (1,1,1), and (2,0,0).
Results at larger momenta tend to have significantly larger
statistical errors, and also suffer from larger momentum-dependent
discretization errors.  In the two-point correlator these momenta are
projected at the sink and in the three-point correlator, at the
current.  In the latter case the three-momentum of the $B$ meson is
set to zero.

\begin{table}
    \centering
    \caption{Parameters of the lattice-gauge-field ensembles. The columns from left to right are 
        the approximate lattice spacing in fm, 
                the bare sea-quark masses in lattice units $ a\hat{m}^\prime/am^\prime_s $, 
        the lightest pseudoscalar in MeV, 
        the root-mean-square (RMS) mass of the pion taste multiplet in MeV, 
        the dimensionless factor $M^P_\pi L $, 
        the dimensions of the lattice in lattice units, 
        the number of configurations in each ensemble (four sources each), 
        and the tadpole-improvement factor~$ u_0 $ (obtained from the average plaquette).}
    \label{tab:params}
    \begin{tabular}{l@{~}r@{/}l@{\quad}c@{\quad}c@{\quad}c@{\quad}c@{\quad}c@{\quad}r@{~}l}
    \hline\hline
$a$~(fm) & $a\hat{m}'$ & $am'_s$ & $M_{\pi}^{P}$~{(MeV)} & $M_{\pi}^{\rm RMS}$~{(MeV)} & ${M^P_\pi} L$ & 
    {Lattice size} & Configs & ~~$u_0$  \\ 
    \hline
${\approx}~0.12\hs$  & $0.02$    & $0.05$  & 560 & 670 & 6.2 & \ $20^3\times 64$ & 2052 & 0.8688   \\ 
          & $0.01$    & $0.05$  & 390 & 540 & 4.5 & \ $20^3\times 64$ & 2259 & 0.8677   \\ 
          & $0.007$   & $0.05$  & 320 & 500 & 3.8 & \ $20^3\times 64$ & 2110 & 0.8678   \\ 
          & $0.005$   & $0.05$  & 270 & 470 & 3.8 & \ $24^3\times 64$ & 2099 & 0.8678   \\ 
    \hline                            
${\approx}~0.09\hs$ & $0.0124$  & $0.031$ & 500 & 550 & 5.8 & \ $28^3\times 96$ & 1996 & 0.8788   \\ 
          & $0.0062$  & $0.031$ & 350 & 420 & 4.1 & \ $28^3\times 96$ & 1931 & 0.8782   \\ 
          & $0.00465$ & $0.031$ & 310 & 380 & 4.1 & \ $32^3\times 96$ & 984  & 0.8781   \\ 
          & $0.0031$  & $0.031$ & 250 & 330 & 4.2 & \ $40^3\times 96$ & 1015 & 0.8779   \\ 
          & $0.00155$ & $0.031$ & 180 & 280 & 4.8 & \ $64^3\times 96$ & 791  & 0.877805 \\ 
    \hline                            
${\approx}~0.06\hs$ & $0.0072$  & $0.018$ & 450 & 470 & 6.3 & \ $48^3\times144$ & 593  & 0.8881   \\ 
          & $0.0036$  & $0.018$ & 320 & 340 & 4.5 & \ $48^3\times144$ & 673  & 0.88788  \\ 
          & $0.0025$  & $0.018$ & 260 & 290 & 4.4 & \ $56^3\times144$ & 801  & 0.88776  \\ 
          & $0.0018$  & $0.018$ & 220 & 260 & 4.3 & \ $64^3\times144$ & 827  & 0.88764  \\ 
    \hline                            
${\approx}~0.045\hs$ & $0.0028$  & $0.014$ & 320 & 330 & 4.6 & \ $64^3\times192$ & 801  & 0.89511  \\ 
    \hline\hline
    \end{tabular}
\end{table}

\begin{table}
\centering
    \caption{ Parameters of the heavy valence quarks.  The approximate
      lattice spacing and bare sea-quark masses in the first two
      columns identify the ensemble.  The remaining columns show the
      coefficient of the clover term in the SW action $c_{\rm SW}$,
      the bare hopping-parameter $\kappa$, and the rotation parameter
      in the current $d_1$. The primes on $\kappa$
      distinguish the simulation from the physical values.}
        \vspace*{4pt}
    \label{tab:params2}
    \begin{tabular}{c@{\quad}r@{/}l@{\quad}l@{\quad}l@{\quad}l@{\quad}l@{\quad}l}
      \hline \hline
       $\approx a$ (fm) & $a\sealight$&$a\seaheavy$ & \multicolumn{1}{c}{$c_\text{SW}$~~} 
           & \multicolumn{1}{c}{$\kappa'_b$~~~} & \multicolumn{1}{c}{$d_{1b}$~~~}    
           & \multicolumn{1}{c}{$\kappa'_c$~~~} & \multicolumn{1}{c}{$d_{1c}$} \\
      \hline
      $0.12\hs$ & $0.02$&$0.05$      & 1.525  & 0.0918 & 0.09439 & 0.1259 & 0.07539 \\
      $0.12\hs$ & $0.01$&$0.05$      & 1.531  & 0.0901 & 0.09334 & 0.1254 & 0.07724 \\
      $0.12\hs$ & $0.007$&$0.05$     & 1.530  & 0.0901 & 0.09332 & 0.1254 & 0.07731 \\
      $0.12\hs$ & $0.005$&$0.05$     & 1.530  & 0.0901 & 0.09332 & 0.1254 & 0.07733 \\
      \hline
      $0.09\hs$ & $0.0124$&$0.031$   & 1.473  & 0.0982 & 0.09681 & 0.1277 & 0.06420 \\
      $0.09\hs$ & $0.0062$&$0.031$   & 1.476  & 0.0979 & 0.09677 & 0.1276 & 0.06482 \\
      $0.09\hs$ & $0.00465$&$0.031$  & 1.477  & 0.0977 & 0.09671 & 0.1275 & 0.06523 \\
      $0.09\hs$ & $0.0031$&$0.031$   & 1.478  & 0.0976 & 0.09669 & 0.1275 & 0.06537 \\
      $0.09\hs$ & $0.00155$&$0.031$  & 1.4784 & 0.0976 & 0.09669 & 0.1275 & 0.06543 \\
      \hline
      $0.06\hs$ & $0.0072$&$0.018$   & 1.4276 & 0.1048 & 0.09636 & 0.1295 & 0.05078 \\
      $0.06\hs$ & $0.0036$&$0.018$   & 1.4287 & 0.1052 & 0.09631 & 0.1296 & 0.05055 \\
      $0.06\hs$ & $0.0025$&$0.018$   & 1.4293 & 0.1052 & 0.09633 & 0.1296 & 0.05070 \\
      $0.06\hs$ & $0.0018$&$0.018$   & 1.4298 & 0.1052 & 0.09635 & 0.1296 & 0.05076 \\
      \hline
      $0.045$   & $0.0028$&$0.014$   & 1.3943 & 0.1143 & 0.08864 & 0.1310 & 0.03842 \\
      \hline \hline
    \end{tabular}

\end{table}

\subsection{Fitting strategy}
\label{formfactor}

We need both two-point and three-point correlation functions to
construct the form factor introduced in
Sec.~\ref{sec:lattice_formalism}.  We use interpolating operators
$\mathcal{O}_{Xa}(\bm{p},t) $ of spatial momentum $\bm{p}$ and time
$t$ with $ X \in \{B, D\}$ and $ a \in \{1S,d\}$.  The notation $d$
signifies a delta function (point) source, while $1S$ denotes a 1S
Richardson wavefunction.  See Ref.~\cite{Bazavov:2011aa} for details.
The correlation functions can be expressed in terms of operator matrix
elements:
\begin{eqnarray}
    C^{2\text{pt},Xa\rightarrow Xb}(\bm{p},t)  &=&
      \langle\mathcal{O}_{Xb}^\dag (\bm{p},0)\mathcal{O}_{Xa}(\bm{p},t)\rangle\, ,\\
    C^{3\text{pt},Xa\rightarrow Yb}_\mu(\bm{p},t) &=&
   \langle\mathcal {O}_{Yb}^\dag (-\bm{p},0)
     V^{\mu}(\bm{p},t)
       \mathcal{O}_{Xa}(\bm{0},T)\rangle\,\,,
\end{eqnarray}
where $T$ is the imaginary time separation between the $B$ and $D$
mesons.  

The spectral decomposition of the two-point correlator is
\begin{equation}
    C^{2\text{pt},Xa\rightarrow Xb}(\bm{p},t) =  \sum_n s_n(t) 
     \frac{\sqrt{Z_{Xa,n}(\bm{p})Z_{Xb,n}(\bm{p})}}{2E_n(\bm{p})}
     \left[\exp(-E_n(\bm{p})t)+\exp(-E_n(\bm{p})(N_t-t))\right]\,\, ,
\end{equation}
where there are either nonoscillating terms with $ s_n(t)= 1 $ or
staggered-fermion opposite-parity oscillating terms $s_n(t) =
-(-1)^t$, $ N_t $ is the lattice extent in time, and $ Z_{Xa,n} $ is
the overlap coefficient. For the three-point function, the
decomposition is similar:
\begin{eqnarray}
    C^{3\text{pt},Xa\rightarrow Yb}_\mu(\bm{p},t) &=& \sum_{n,m} s_n(t) s_m(T-t)
    \sqrt{Z_{Yb,n}(\bm{p})}
    \frac{e^{-E_n(\bm{p}) t}}{\sqrt{2E_n(\bm{p})}} 
    \langle Yb,n(\bm{p})|V^\mu|Xa,m(\bm{0})\rangle \\
   &\times&  \frac{e^{-M_m(T-t)}} {\sqrt{2M_m}}
    \sqrt{Z_{Xa,m}(\bm{0})}     \,\,, \nonumber
\end{eqnarray}
where we have assumed $t < T \ll N_t$, so we may neglect wraparound terms
with $t \rightarrow N_t - t$ and $T-t \rightarrow N_t - (T-t)$.

The double ratio $R_+$ can be calculated very precisely from
\begin{equation}
  R_{+,b}(t,T) = 
\frac{
    C^{3\text{pt},B,1S\rightarrow Db}_4(\bm{0},t)
    C^{3\text{pt},Db \rightarrow B,1S}_4(\bm{0},t)
    }{
    C^{3\text{pt},Db\rightarrow Db}_4(\bm{0},t)
    C^{3\text{pt},B,1S\rightarrow B,1S}_4(\bm{0},t)} \,.
 \label{eq:doubleratio}
\end{equation}
This quantity depends on $t$, $T$ and the $D$-meson interpolating
operator, labeled by $b$.  The dependence arises from contributions
from excited states and opposite-parity oscillating states. 
As in Refs.~\cite{B2Dstar,Bernard:2008dn,Bailey:2008wp} we suppress
contributions from oscillating states by averaging
\begin{equation}
  \bar R_{+,b}(t,T) \equiv \frac{1}{2} R_{+,b}(t,T) +
     \frac{1}{4} R_{+,b}(t,T+1) + \frac{1}{4} R_{+,b}(t+1,T+1) \,\, .
    \label{eq:R+_avg}
\end{equation}
We drop the bar henceforth.  We use a similar method for the other
three-point correlation functions. We find that the suppression of
oscillating states for $B\to D$ correlators is similar to that of our
previous work on $B\to D^* \ell \nu$~\cite{B2Dstar}.  In particular,
the contribution from the first oscillating $B$- and $D$-meson excited
states, which does not itself oscillate in time, is reduced by a
factor of $\sim 5$--12 using the average in Eq.~(\ref{eq:R+_avg}),
where greater suppression occurs for finer lattice spacings.

For large $t$ and $T - t$, excited-state
contributions are negligible, giving the desired result,
\begin{equation}
  R_{+,b}(t,T) \rightarrow R_+ \,\, ,
\end{equation}
as a plateau in the ratio {\it vs.} $t$, as illustrated in
Fig.~\ref{fig:Rplussample}. 
The leading corrections to the plateau arise from contributions from
the first excited $D$- and $B$-meson states.  For large $t$ and $T -
t$, their contributions to the correlator double ratio fall off as
$\exp[-\Delta M t]$ and $\exp[-\Delta M(T-t)]$, where $\Delta M =
\Delta M_B$ or $\Delta M_D$, the splitting between the ground state
and first excited state of the $B$- and $D$-mesons, respectively.
Since they are both small, for fitting the ratio, we use the
approximation
\begin{eqnarray}
   R_{+,b}(t,T) &\approx& R_+
    + A_{R_+,b} \exp(-\Delta M_D t) + B_{R_+,b}\exp[-\Delta M_B (T-t)] \nonumber \\
   &+&  C_{R_+,b}\exp(-\Delta M_B t) + D_{R_+,b} \exp[-\Delta M_D (T-t)]
    \exp(\Delta M_B t)\, ,
\label{eq:Rplusratiofit}
\end{eqnarray}
However, since $\Delta M_D \approx \Delta M_B$ we construct the fit
model from only the $R_+$, $A$, and $B$ terms.

Similarly, we introduce a time- and interpolating-operator-dependent
ratio
\begin{equation}
   Q_{+,b}(\bm{p},t,T) \equiv 
    \frac{C^{3\text{pt},B,1S\rightarrow Db}_4(\bm{p},t)}
         {C^{3\text{pt},B,1S\rightarrow Db}_4(\bm{0},t)}
    \frac{E_D Z_{Db}(\bm{0})}{M_D Z_{Db}(\bm{p})}
    e^{(E_D-M_D)t} \,.
 \label{eq:Rplusratio2}
\end{equation}
In this ratio the plateau takes on the value $Q_+(\bm{p})$
introduced in Eq.~(\ref{eq:Qplus}).  Again, the leading corrections to
the plateau arise from contributions from the first excited $D$- and
$B$-meson states.  For large $t$ and $T - t$, their contributions to
the correlator ratio fall off as $\exp[-\Delta E_D t]$ and
$\exp[-\Delta M_B(T-t)]$.  Where they are both small, for fitting the
ratio, we use the approximation
\begin{eqnarray}
   Q_{+,b}(\bm{p},t,T) &\approx& Q_+(\bm{p})
    \exp(\delta m\, t) + A_{Q_+,b}(\bm{p}) \exp(-\Delta E_D t) \nonumber \\
    &+&  B_{Q_+,b}(\bm{p}) \exp[-\Delta M_D t] + C_{Q_+,b}(\bm{p}) \exp[-\Delta M_B (T-t)]  \, .
\label{eq:Qplusfit}
\end{eqnarray}
The parameter $\delta m$ vanishes when the exponential factor in
Eq.~(\ref{eq:Rplusratio2}) cancels the time dependence in the
three-point functions, as it should.  Since there may be slight
differences in the determination of the masses from the three-point
and two-point functions, the cancellation might not be perfect.
Therefore, we introduce $\delta m$ as a constrained fitting
parameter. The prior constraint is centered at zero and it has a width
determined from the small statistical error in the two-point-fitted
energies. In practice, the values of $a \delta m$, are typically of
order $10^{-4}$.

For $R_{-,b}^i(\bm{p},t,T)$ we form the ratio
\begin{equation}
   R_{-,b}^i(\bm{p},t,T) = \frac{
    C^{3\text{pt},B,1S\rightarrow Db}_i(\bm{p},t)
    }{
    C^{3\text{pt},B,1S\rightarrow Db}_4(\bm{p},t)
    }   \,\,,
\end{equation}
so that for large $t$ and $T-t$ we have $ R_{-,b}(\bm{p},t,T)
\rightarrow R_{-}(\bm{p}) $.
Similarly, for $\bm{x}_f(\bm{p},t,T)$, we use the ratio
\begin{equation}
   x_{f,b}^i(\bm{p},t,T) = \frac{
    C^{3\text{pt},Db\rightarrow Db}_i(\bm{p},t)
    }{
    C^{3\text{pt},Db\rightarrow Db}_4(\bm{p},t)
    }  \,\,,
\end{equation}
so that $\bm{x}_{f,b}(\bm{p},t,T) \rightarrow \bm{x}_{f}(\bm{p})$.  To
fit the time dependence of $\bm{R}_{-,b}(\bm{p},t,T)$ we use 
\begin{equation}
  \bm{R}_{-,b}(\bm{p},t,T) \approx \bm{R}_-(\bm{p}) + 
  \bm{A}_{R_-,b}(\bm{p}) \exp(-\Delta E_D t) + 
  \bm{B}_{R_-,b}(\bm{p}) \exp[-\Delta M_B (T-t)]  \, ,
\label{eq:xfRminusfit}
\end{equation}
and for the time dependence of $\bm{x}_f(\bm{p},t,T)$, we use the same
form, except replacing $\Delta M_B$ with $\Delta M_D$.

\subsection{Correlator Fitting}

We obtain the lattice form factors via a two-step procedure. First, we
fit the $ B $- and $ D $-meson two-point correlators to obtain the
energies and overlap factors.  Then we use these results as
constraints with Bayesian priors in the three-point fits.  Errors in
the resulting form factors $h_+$ and $h_-$ are determined from a
complete single-elimination jackknife procedure.

\begin{table}
    \centering
\caption{Comparison of ground-state energies $E_D$, excited-state
  energies $E_D^\prime$, and ground-state overlap factors $Z_d$ and
  $Z_{1S}$ for the $(2+2)$-state and $(3+3)$-state
  two-point-correlator fits for the $D$ meson on the $a \approx 0.12$
  fm, $\hat m^\prime = 0.14m_s^\prime$ ensemble. In all cases the
  $(2+2)$-state fitting range is [6, 16] and the $(3+3)$-state fitting
  range is [4, 23].  We use (3+3)-state fits for the analysis; the
  (2+2)-state fits just provide a check of systematic effects.}
    \label{tab:coarse_2pts}
    \begin{tabular}{lll@{\quad}llllll}
        \hline\hline
             & \multicolumn{2}{c}{Ground state ($aE_D$)}  & \multicolumn{2}{c}{$1^{st}$ excited state ($aE_D^\prime$)} &   &   & 
             \multicolumn{2}{c}{$\chi^2/df$}  \\
         $p$~$(2\pi/L)$ & \multicolumn{1}{c}{$2+2$}  & \multicolumn{1}{c}{$3+3$} & \multicolumn{1}{c}{$2+2$} & 
            \multicolumn{1}{c}{$3+3$} & \multicolumn{1}{c}{$Z_{1S,1S}$} & \multicolumn{1}{c}{$Z_{d,d}$} & 
            \multicolumn{1}{c}{$2+2$} & \multicolumn{1}{c}{$3+3$}  \\
        \hline
        000  &  0.9566(6)  & 0.9566(7)  & 1.54(3) & \qquad 1.41(4) & 4.045(29) & 0.0785(7)  & 18.9/18 & 35.5/37\\
        100  &  1.0013(10) & 1.0017(9)  & 1.55(2) &  \qquad 1.39(4) & 2.912(36) & 0.0741(10) & 16.5/18 & 46.6/37\\
        110  &  1.0436(15) & 1.0433(12) & 1.56(2) & \qquad  1.41(3) & 2.149(38) & 0.0704(13) & 16.5/18 & 36.9/37\\
        111  &  1.0838(21) & 1.0831(15) & 1.60(2) & \qquad  1.45(3) & 1.628(40) & 0.0673(18) & 22.6/18 & 39.3/37\\
        200  &  1.1206(31) & 1.1172(23) & 1.60(3) & \qquad  1.48(4) & 1.279(46) & 0.0658(25) & 17.4/18 & 46.9/37\\
        \hline\hline
    \end{tabular}
\end{table}

\begin{table}
    \centering
    \caption{Fit ranges $[t_{\rm min}, t_{\rm max}]$ for two-point and
        three-point functions.  They are chosen to be approximately
        similar in physical units and independent of sea-quark masses
        with one exception: for the case $a \approx 0.12$ fm and $\hat
        m^\prime/m_s^\prime = 0.1$, the two-point range was [3,23].}
    \label{tab:fit_ranges}
        \begin{tabular}{lcc}
            \hline
            \hline
  $\approx a$~(fm)  & two-point & three-point \\
            \hline
 $0.12\hs$  & [4,23] & [2,10] \\
 $0.09\hs$  & [5,33] & [2,15] \\
 $0.06\hs$  & [7,45] & [4,18] \\
 $0.045\hs$ & [11,80]& [7,24] \\
            \hline
            \hline
        \end{tabular}
\end{table}

\subsubsection{Two-point correlator fits}%
\label{sec:2ptfits}

The two-point functions are constructed from both a local and a
smeared interpolating operator.  They are fit simultaneously to
determine the ground- and excited-state energies.  We include
oscillating and nonoscillating states in pairs and test the stability
of the fits by comparing results with 2+2 and 3+3 states.  An example
is shown in Table~\ref{tab:coarse_2pts} for the $a \approx 0.12$ fm,
$\hat m^\prime = 0.14m_s^\prime$ ensemble.  For this case we choose a
fit range of [4,23] with 3+3 states.  Results for the energy and
overlap factor for that range agree with fits in the range [6,16] with
2+2 states.  For the analysis, we use fits with 3+3 states; the
(2+2)-state fits provide a check of systematic effects from
excited-state contamination.  We select approximately the same fit
ranges in physical units for all ensembles, as shown in
Table~\ref{tab:fit_ranges}.

\subsubsection{Three-point correlator fits}%
\label{sec:3ptfits}

\begin{figure}
    \includegraphics[width=0.32\textwidth]{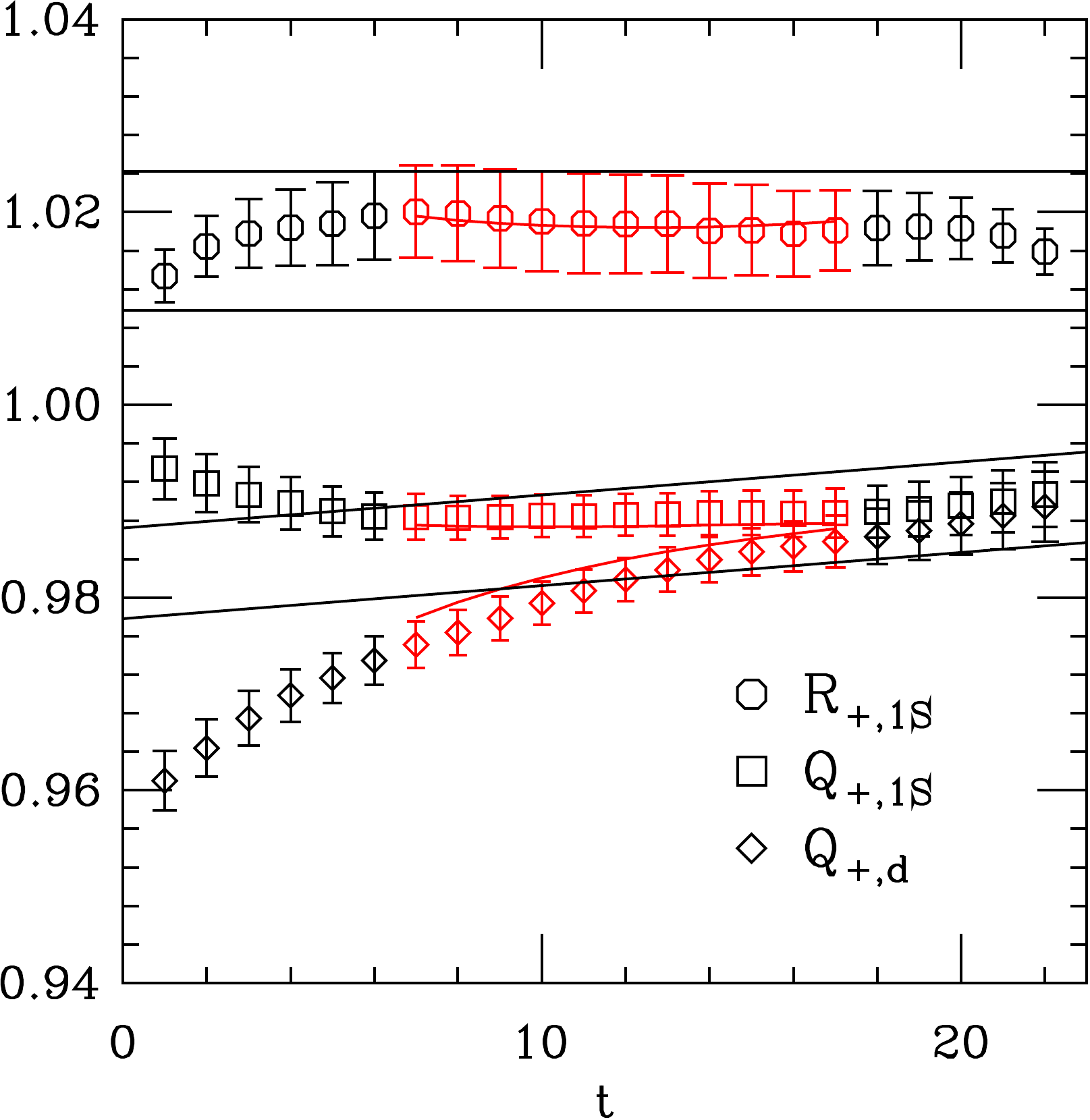} \hfill
        \includegraphics[width=0.32\textwidth]{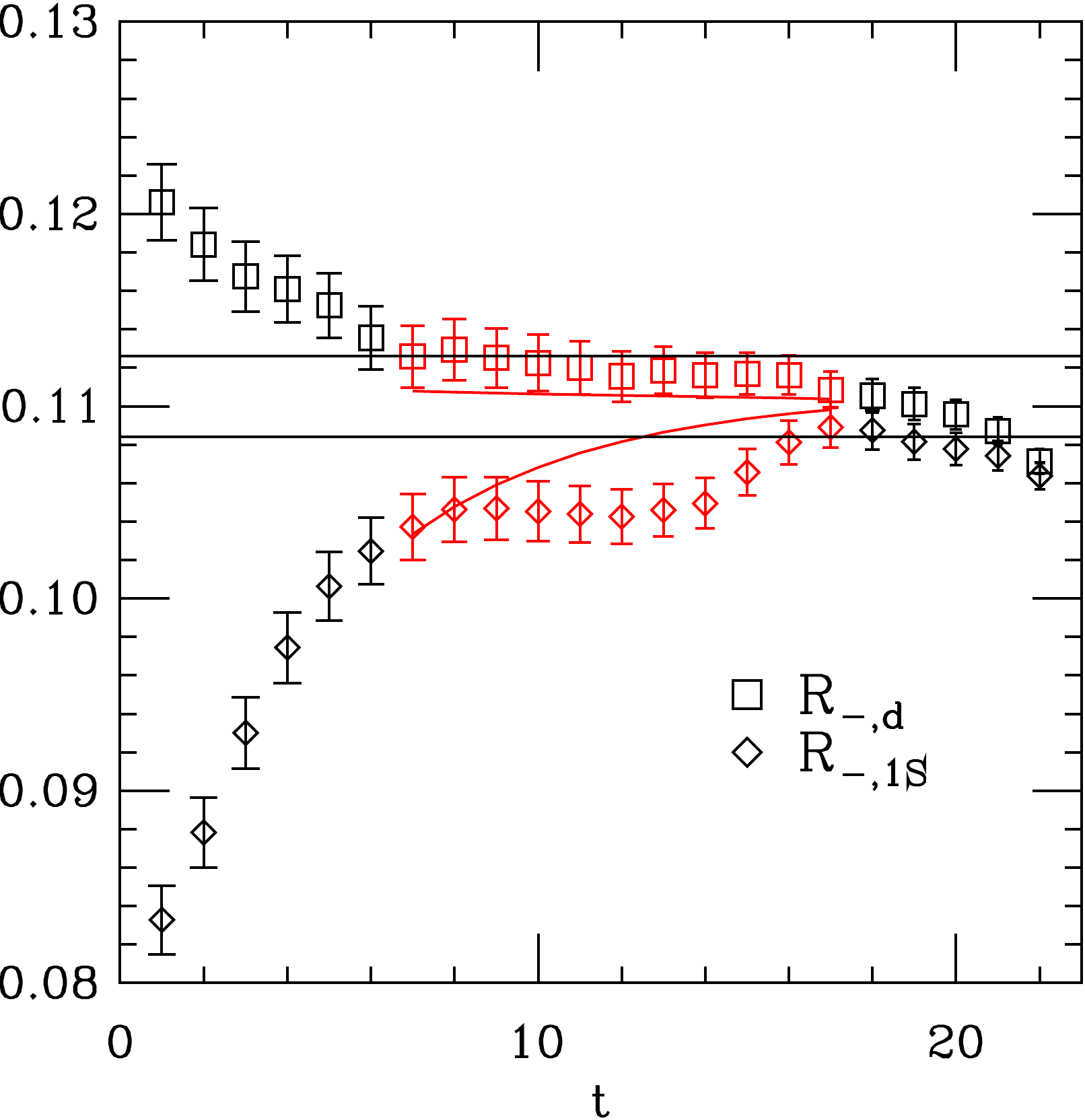} \hfill
    \includegraphics[width=0.32\textwidth]{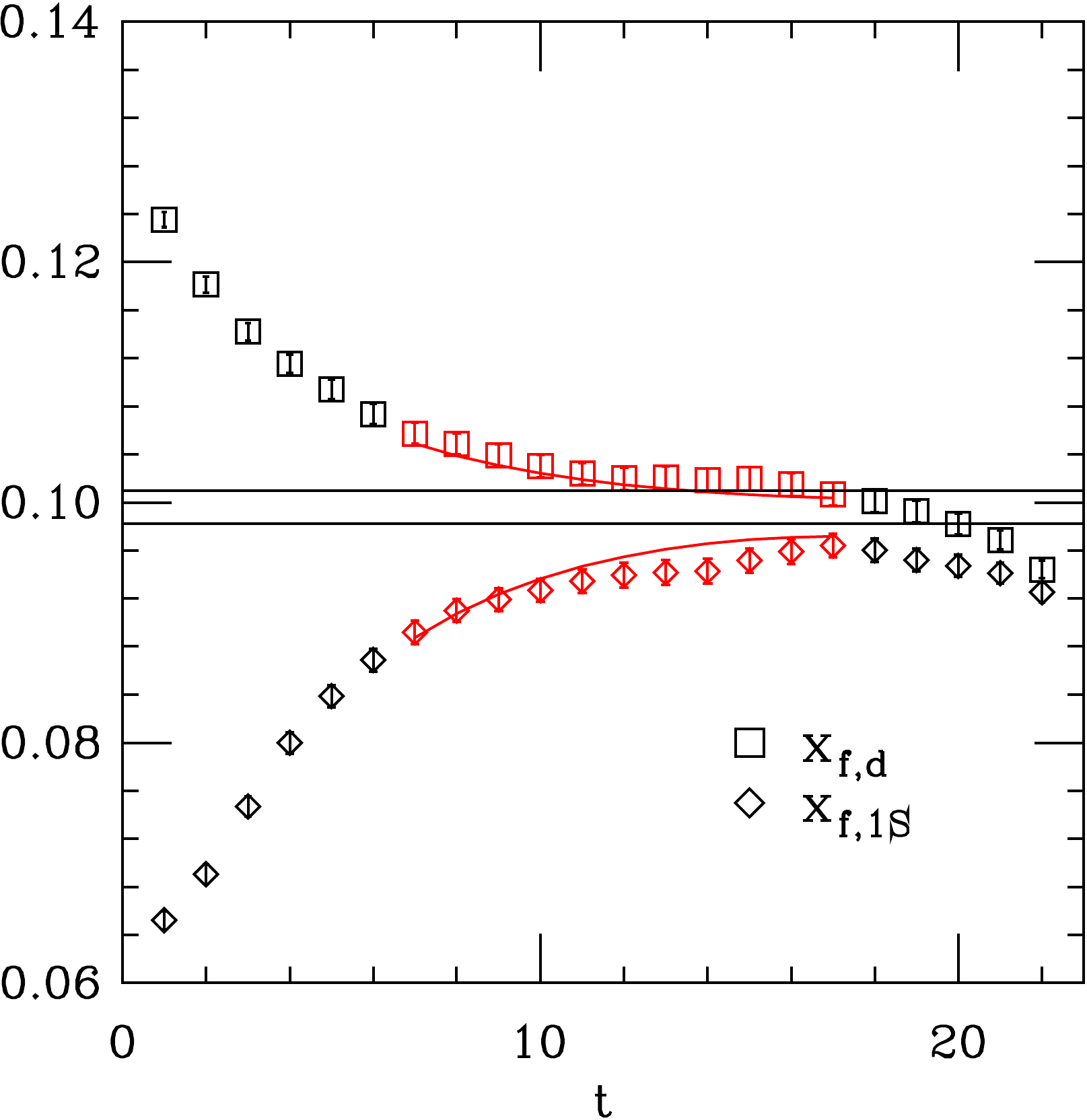}
    \caption{Sample joint three-point function fits for determining
      the ratios $R_+$ and $Q_+(\bm{p})$ (left), $R^1_-(\bm{p})$
      (middle) and $x^1_f(\bm{p})$ (right), for lattice momentum
      $\bm{p} = (1,0,0)$.  Data shown are for the $a \approx 0.06$ fm,
      $\hat m^\prime = 0.14m_s^\prime$ ensemble with $B$-$D$
      separation $T=24,25 $.  Values are plotted against the time $t$
      of the vector-current insertion.  Data points at the left and
      right extremities are not included in the fit.  A color
      (gray-scale) change indicates which points are included in the
      fit.  Black lines indicate the upper and lower $1\sigma$ range
      of the ground-state contribution.  Best fit lines are shown in
      red (gray).  For $Q_+({\bf p})$ the ``plateau'' is slanted
      because of the factor $\exp(\delta m\, t)$ in
      Eq.~(\ref{eq:Qplusfit}).
    \label{fig:Rplussample}}
\end{figure}

To determine the nonzero-recoil form factor $R_+(\bm{p})$, we fit
three ratios simultaneously: the double ratio for the 1S source from
Eq.~(\ref{eq:doubleratio}) and the local- and smeared-source ratios
$Q_{+,d}(\bm{p},t,T)$ and $Q_{+,1S}(\bm{p},t,T)$ from
Eq.~(\ref{eq:Rplusratio2}).  Because the fit model
[Eq.~(\ref{eq:Qplusfit})] includes effects of the same first-excited
states that occur in the two-point functions, we use the
two-point-fit values for these states to set priors for $\delta m$,
$\delta E_D$, and $\delta M_B$.  The best-fit values are used as the
central values and their errors as the widths of the Gaussian
priors. Fit ranges are chosen for stability. We use the same range
for all three-point correlators in a given ensemble.  The ranges are
listed in Table~\ref{tab:fit_ranges}, and a sample three-point fit
is plotted in Fig.~\ref{fig:Rplussample}, left.

For $\bm{x}_{f,b}(\bm{p},t,T)$ and $\bm{R}_{-,b}(\bm{p},t,T)$, we fit
values for both local and smeared sources jointly with the fitting
form of Eq.~(\ref{eq:xfRminusfit}).  Sample three-point fits are
plotted in the middle and right panels of Fig.~\ref{fig:Rplussample}.
Then, having determined all the needed quantities, we calculate $w$,
$h_+(w)$ and $h_-(w)$ from Eqs.~(\ref{eq:w}) and
(\ref{eq:hplusQ})--(\ref{eq:hminusQ}) for each momentum $\bm{p}$
and ensemble.

\subsection{Heavy-quark-mass adjustment}%
\label{sec:kappa}

We adjust the bare masses of the $b$ and $c$ quarks so that the
kinetic masses of the $D_s$ and $B_s$ mesons obtain their physical
values.  When computing the two-point and three-point correlators, we
used good estimates of these quark masses.  By the end of the data
generation, we could obtain better estimates via the procedure
described in \cite{B2Dstar}.  

Because there are small differences between the simulation values and
final, tuned values, an adjustment of the form factors is
required. Details are given in Appendix \ref{app:kappa}. To obtain the
adjustment we computed a full set of correlation functions on one of
our ensembles with a few heavy-quark masses close to the tuned value
and use these results to calculate the slopes of the form factors with
respect to the quark masses.  These results and the known corrections
then give the needed small adjustments tabulated in Table
\ref{tab:tableformfactor}.  The size of the heavy-quark mass
corrections to $h_+$ ($h_-$) range from 0 to 0.2\% (0 to 2\%). Small
errors arise both from uncertainties in the tuned quark masses and
uncertainties in the determination of the slopes.

%%%%%%%%%%%%%%%%%%%%%%%%%%%%%%%%%%%%%%%%%%%%%

\subsection{Current renormalization}
\label{sec:PT}

Here we summarize the procedure for matching the lattice matrix
elements to the continuum.  The three-point fits yield ratios in which
the flavor-diagonal factors $\ZVcc\ZVbb$ from
Eq.~(\ref{eq:rhoV_ratio}) cancel.  Thus, to normalize the form factors
to continuum conventions, we only have to apply the flavor
off-diagonal factors $\rhoV{\mu}$ as in
Eqs.~(\ref{eq:renormR+})--(\ref{eq:renormR-}).  Matching factors with
two heavy quarks depend on the recoil $w$, but the $w$ dependence is
not available.  Even so, we can obtain some information by considering
the limit $m_{2c}a\ll1$, where the $w$ dependence goes away.  For this
reason, each of the matching factors in
Eqs.~(\ref{eq:renormR+})--(\ref{eq:renormR-}) requires somewhat
different treatment.  Appendix~\ref{app:hqerror} provides further
details on the matching calculations.

The calculation of the zero-recoil matching factor $\rhoV{4}(1)$
needed to renormalize $R_+$ is completely analogous to that of the
axial-vector matching factor used in Ref.~\cite{B2Dstar}.  Following
Ref.~\cite{Harada:2001fj}, we compute it to one-loop order in
perturbation theory,
\begin{equation}
    \rhoV{4}(1) = 1 + \alpha_V(q^*)\rhoV{4}^{[1]}(1),
    \label{eq:rho-one-loop}
\end{equation}
where $\alpha_V(q^*)$ is the QCD coupling in the $V$
scheme~\cite{Lepage:1992xa}, evaluated here at the scale $q^*=2/a$.
The result for each ensemble is listed in Table~\ref{tab:rho}.
\begin{table}
  \caption{One-loop estimates of the matching factors for the lattice
  ensembles in this study.  Shown are the approximate lattice spacing
  in fm, the sea-quark mass ratio $\hat m^\prime/m^\prime_s$, the
  tuned $\kappa$ values of the charm and bottom
  quarks~\protect\cite{B2Dstar}, the strong coupling in the $V$-scheme
  evaluated at $q^*=2/a$, and the zero-recoil factors $\rhoV{4}(1)$
  and $\rhoV{i}(w)/\rhoV{4}(w)$ on that ensemble.  The first error in
  each tuned $\kappa$ value is statistical, and the second reflects
  the uncertainty in the lattice scale
  determination~\protect\cite{B2Dstar}.  The correction factors are
  evaluated at the tuned heavy-quark masses except for
  $\rhoV{i}/\rhoV{4}$, which is evaluated at $m_ca=0$.  The systematic
  uncertainties in the $\rho$ factors are discussed in
  Sec.~\ref{sec:err} and Appendix~\ref{app:hqerror}.}
    \label{tab:rho}
    \begin{tabular*}{\textwidth}{@{\extracolsep{\fill}}llcccccc}
            \hline
            \hline
    $\approx a$ (fm) & $\hat m^\prime/m_s^\prime$ & 
        \multicolumn{1}{c}{$\kappa_c$}  & \multicolumn{1}{c}{$\kappa_b$}  & 
         \multicolumn{1}{c}{$\alpha_V(q^*=2/a)$} & 
        \multicolumn{1}{c}{$\rhoV{4}(1)$} &
        \multicolumn{1}{c}{$\rhoV{i}(w)/\rhoV{4}(w)$} \\ 
            \hline
~~0.12  & 0.4   & 0.12452(15)(16) & 0.0879(9)(3) & 0.3047 & 1.025105 & 0.892347 \\
~~0.12  & 0.2   & 0.12423(15)(16) & 0.0868(9)(3) & 0.3108 & 1.026472 & 0.888051 \\
~~0.12  & 0.14  & 0.12423(15)(16) & 0.0868(9)(3) & 0.3102 & 1.026395 & 0.888248 \\
~~0.12  & 0.1   & 0.12423(15)(16) & 0.0868(9)(3) & 0.3102 & 1.026388 & 0.888241 \\ \hline
~~0.09  & 0.4   & 0.12737(9)(14)  & 0.0972(7)(3) & 0.2582 & 1.015603 & 0.924664 \\
~~0.09  & 0.2   & 0.12722(9)(14)  & 0.0967(7)(3) & 0.2607 & 1.016080 & 0.923051 \\
~~0.09  & 0.15  & 0.12718(9)(14)  & 0.0966(7)(3) & 0.2611 & 1.016160 & 0.922757 \\
~~0.09  & 0.1   & 0.12714(9)(14)  & 0.0965(7)(3) & 0.2619 & 1.016259 & 0.922319 \\
~~0.09  & 0.05  & 0.12710(9)(14)  & 0.0964(7)(3) & 0.2623 & 1.016340 & 0.922022 \\ \hline
~~0.06  & 0.4   & 0.12964(4)(11)  & 0.1054(5)(2) & 0.2238 & 1.008792 & 0.947870 \\
~~0.06  & 0.2   & 0.12960(4)(11)  & 0.1052(5)(2) & 0.2245 & 1.008945 & 0.947361 \\
~~0.06  & 0.14  & 0.12957(4)(11)  & 0.1051(5)(2) & 0.2249 & 1.009017 & 0.947085 \\
~~0.06  & 0.1   & 0.12955(4)(11)  & 0.1050(5)(2) & 0.2253 & 1.009098 & 0.946829 \\ \hline
~~0.045 & 0.2  & 0.130921(16)(70)& 0.1116(3)(2) & 0.2013 & 1.004566 & 0.962520 \\
            \hline
            \hline
    \end{tabular*}
\end{table}

For the matching factor $\rhoV{4}(w)/\rhoV{4}(1)$, we note that, by
construction, the one-loop coefficient must be proportional to $w-1$.
Moreover, for $m_{2c}a\ll1$, which holds on the two finest lattices,
one may treat the charm quark as a light quark~\cite{Harada:2001fj},
using the HQET formalism for heavy-light
currents~\cite{Harada:2001fi}.  The $w$ dependence goes away in this
limit, so the one-loop coefficient must also be proportional to
$m_{2c}a$.  Thus,
$\rhoV{4}(w)/\rhoV{4}(1)=1+\mathcal{O}\left(\alpha_s(w-1)m_{2c}a\right)$,
where the coefficient of the one-loop correction is not known.  In our
analysis, we take $\rhoV{4}(w)/\rhoV{4}(1)=1$ and include the
estimated size of the one-loop correction as a $w$-dependent
uncertainty.

For the matching factor $\rhoV{i}(w)/\rhoV{4}(w)$, we can take the
heavy-light theory a step further and calculate the matching explictly
for $m_{2c}a\ll1$.  The calculation does not depend on $w$.  The
resulting values for $\rhoV{i}(w)/\rhoV{4}(w)$ as in
Eq.~(\ref{eq:rho-one-loop}) are listed in Table~\ref{tab:rho}.  The
error in the one-loop coefficient introduced by taking the limit
$m_{2c}a \to 0$ is proportional to $\alpha_s m_{2c}a$ with a,
presumably, mild $w$ dependence, and is again included as an
uncertainty.

%%%%%%%%%%%%%%%%%%%%%%%%%%%%%%%%%%%%%%%%%%%%%

\begin{figure}
    \includegraphics[width=0.48\textwidth]{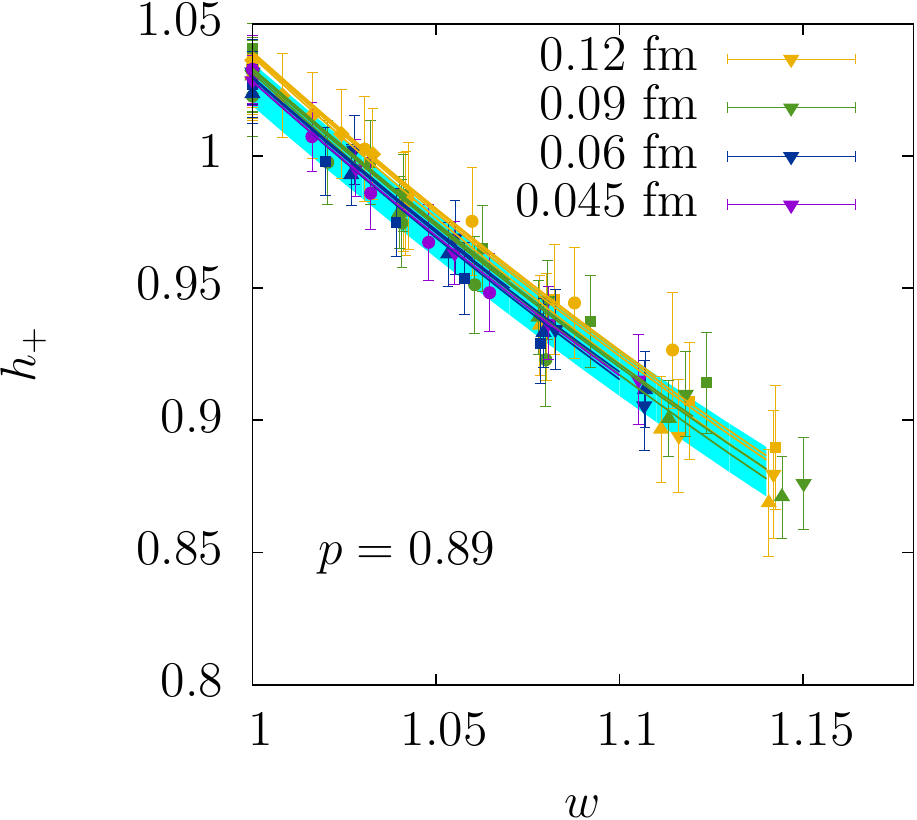} \hfill
    \includegraphics[width=0.48\textwidth]{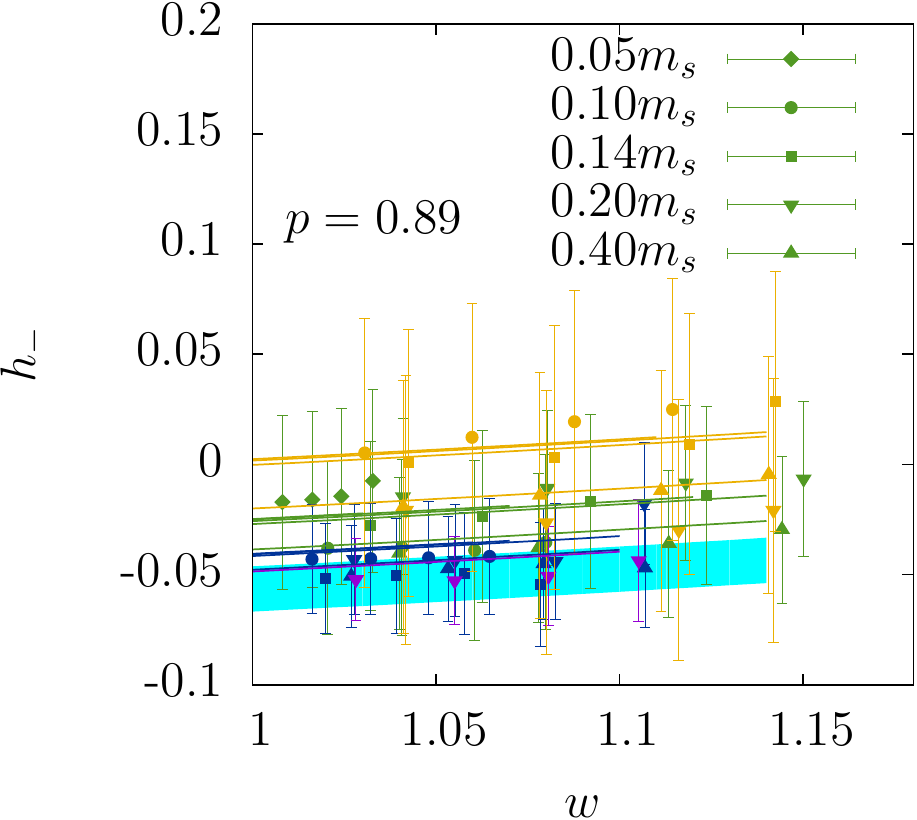}
    \caption{Global fit of all data for the form factors $ h_+ $
      (left) and $ h_- $ (right) {\it vs.} recoil $w$. The blue
      (shaded) band gives the $1 \sigma$ confidence range for the
      continuum extrapolation at physical quark masses.  Fit errors
      include statistics, matching and truncation of the chiral
      expansion. The legend in the left figure gives the color
      convention for the lattice spacing, and, in the right, it gives
      the shape convention for the sea-quark mass ratio.}
    \label{fig:hplushminus}
\end{figure}

\subsection{Chiral-continuum extrapolation}%
\label{sec:chiral}

The resulting form factors $ h_+ $ and $ h_- $, after applying the
$\kappa$ corrections and renormalization factors, are shown in
Fig.~\ref{fig:hplushminus}. As can be seen, the dependence of $ h_+ $
on lattice spacing $ a $ and light-quark-mass ratio $ \hat
m^\prime/m^\prime_s $ is quite mild.  The form factors must be
extrapolated to the physical average value of the up and down quark
mass $r_1 \hat m $ and zero lattice spacing ($a \rightarrow 0$) (the
physical point).

To this end we fit both form factors to the following expressions:
\begin{eqnarray}
    h_+(a,\hat m^\prime, m_s^\prime ,w)     & =     & 1 + \frac{ X_+(\Lambda_\chi)}{m_c^2}
     - \rho_+^2 (w-1) + k_+ (w-1)^2
     + c_{1,+} x_l + c_{a,+} x_{a^2} + c_{a,a,+} x_{a^2}^2 \nonumber \\
   & + & 
    c_{a,m,+} x_l x_{a^2} + c_{2,+} x_l^2 + 
    \frac{g_{D^*D\pi}^2}{16 \pi^2 f_\pi^2 r_1^2} {\rm logs}_{\rm SU(3)}(a, \hat m^\prime,  m_s^\prime ,w, \Lambda_\chi)
    \label{eq:h+chiral}\\
    h_-(a,\hat m^\prime,  m_s^\prime, w)     &=      & \frac{X_-}{m_c} - \rho_-^2 (w-1) + k_-
    (w-1)^2 + c_{1,-} x_l + c_{a,-} x_{a^2} \nonumber \\
    & + &
    c_{a,a,-} x_{a^2}^2 + c_{a,m,-} x_l x_{a^2} + c_{2,-} x_l^2 \label{eq:h-chiral} \,\, ,
\end{eqnarray}
which contain the correct dependence on the light and strange-quark
masses, lattice spacing, and recoil $w$ at next-to-leading order (NLO)
in chiral perturbation theory.  The chiral logarithm term, denoted
``${\rm logs}_{\rm SU(3)}$'', contains non-analytic dependence upon
the pion and kaon masses (or equivalently $\hat m^\prime$ and
$m_s^\prime$).  It comes from a staggered-fermion version of the
one-loop continuum result of Chow and Wise \cite{Chow:1993hr} that
includes taste-breaking discretization effects \cite{Laiho:2005ue}.
The explicit expression for ${\rm logs}_{\rm SU(3)}$ is given in the
Appendix of Ref.~\cite{Bailey:2012rr}.  The coefficient of the
logarithm term is predicted in $\chi$PT, but depends upon the value of
the $D^*$-$D$-$\pi$ coupling, $g_{D^*D\pi}$, which is not known
precisely.  We allow $g_{D^*D\pi}$ to vary in the fit, but constrain
it with a Gaussian prior $0.53 \pm 0.08$, motivated by the spread of
experimental~\cite{Anastassov:2001cw,Lees:2013uxa,Lees:2013zna} and
recent lattice-QCD
results~\cite{Detmold:2011bp,Can:2012tx,Becirevic:2012pf,Detmold:2012ge,Flynn:2013kwa,Bernardoni:2014kla}.
The analytic terms depend on the light spectator-quark mass through
$x_l = 2 B_0 \hat m^\prime /(8 \pi^2 f_\pi^2)$ and on the lattice
spacing through $x_{a^2} = [a/(4 \pi f_\pi r_1^2)]^2$, which,
according to $\chi$PT power counting, are expected to have
coefficients of order~1 \cite{Bazavov:2011aa}.  The NLO expression is
supplemented by next-to-next-to-leading order (NNLO) analytic terms in
the light-quark mass $\hat m^\prime$ and lattice spacing to
incorporate the error from the truncation of the chiral expansion, as
explained below, and by terms analytic in $(w-1)$ to allow
interpolation in $w$ at nonzero recoil.  We do not include analytic
functions of the strange sea-quark mass because (1) we do not have
sufficiently varied values of $m_s^\prime$ to be able to resolve any
strange-quark mass dependence, (2) from $\chi$PT we expect the
sea-quark mass dependence of the form factors to be significantly
smaller than the light spectator-quark mass dependence, and (3) as
discussed in Sec.~\ref{subsection:light-sea-tuning}, we do not observe
any strange sea-quark mass dependence within our current statistical
precision.

The statistical errors and correlations from the two-point and
three-point ratio fits are propagated to the chiral fits using a
single-elimination jackknife procedure.  The strongest correlations
are between the data for $h_{+}(w)$ (or $h_{-}(w)$) at different $w$
values on the same ensemble.  The data for $h_+(w)$ and $h_-(w)$ on
the same ensemble are only weakly correlated.  Results from different
ensembles are statistically independent.  The fits to
Eqs.~(\ref{eq:h+chiral}) and (\ref{eq:h-chiral}) are done taking fully
into account all statistical correlations.

\begin{figure}
    \includegraphics[width=0.48\textwidth]{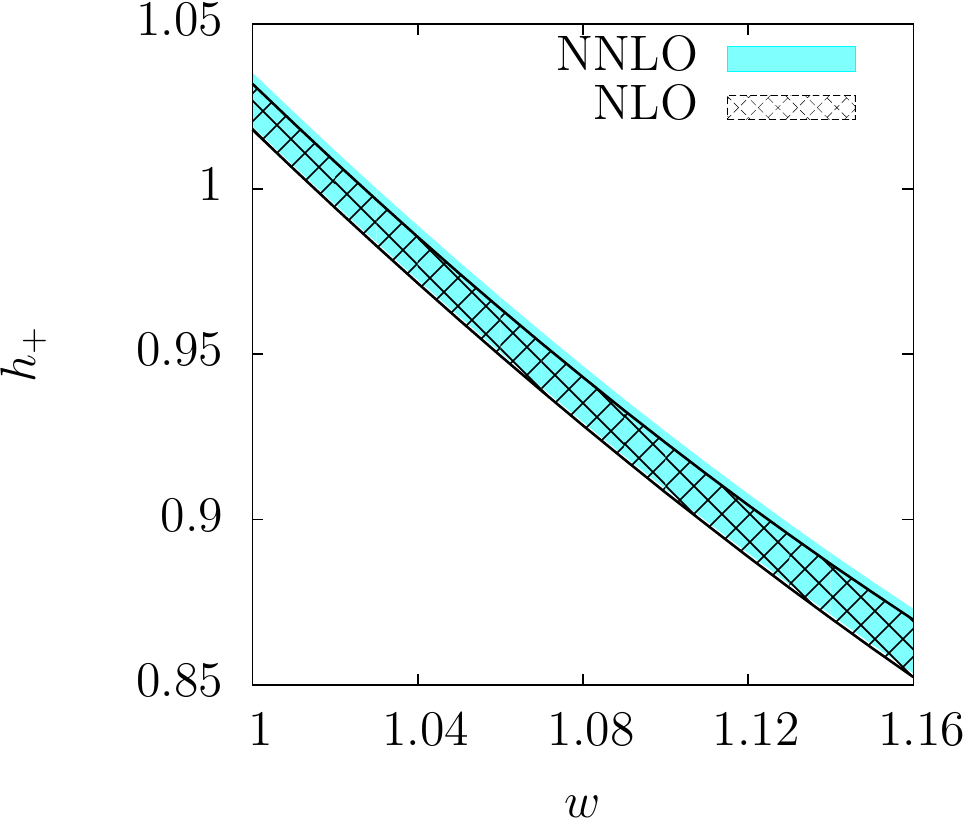} \hfill
    \includegraphics[width=0.48\textwidth]{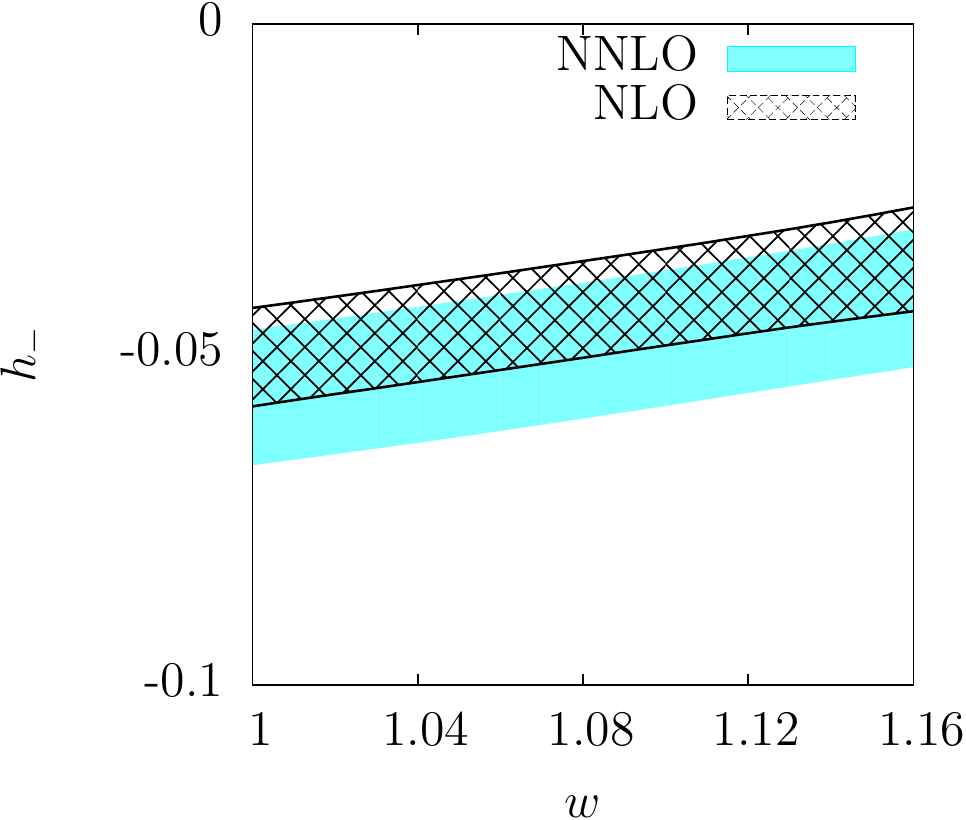}
    \caption{Comparison of NLO (hatched) and NNLO (solid)
      chiral-continuum fits for $ h_+ $ (left) and $ h_- $ (right)
      {\it vs.} recoil $w$.}
    \label{fig:NLOvsNNLO}
\end{figure}

To test the applicability of NLO chiral perturbation theory to our
data, we first fit without the analytic NNLO terms. The $p$ value,
$p=0.93$ of the joint, exclusively NLO fit to $h_+$ and $h_-$ is
satisfactory.\footnote{With Gaussian priors our $p$ value is
  determined from the augmented $\chi^2$.  We count degrees of freedom
  as the number of data points minus adjustable parameters plus the
  number of theoretically-motivated priors. Very loose priors that
  have no impact but to stabilize the fits are not counted.}  
Next we include the analytic NNLO terms with priors $0\pm 2$ based on
expectations from $\chi$PT power-counting in order to test for the
effect of truncating the chiral perturbation series.  The $p$ value
decreases slightly to 0.87.  Including these terms increases the
standard deviation by at most $\sim$10\% for $h_+$ and $\sim$30\% for
$h_-$, and shifts the central values by much less than the final
standard deviation, as shown in Fig.~\ref{fig:NLOvsNNLO}.  The
statistical errors then can be safely assumed to include the
systematic error of the truncation. We therefore use this fit
including NNLO terms to obtain our preferred value for the form
factors at the physical point.  The results of the extrapolation with
propagated statistical errors are shown as bands in
Fig.~\ref{fig:hplushminus}.

In heavy-quark effective theory, Luke's theorem states that $h_+(w=1)$
has leading corrections only at second order in the inverse
heavy-quark masses, namely $1/m_c^2$ and $1/m_b^2$, whereas $h_-$ has
corrections at first order.  Appendix \ref{app:hqerror} and
Ref.~\cite{Kronfeld:2000ck} show how Luke's theorem applies in lattice
gauge theory and, hence, that one expects $ h_- $ to have larger
heavy-quark discretization errors than $ h_+ $.  Indeed, we see that
$h_-$ does have a stronger dependence on lattice spacing than $h_+$.
For the determination $ f_+$, and therefore $ |V_{cb}| $, the
contribution of $ h_- $ over the entire kinematic range is small, so
the larger errors in $ h_- $ do not increase the overall error much.
These trends in lattice spacing with fourteen ensembles are consistent
with our previous findings with four ensembles \cite{Bailey:2012rr}.

We build the systematic errors from $\kappa$ tuning and from the
matching factors into the chiral-continuum extrapolation by forming
the combined covariance matrix for the data as follows:
\begin{equation}
    C_{ij} = C_{ij}^\text{stat} + \delta^{(\rho)}_i \delta^{(\rho)}_j + 
     \delta^{(\kappa)}_i \delta^{(\kappa)}_j ,
    \label{eq:Cstat-syst}
\end{equation}
where the first term is the statistical covariance, and the index $i$
runs over all data (ensembles, momenta, and $h_+$ and $h_-$). We
denote by $\delta^{(\rho)}_i$ and $\delta^{(\kappa)}_i$ the shift on
the $i$th datum due to the matching and $\kappa$-tuning errors,
respectively.  Equation~(\ref{eq:Cstat-syst}) conservatively assumes
that the matching-factor errors (or $\kappa$-tuning errors) are 100\%
correlated between all data points.  For the systematic errors due to
our matching procedure, we have estimates for the uncertainty in
$\rho_{V^4}(1)$, $\rhoV{4}(w)/\rhoV{4}(1)$ and
$\rhoV{i}(w)/\rhoV{4}(w)$ in Eqs.~(\ref{eq:rhov4error1}),
(\ref{eq:rhov4error}), and~(\ref{eq:rhoVierror}), respectively.  The
form factors $h_\pm(w)$ change the most when $\rhoV{4}(w)$ and
$\rhoV{i}(w)/\rhoV{4}(w)$ are simultaneously shifted in opposite
directions.  We take the average of these two shifts as an estimate of
the $\rho$-factor error for all $h_\pm(w)$ on all ensembles.  For the
$\kappa$-tuning error, we take the same approach in principle,
propagating the uncertainties of the intercepts and slopes in
Appendix~\ref{app:kappa} to shifts $\delta^{(\kappa )}_i$ of the
form-factor data.  However, we find that the resulting
$\delta^{(\kappa )}_i$ are negligibly small, and we therefore set them
to zero in Eq.~(\ref{eq:Cstat-syst}).

\begin{figure}
   \centering{
     \includegraphics[width=0.50\textwidth]{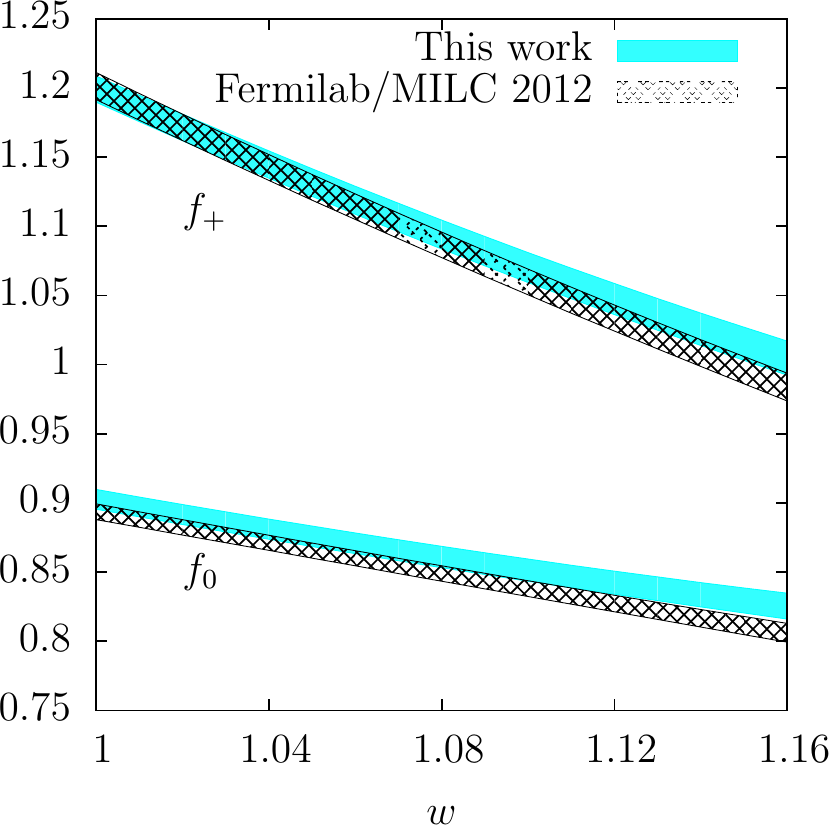}}
    \caption{The form factors $f_+$ and $f_0$ as a function of
        the recoil $w$ resulting from the chiral-continuum
        fit in this study (cyan band), compared with the results
        from~\cite{{Bailey:2012rr}} (cross-hatched band). The width of
        each band indicates the 1$\sigma$ error from the
        chiral-continuum fit, but uncertainties from the matching
        factors are included only in the cyan bands.  See the text for
        additional details. }
    \label{fig:chipt_compare}
\end{figure}

Given the chiral-continuum fit results for $h_+$ and $h_-$, we
construct the vector and scalar form factors $f_+$ and $f_0$ using
Eqs.~(\ref{eq:f+fromh}) and~(\ref{eq:f0fromh}).
Figure~\ref{fig:chipt_compare} compares our new $B \to D \ell\nu$
form-factor results with those from our earlier
work~\cite{Bailey:2012rr} in the $w$ range where we have simulation
data.  The curves shown are output from the chiral-continuum
extrapolation, and therefore include the uncertainties from
statistics, the chiral-continuum extrapolation, and matching (for the
current work); they do not include the remaining systematic
uncertainties, which we add in quadrature {\it a posteriori} in both
works.  We expect that the two results are largely independent because
they have only a small subset of overlapping data (the earlier work
included only four ensembles), and the new work includes NNLO analytic
terms in the $\chi$PT fit function.  The results are consistent for
both form factors over almost all simulated $w$ values, and diverge
only slightly for $f_0$ for $w > 1.13$.  The central values of the
new form factors are slightly higher than in~\cite{Bailey:2012rr},
primarily due to explicit inclusion of the perturbative correction
factors $\rhoV{i}(w)/\rhoV{4}(w)$ which have a bigger effect on the
form factor $f_0$ than on $f_+$.  The total errors on the form factors
in this work are similar in size to those in
Ref.~\cite{Bailey:2012rr}, but the additional ensembles used in this
work enable a more detailed and reliable systematic error analysis as
described in Sec.~\ref{sec:err}. (Reference~\cite{Bailey:2012rr}
focused on form-factor ratios in which most of the systematic errors
are suppressed.)

%%%%%%%%%%%%%%%%%%%%%%%%%%%%%%%
\section{Systematic errors}%
\label{sec:err}
%%%%%%%%%%%%%%%%%%%%%%%%%%%%%%%

\begin{figure}[tb]
    \includegraphics[width=0.48\textwidth]{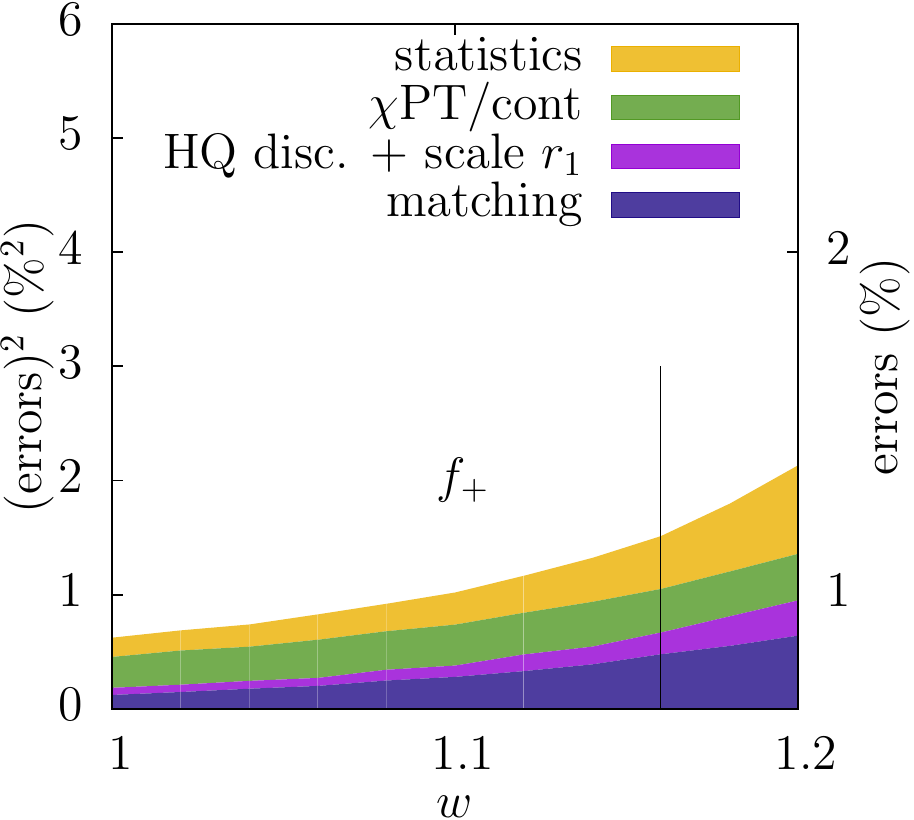} \hfill
    \includegraphics[width=0.48\textwidth]{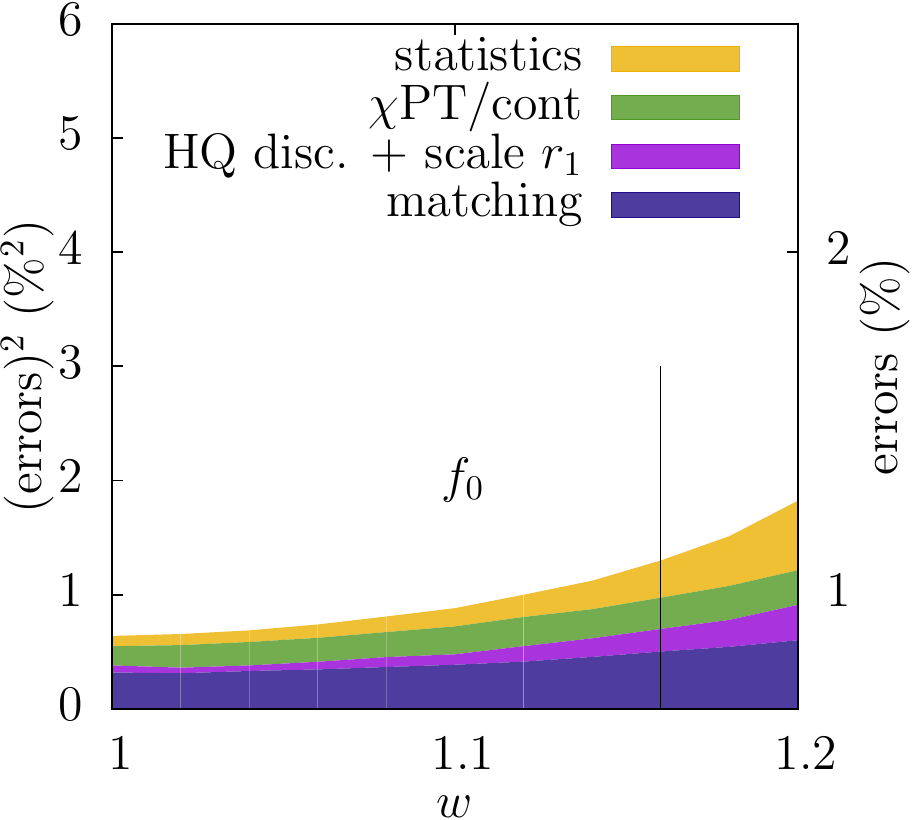}
    \caption{Error budgets for $f_+$ and $f_0$ as a function of the
      recoil $w$.  The colored bands show the error contribution of
      each uncertainty source to the quadrature sum.  The
      corresponding error is provided on the right $y$-axis. Our
      lattice simulation results are for $w \in [0, 1.16]$, {\it
        i.e.}, to the left of the vertical line.}
    \label{fig:fpf0errbudget}
\end{figure}

In this section we discuss the sources of systematic error in the
lattice determinations of $h_+$ and $h_-$ and their propagation to the
form factors $f_+$ and $f_0$.  As can be seen from
Fig.~\ref{fig:NLOvsNNLO}, the magnitude of $h_-$ is about 5\% of $h_+$
for the entire range of simulated $w$ values.  Further, the
contribution of $h_-$ to the vector form factor $f_+$ is suppressed
relative to the contribution from $h_+$ by the factor $(1-r)/(1+r) =
0.477$, while the contribution of $h_-$ to the scalar form factor
$f_0$ is exactly zero at $w=1$ and grows linearly with recoil as
$(w-1)$.  Thus even large percentage systematic errors in $h_-$ lead
to only small uncertainties in $f_+$ and $f_0$.
Figure~\ref{fig:fpf0errbudget} shows the momentum-dependence of the
error contributions to $f_+(w)$ and $f_0(w)$, while
Table~\ref{tab:fpf0errbudget} provides numerical values for a
representative recoil $w=1.16$.

\begin{table}[tb]
    \caption{Error budget (in percent) for $f_+$ and $f_0$ at $w =
      1.16$, which is the largest recoil value used in our momentum
      extrapolation to the full kinematic range and determination of
      $|V_{cb}|$ (see Sec.~\ref{sec:Vcb}).  The first row includes the
      combined error from statistics, matching, and the error from
      truncating the chiral expansion resulting from the
      chiral-continuum fit: errors in parentheses are approximate
      sub-parts estimated as described in the text.  The total error
      is obtained by adding the individual errors in quadrature.  Not
      explicitly shown because they are negligible are finite-volume
      effects, isospin-breaking effects, and light-quark mass tuning.}
    \label{tab:fpf0errbudget}
    \begin{center}
        \begin{tabular}{l@{\quad}c@{\quad}c}
            \hline
            \hline
            Source                                & $f_+$(\%) & $f_0$(\%)  \\
            \hline
	Statistics+matching+$\chi$PT cont. extrap.            & 1.2     &  1.1 \\
             \quad (Statistics)                               & (0.7)   &  (0.7) \\
             \quad (Matching)                                 & (0.7)   &  (0.7) \\
             \quad ($\chi$PT/cont. extrap.)                   & (0.6)   &  (0.5) \\
             Heavy-quark discretization                       & 0.4 &      0.4 \\
             Lattice scale $r_1$                              & 0.2 &      0.2 \\
            \hline
            Total error                                       & 1.2 &      1.1 \\
            \hline
            \hline
        \end{tabular}
    \end{center}
\end{table}

\subsection{Overview of systematic errors in $f_+$ and $f_0$}

As can be seen from Fig.~\ref{fig:fpf0errbudget}, the dominant
uncertainty in both form factors arises from the
chiral-continuum fit, which includes contributions from statistics,
matching factors, and higher-order terms in the chiral expansion.
Although we cannot strictly disentangle the contributions to the error
from these sources, we can estimate their sizes by repeating the
chiral-continuum fit omitting either the errors in the matching
factors or the NNLO terms in the chiral expansion, and take the
quadrature difference of the resulting error estimates.  The
contribution from ``statistics'' is defined to be the error in the NLO
chiral-continuum fit to data with no matching-factor uncertainties
included. This imprecise scheme does not guarantee that the individual
errors sum to the total fit error, but, roughly speaking, we find that
the statistics, matching, and truncation uncertainties in the
chiral-continuum expansion contribute approximately equally to the
error in the full NNLO fit.  Despite our incomplete knowledge of the
matching factors, we find their contributions to the uncertainty in
$f_+$ and $f_0$ to be modest.  The errors from the chiral-continuum
fit are under good control for the range of simulated lattice recoil
values, but grow rapidly for $w\gtrsim1.16$ where we do not have data.

We add the remaining systematic uncertainties {\it a posteriori} to
the chiral-continuum fit error.  We estimate the individual
contributions to the form-factor error budget in the following
subsections, discussing each source in a separate subsection for
clarity.  In practice, only the heavy-quark discretization errors
(Sec.~\ref{subsubsection:hqerr}) and lattice-scale uncertainty
(Sec.~\ref{subsubsection:roneerr}) turn out to be significant.

We assume that systematic uncertainties from heavy-quark
discretization effects and the lattice-scale uncertainty are
uncorrelated, and therefore add them in quadrature.  We then propagate
them to $f_+$ and $f_0$ according to the linear transformation
Eqs.~(\ref{eq:f+fromh}) and (\ref{eq:f0fromh}), which depends on the recoil $w$,
taking them to be 100\% correlated between $w$ values and between
$h_+$ and $h_-$.  Both the lattice-scale and heavy-quark
discretization errors are substantially smaller than the
chiral-continuum fit error, and increase only slowly with $w$.

\subsection{Matching}

The $\rho$ factors in Eq.~(\ref{eq:rhoV_ratio}) enter in the
renormalization of the components of the transition vector current
$V_{cb}^{\mu}$.  As explained in Sec.~\ref{sec:PT} these factors are
estimated in one-loop lattice perturbation theory to the extent that
such calculations are available.  As discussed near the end of
Sec.~\ref{sec:chiral}, we build the uncertainty estimates of
Eqs.~(\ref{eq:rhov4error1}), (\ref{eq:rhov4error-quad})
and~(\ref{eq:rhoVierror}) into the chiral-continuum fit via
Eq.~(\ref{eq:Cstat-syst}).

A noteworthy feature of Table~\ref{tab:fpf0errbudget} is the size of
the matching error after the chiral-continuum fit.  Had we omitted the
errors in Eqs.~(\ref{eq:rhov4error1}), (\ref{eq:rhov4error-quad}),
and~(\ref{eq:rhoVierror}) from the fitting function, we would have to
add them \emph{a posteriori}, as we did for $B\to D^*$ at zero
recoil~\cite{B2Dstar}.  Following the procedure used in
Ref.~\cite{B2Dstar}, we would assign errors of 1.4\% and 1.1\% for
$f_+$ and $f_0$, respectively, at $w=1.16$, based on the second-finest
lattice with $a\approx0.06$~fm and its value of $\alpha_s=0.225$.
Incorporating the matching errors into the chiral-continuum fit,
however, allows them to vary with lattice spacing and to be informed
by the data.  It is reasonable that the additional information reduces
the uncertainty to about 0.7\% for both $f_+$ and $f_0$ at $w=1.16$,
as shown in Table~\ref{tab:fpf0errbudget}.

\subsection{Light-quark and gluon discretization errors}

Our improved actions have light-quark and gluon discretization errors
of order $\alpha_s a^2$ and $\alpha_s^2 a^2$~\cite{B2Dstar}.  As
discussed in Sec.~\ref{sec:chiral}, they are already included in the
fit model of Eqs.~(\ref{eq:h+chiral}) and (\ref{eq:h-chiral}).  From
Table~\ref{tab:fpf0errbudget}, the errors due to the truncation of the
chiral expansion and extrapolation to the continuum limit are about
0.6\% and 0.5\% for $f_+$ and $f_0$, respectively, at $w=1.16$.  Using
simple power-counting, we would conservatively estimate the size of
generic light-quark and gluon discretization errors on the $a \approx
0.06$~fm lattice to be about 1\%.  The data for $h_+$, which give the
dominant contribution to $f_+$ and $f_0$, do not display significant
lattice-spacing dependence.  Therefore, allowing the data to constrain
the possible size of light-quark and gluon discretization effects
reduces the error.

\begin{figure}
    \centerline{\includegraphics[width=0.33\textwidth]{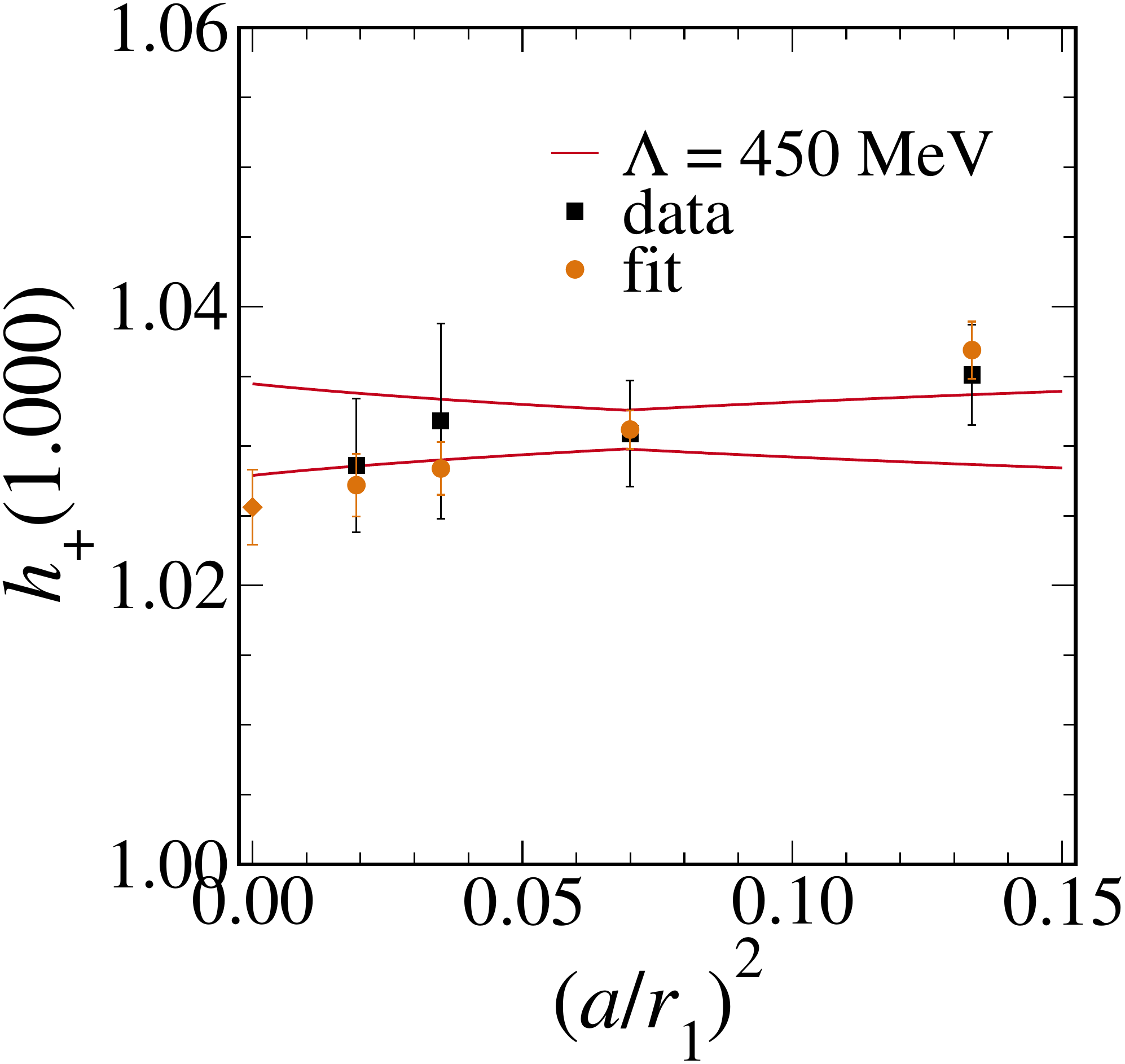}
    \includegraphics[width=0.33\textwidth]{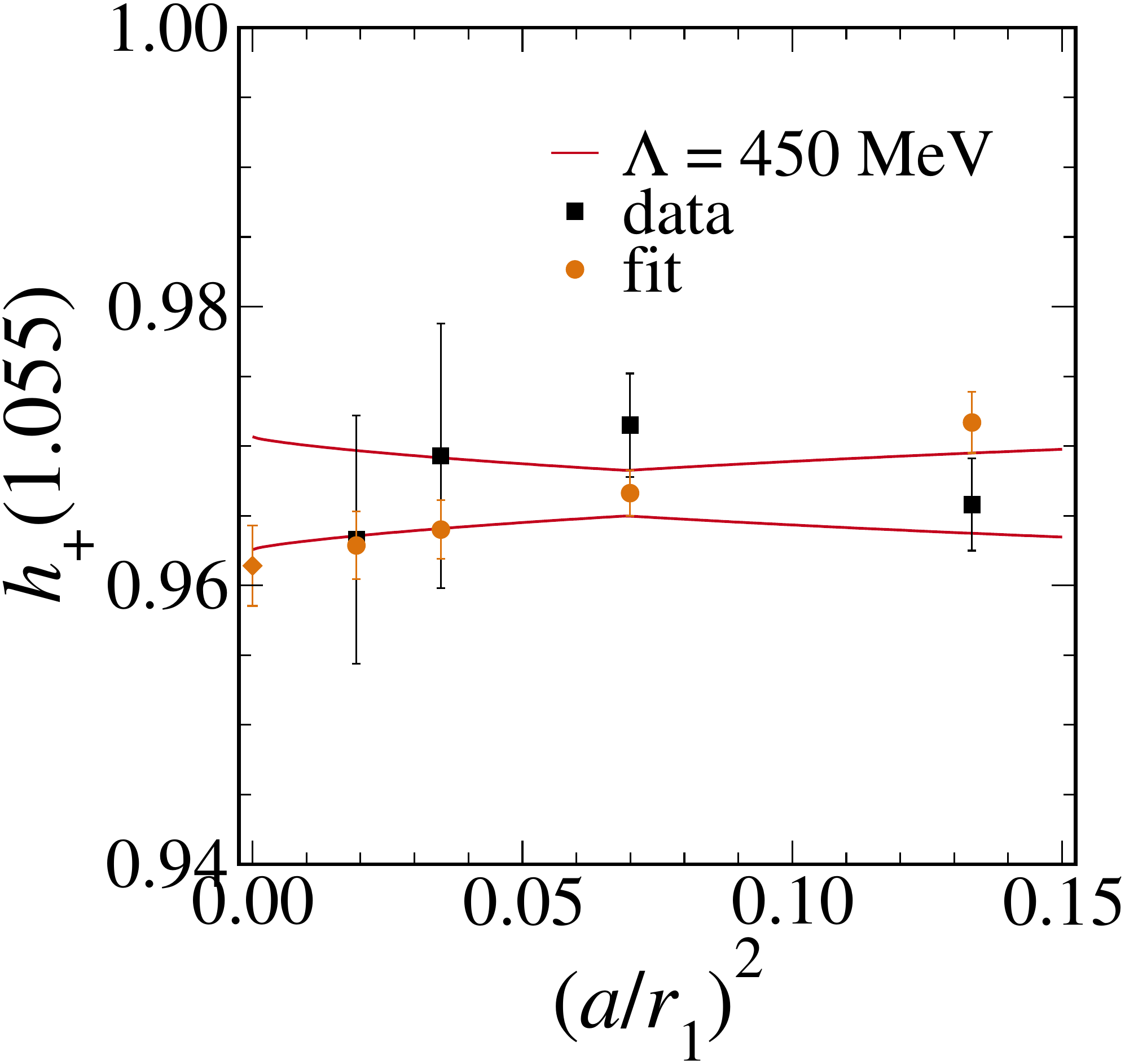}
    \includegraphics[width=0.33\textwidth]{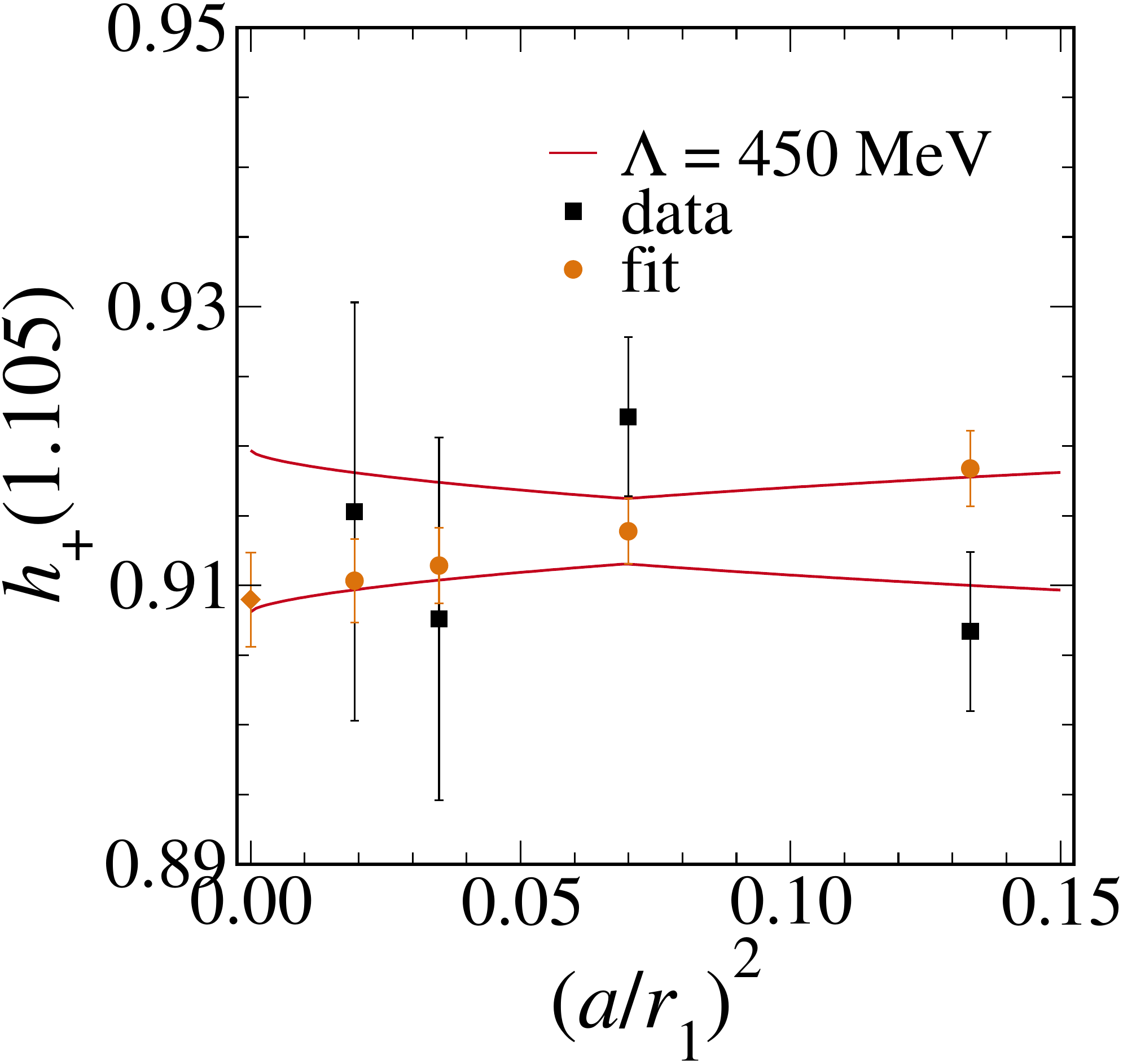}}
    \caption{The form factor $h_+(w)$ at three representative values
      of the recoil~$w$ as a function of the squared lattice spacing
      $(a/r_1)^2$, for all ensembles with
      $\hat{m}^\prime=0.2{m^\prime_s}$.  Black squares denote data
      points interpolated to the same recoil value, while orange
      circles denote fit values interpolated further, so that the
      light-quark masses $\hat{m}^\prime$ and $m_s^\prime$ correspond to the values
      on the lattice with $a\approx0.09$~fm; the orange diamonds
      denote the continuum limit in this case.  The solid curves show
      the $a$~dependence predicted by the HQET description of cutoff
      effects, with $\bar{\Lambda}=450$~MeV.  These trends are shown as
      deviations from the $a\approx0.09$~fm lattice.  For
      details, see the discussion of Tables~\ref{tbl:diffs}
      and~\ref{tbl:errors} in Appendix~\ref{app:hqerror}.  Note that
      the data and fit points reflect discretization errors from light
      quarks and gluons, as well as those from the heavy quarks.}
    \label{fig:hplusvsa}
\end{figure}

\subsection{Heavy-quark discretization errors}
\label{subsubsection:hqerr}

An important uncertainty comes from discretization errors in the
lattice treatment of the heavy quarks.  Applying the theory of
heavy-quark cutoff effects developed in
Refs.~\cite{Kronfeld:2000ck,Harada:2001fj}, we estimate the size of
these errors in Appendix~\ref{app:hqerror}, providing in
Tables~\ref{tbl:diffs} and~\ref{tbl:errors} numerical results for the
errors on $h_{\pm}(w)$ from mismatches in the lattice action and
currents for a heavy-quark scale of $\bar{\Lambda}=450$~MeV.  The value is
the same as that used in Ref.~\cite{B2Dstar}, and here we explain why
this choice is reasonable in this case too.

In Fig.~\ref{fig:hplusvsa}, we show the observed lattice-spacing
dependence of our simulation data for $h_+(w)$ at three recoil values
on the $\hat{m}'=0.2m'_s$ ensembles.  The raw data (black squares) are
adjusted slightly to obtain the same $w$ values for all~$a$ using a
chiral-continuum fit with the $\rho$-factor and $\kappa$-tuning errors
turned off.  Thus, the error bars shown here are statistical only.  We
also use this fit to adjust the light-quark masses to those on the
$\hat m^\prime = 0.2m^\prime_s$, $a\approx0.09$~fm ensemble (orange
circles).  (In practice, shifting the strange sea-quark mass has
little impact on the fit points.)  To compare the trend with the
expected heavy-quark discretization error, we draw the size of the
effect --- defined as the difference from $a\approx0.09$~fm ---
predicted in Appendix~\ref{app:hqerror} for $\bar{\Lambda}=450$~MeV.

For all values of $w$, Fig.~\ref{fig:hplusvsa} shows that this
estimate captures most of the discretization effect, given the
statistical scatter.  Note that the fit-interpolated (orange) points
make clear that the trend is predominantly linear in~$a^2$.  This
dependence is characteristic of generic discretization effects of the
light quarks and gluons, which are already included in the
chiral-continuum fit model, Eqs.~(\ref{eq:h+chiral}) and
(\ref{eq:h-chiral}).  Moreover, the heavy-quark discretization effects
turn out to be nearly linear in $a^2$, so they, too, are mostly
absorbed by the fit model.  It does not make sense to count this
well-modeled $a$~dependence twice by, say, inflating $\bar{\Lambda}$
to encompass all of the variation seen in Fig.~\ref{fig:hplusvsa}.
That said, we do not have an argument to reduce the value of
$\bar{\Lambda}$ used in Tables~\ref{tbl:diffs} and~\ref{tbl:errors}
below 450~MeV.  Following Ref.~\cite{B2Dstar}, we base our final
estimate on our next-to-smallest lattice spacing, $a\approx0.06$~fm,
leading to the error estimates in Table~\ref{tab:fpf0errbudget}. The
heavy-quark discretization error is found to be small compared with
the chiral-continuum extrapolation error.  For $h_+(w)$ it ranges from
approximately 0.15\% at $w = 1$ to 0.35\% at our largest $w$ values.
For $h_-(w)$ it is approximately 20\%.

\subsection{Lattice-scale error}\label{subsubsection:roneerr}

We use the distance scale $r_1$ and the relative lattice spacing
$a/r_1$ to determine the lattice scale $a$.  The ratio $a/r_1$ for the
ensembles in this study is known quite precisely from a fit to a wide
range of data for the heavy-quark potential \cite{Bazavov:2009bb}.
For this study we use the values of $r_1/a$ presented in Table~III
of~\cite{B2Dstar}.  The continuum, physical quark-mass value of $r_1$
is determined from studies of the light pseudoscalar-meson spectrum
and decay constants.  For this study we use $r_1 = 0.3117(22)$ fm,
based on the PDG value of $f_\pi$ \cite{Bazavov:2011aa}.

Because the form factors are dimensionless, the lattice scale enters
only weakly into their determination via: (1) tuning the heavy-quark
masses, (2) setting light-meson masses in the chiral logarithms, and
(3) fixing the location of the continuum limit.  To determine the
error due to uncertainties in $r_1$ we see how much our results shift
when we change $r_1$ by one standard deviation.  We find that the
changes in the form factors are smaller than 0.2$\%$.

\subsection{Finite-volume corrections}

The finite-volume effects can be estimated within NLO heavy-light meson $\chi$PT
by replacing the loop integrals with discrete sums.  The corrections to the
integrals in the formulas appearing in $B\to D$ decays at zero recoil were
worked out by Arndt and Lin \cite{Arndt:2004bg}.  At the values of
quark masses and volumes at zero recoil where we have data, the
effects predicted by $\chi$PT are less than one part in $10^4$.  This
is not a result of cancellation, but is due to the fact that the
chiral logarithms make only a very small contribution to the form
factor.  We did not calculate the finite-volume corrections at
nonzero recoil because the integrals appearing in those formulas are
much more complicated, but there is no reason to expect these effects
to be significantly enhanced away from the zero-recoil point.  Thus,
finite-size effects are expected to be negligible compared with our
other errors, and we do not assign any additional error due to them.

\subsection{Light-quark-mass tuning}
\label{subsection:light-sea-tuning}

We extrapolate the form factors to the physical average of the up- and
down-quark masses $r_1 \hat m = 0.003612(126)$, determined from an
analysis of the light pseudoscalar-meson spectrum and decay constants
on the same ensembles~\cite{Bazavov:2010hj}.  Varying $r_1
\hat m$ by plus and minus 1$\sigma$ in our chiral-continuum fit
leads to relative changes of order $10^{-5}$ for both form
factors in the range of simulated recoil values.

On some ensembles the strange sea-quark mass deviates by as much as
30$\%$ from its physical value.  From heavy-light meson $\chi$PT, we
expect the $B\to D$ form factors to be largely insensitive to
sea-quark masses.  Nevertheless we study the impact of the strange
sea-quark mass by calculating the ratios in
Eqs.~(\ref{eq:Rplus})--(\ref{eq:xf}) on an $a \approx 0.12$~fm
ensemble with an unphysically-light strange sea quark,
$a\hat{m}^\prime/am_s^\prime = 0.005/0.005$. We do not observe any
statistically-significant differences in these ratios from those on
the $a\hat{m}^\prime/am_s^\prime = 0.005/0.05$ ensemble.  We therefore
conclude that errors from mistuning the strange sea-quark mass are
negligible within our current precision.

\subsection{Heavy-quark-mass tuning}

As described in Sec.~\ref{sec:kappa}, we adjust the simulation data
before the chiral-continuum fit to account for the slight difference
between the simulated bottom and charm $\kappa$ values and the
physical ones, using the corrections estimated in
Appendix~\ref{app:kappa}.  The size of these corrections is quite
small, ranging from 0 to 0.2\% for $h_+$, and from 0 to 2\% for $h_-$.
Repeating the chiral-continuum fit omitting the $\kappa$ corrections
does not appreciably change the chiral-continuum fit result.  Thus we
conclude that the uncertainty in the form factors due to errors in the
heavy-quark masses is negligible.

\subsection{Isospin correction}
\label{subsubsection:Ic}

In our calculation we have assumed that the up- and down-quark masses
are equal, although in nature, they are not. Therefore, if we
distinguish between them in calculating the value of the form factors
at the physical point, we get a slightly different result. To estimate
the sensitivity of $f_+$ and $f_0$ to isospin splitting of the
light-quark masses, we use our best-fit parameters in the
chiral-continuum model, and evaluate the fit function at the physical
values of $r_1m_u = 0.002236$ and $r_1 m_d = 0.004988$, instead of
$r_1 \hat m$ given above.  These values are obtained by combining $r_1
\hat m$ obtained on the asqtad ensembles with the ratio $m_u/m_d =
0.4482\left(^{+173}_{-207} \right)$ obtained from the MILC
Collaboration's study of electromagnetic effects on the pion and kaon
mass-splittings on the $(2+1+1)$-flavor HISQ
ensembles~\cite{Basak:2014vca}. The relative shifts in both form
factors for all simulated recoil values are of order $10^{-4}$, and
therefore negligible.  Although this method varies the light valence-
and sea-quark masses together, the shifts are primarily due to the
different valence-quark mass.
  
%%%%%%%%%%%%%%%%%%%%%%%%%%%%%%%
\section{Determination of $|V_{cb}|$}
\label{sec:Vcb}
%%%%%%%%%%%%%%%%%%%%%%%%%%%%%%%

\begin{table}
  \caption{ Selected values of the form factors $f_+(w)$ and $f_0(w)$
    at the physical point (synthetic data) and their correlations.
    Errors shown include statistics and all systematics added in quadrature.}
\label{tab:synthetic}  
\begin{tabular}{r|c|rrrrrr}
\hline\hline
                 && \multicolumn{6}{c}{Correlation matrix} \\
               & value     & $f_+(1)$ & $f_+(1.08)$ & $f_+(1.16)$  & $f_0(1)$ & $f_0(1.08)$ & $f_0(1.16)$ \\ 
  \hline
  $f_+(1)$     & 1.1994(095)  & 1.0000 & 0.9674 & 0.8812 & 0.8290 & 0.8533 & 0.8032 \\
  $f_+(1.08)$  & 1.0941(104)  &        & 1.0000 & 0.9523 & 0.8241 & 0.8992 & 0.8856 \\
  $f_+(1.16)$  & 1.0047(123)  &        &        & 1.0000 & 0.7892 & 0.8900 & 0.9530 \\
  $f_0(1)$     & 0.9026(072)  &        &        &        & 1.0000 & 0.9650 & 0.8682 \\
  $f_0(1.08)$  & 0.8609(077)  &        &        &        &        & 1.0000 & 0.9519 \\
  $f_0(1.16)$  & 0.8254(094)  &        &        &        &        &        & 1.0000 \\ 
  \hline
  \hline
\end{tabular}
\end{table}  

\subsection{Synthetic data}

The preferred chiral-continuum fit results for $h_+(w)$ and $h_-(w)$
are continuous functions of~$w$ at zero lattice spacing and physical
quark masses.  Via Eqs.~(\ref{eq:f+fromh}) and~(\ref{eq:f0fromh}), we
can obtain the corresponding functions for $f_+(w)$ and~$f_0(w)$.  As
discussed above, the errors are under control for $w<1.2$, {\it i.e.},
where we have lattice measurements. Following
Refs.~\cite{Bailey:2008wp,Bailey:2012rr}, we proceed to extend our
results to the full kinematic range by generating synthetic data
$f_+(w_j)$ and $f_0(w_j)$ for a finite set of $w$ values, $w_j$.
Because the functions are described by only six independent functions
(in the physical limit), we can only generate six such data points.
Generating more would just lead to a covariance matrix of low rank.
We choose the values of $w_j = 1$, 1.08, and 1.16, for $f_+$ and
$f_0$, to cover the kinematic range of the lattice-QCD calculation.
The values of $f_+(w_j)$ and $f_0(w_j)$, as well as the matrix of
correlations among them, are given in Table~\ref{tab:synthetic}.

\subsection{$ z $ expansion}%
\label{sec:zexp}

Experimental measurements of the form factor are available over a
larger kinematic range of $w$ [1, 1.58] than the lattice values
[1,1.16], but experimental errors are largest where lattice errors are
small and vice versa.  Although the value of $f_+$ at a single
$w$-value suffices for obtaining $|V_{cb}|$, a better strategy is to
fit both sets of data simultaneously to a common fitting function in
which $|V_{cb}|$ is a free parameter that multiplies all the lattice
values and is determined in the fit
\cite{Bailey:2008wp,Bailey:2012rr}.  This approach minimizes the
uncertainty in $|V_{cb}|$ by combining all of the available
experimental and lattice information.  Further, a comparison of the
shapes of the experimental and lattice results as a function of $w$
provides a valuable consistency check that is not available when using
only a single recoil point.

For this purpose we need a model-independent parameterization to carry
out the necessary interpolation/extrapolation.  The $ z $~expansion of
Boyd, Grinstein and Lebed (BGL) \cite{Boyd:1994tt} is just such a
parameterization.  It builds in constraints from analyticity and
unitarity.  It is based on the conformal map
\begin{equation}
    z(w) = \frac{\sqrt{1+w}-\sqrt{2}}{\sqrt{1+w}+\sqrt{2}} \, ,
\end{equation}
which takes the physical region $ w \in [1, 1.59] $ to $ z \in
[0,0.0644]$.  It pushes poles and branch cuts relatively far away to
$|z| \approx 1 $.  Form factors are then parameterized as
\begin{equation}
    f_i(z) = \frac{1}{P_i(z) \phi_i(z)} \sum_{n=0}^{\infty} a_{i,n} z^n  \, ,
\label{eq:zexp}
\end{equation}
where the $ P_i(z) $ are the ``Blaschke factors'' containing explicit
poles ({\it e.g.}, a $B_c$ or $B_c^*$ meson) in the channel variable
$q^2$, and the $ \phi_i $ are the ``outer functions'', whose purpose
is described below.  The only unknown parameters are the polynomial
coefficients $a_{i,n}$.  In this work, we do not introduce any pole,
so $ P_i(z)=1 $.\footnote{We have checked that including a pole
  located at the theoretically-predicted $B_c^*$
  mass~\cite{Gregory:2009hq} does not appreciably change the $z$-fit
  result.}
The choice of outer functions is arbitrary as long as they are
analytic functions that do not introduce poles or branch cuts; the
$\phi_i$ just affect the numerical values of the series coefficients,
$a_i$.  For $ f_+ $ and $ f_0 $, we use 
\begin{eqnarray}
    \phi_+(z)   &=      &\Phi_+(1+z)^2(1-z)^{1/2}[(1+r)(1-z)+2\sqrt{r}(1+z)]^
    {-5}\,, \\
    \phi_0(z)   &=      &\Phi_0(1+z)(1-z)^{3/2}[(1+r)(1-z)+2\sqrt{r}(1+z)]^
    {-4}\,,
\end{eqnarray}
such that, numerically, $ \Phi_0=0.5299 $ and $ \Phi_+=1.1213 $
\cite{Boyd:1994tt}.  With this choice, the bound on the series
coefficients from unitarity takes a particularly simple form:
\begin{equation}
    \sum_{n=0}^N |a_{i,n}|^2 \le 1 \, ,
\label{eq:unitarity}
\end{equation}
where this bound holds for any $N$.  This bound, in combination with
the small range of $|z|$, ensures that only a small number of
coefficients is needed to parameterize the form factors over the
entire kinematic range to high precision.

\begin{figure}
    \centerline{
    \includegraphics[width=0.5\textwidth]{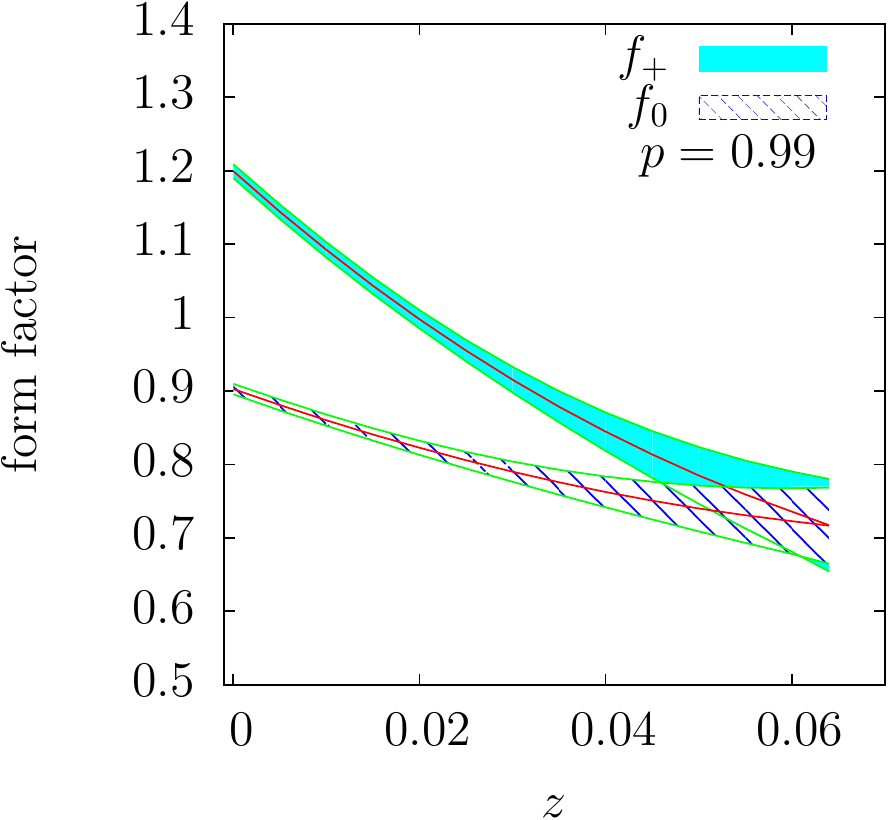}
    \includegraphics[width=0.5\textwidth]{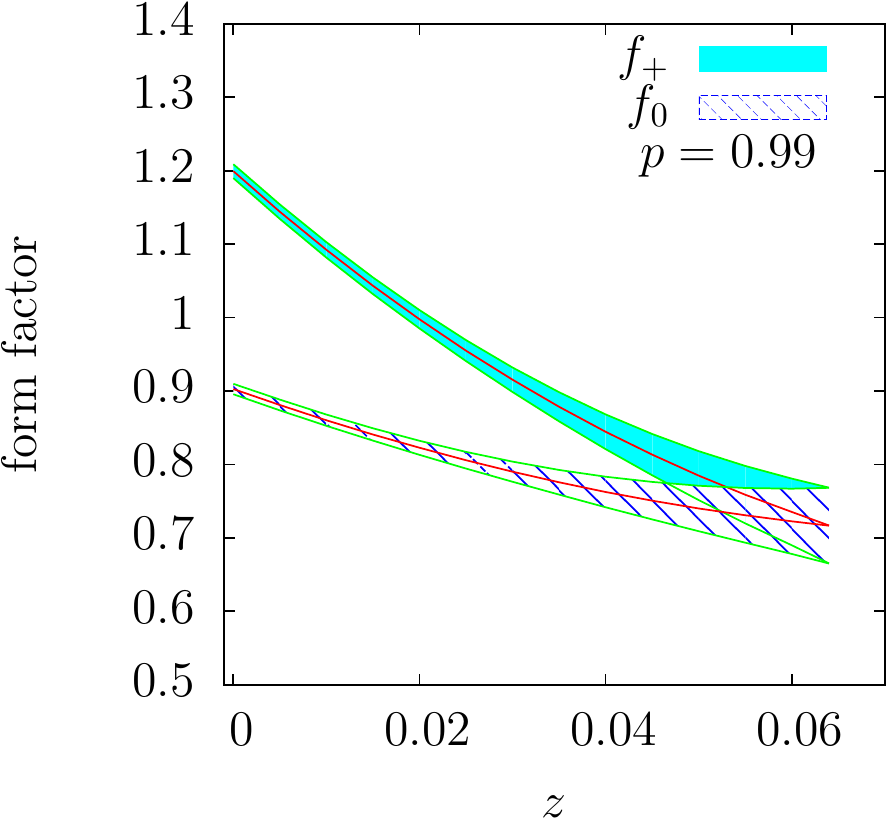}
    }
    \caption{Result of the $ z $-expansion fit of the lattice
      form-factor values without (left) and with (right) the kinematic
      constraint $f_+(q^2=0) = f_0(q^2=0)$.  The expansion is
      truncated after the cubic term.  The solid error band is for $
      f_+ $, while the slashed band is for $ f_0 $. Without imposing
      the constraint, we find that it is nonetheless satisfied to a
      high accuracy.}
    \label{fig:sysfit}
\end{figure}

To implement the $ z $ expansion, we start from the synthetic data for
$ f_+ $ and $ f_0 $ at $ z $ values corresponding to $w_j = 1$, 1.08,
and 1.16, choose a truncation $N$ and fit to determine the
coefficients $ a_{i,n} $ for $n = 0,\ldots{},N$.  These coefficients
are then used to parameterize the form factors over the full kinematic
range.  We find we need only the first few coefficients in the
expansion to obtain a stable fit with a good $p$ value.  The kinematic
constraint requires $ f_+ = f_0 $ at $ q^2 = 0 $ where $ z \approx
0.0644 $.  It is interesting to fit the data without the constraint to
see to what extent it is automatically satisfied.  The result for
$N=3$ in the left panel of Fig.~\ref{fig:sysfit} shows that the data
satisfy the constraint much better than our statistics would suggest.
Nonetheless, in subsequent fits, we include the constraint to reduce
the form-factor errors at $q^2=0$.  The constraint is imposed by
expressing the parameter $a_{0,0}$ in Eq.~(\ref{eq:zexp}) in terms of
the other series coefficients.  Table~\ref{tab:z_expansion_coeff1}
shows the series coefficients and goodness-of-fit obtained for fits of
the lattice form-factor data imposing the kinematic constraint with
$N=2$--4.  For the fits at cubic and quartic order in the $z$
expansion, we have more parameters than data, but the unitarity bound
in Eq.~(\ref{eq:unitarity}) justifies imposing a prior with central
value 0 and width 1 on the coefficient(s) of the cubic (and quartic)
term(s).

\begin{table}
    \caption{Coefficients of the $z$ expansion for fits to the lattice
      form factors including the kinematic constraint
      $f_+(q^2=0)=f_0(q^2=0)$. For completeness, the inferred value
      and error in $a_{0,0}$ is quoted.  We also show the zero-recoil
      form factor $\mathcal{G}(1)$.  The results for different
      truncations $N$ are virtually identical.  The unusually low
      (augmented) $\chi^2$ comes about because these fits essentially
      behave like solves.  This happens because the kinematic
      constraint is so nearly perfectly satisfied already at the
      quadratic level, $N=2$.  Higher-order terms with $N = 3$ and~4
      provide no further improvement and, hence, no change.}
    \label{tab:z_expansion_coeff1}
    \begin{center}
         \begin{tabular}{cr@{.}lr@{.}lr@{.}l}
         \hline\hline
        & \multicolumn{2}{c}{$N=2$} &\multicolumn{2}{c}{$N=3$} &\multicolumn{2}{c}{$N=4$} \\
              \hline
   $a_{+,0}$          & 0&01262(10)  & 0&01262(10) & 0&01262(10)    \\
   $a_{+,1}$          & $-$0&097(3)  & $-$0&097(3) & $-$0&097(3)    \\
   $a_{+,2}$          & 0&50(14)     & 0&50(17)    & 0&50(17)       \\
   $a_{+,3}$          & \NA          & $-$0&06(90) & $-$0&06(90)    \\
   $a_{+,4}$          & \NA          & \NA	   & $-$0&0(1.0)    \\\hline
   $a_{0,0}$          & 0&01142(14)  & 0&01142(14) & 0&01142(10)    \\
   $a_{0,1}$          & $-$0&060(3)  & $-$0&060(3) & $-$0&060(3)    \\
   $a_{0,2}$          & 0&31(15)     & 0&31(15)    & 0&31(15)       \\
   $a_{0,3}$          & \NA          & 0&06(91)    & 0&06(91)       \\
   $a_{0,4}$          & \NA          & \NA         & 0&0(1.0)     \\\hline
   $\mathcal{G}(1)$   & 1&0541(83)   & 1&0541(83)  & 1&0541(83)     \\\hline
   $\chi^2/df$        & 0&1/1        & 0&0/1       & 0&0/1          \\
             \hline
             \hline
          \end{tabular}
    \end{center}
\end{table}

\begin{table}
  \caption{Central values, errors, and correlation matrix for the
    parameters of the cubic fit to $f_+$ and $f_0$ including the
    kinematic constraint at $q^2=0$.}
  \label{tab:N3corr}
  \begin{tabular}{r|r@{.}l|rrrrrrr}
  \hline\hline
                 &&& \multicolumn{7}{c}{Correlation matrix} \\
 & \multicolumn{2}{c|}{value} & $a_{+,0}$    & $a_{+,1}$    & $a_{+,2}$    & $a_{+,3}$ &    $a_{0,1}$    & $a_{0,2}$    & $a_{0,3}$ \\
   \hline
 $a_{+,0}$ & $ 0$&01262(10) & $  1.00000 $ & $ 0.21726 $ & $  0.07203 $ & $  0.00387 $ & $  0.19347 $ & $  0.15590 $ & $ -0.00364 $ \\
 $a_{+,1}$ & $-$0&0969(34)  & $          $ & $ 1.00000 $ & $ -0.47505 $ & $  0.25544 $ & $  0.80946 $ & $ -0.26302 $ & $ -0.18212 $ \\
 $a_{+,2}$ & $ 0$&50(17)    & $          $ & $         $ & $  1.00000 $ & $ -0.45415 $ & $ -0.43845 $ & $  0.85491 $ & $  0.25116 $ \\
 $a_{+,3}$ & $-$0&06(90)    & $          $ & $         $ & $          $ & $  1.00000 $ & $  0.11415 $ & $ -0.15582 $ & $  0.21768 $ \\
 $a_{0,1}$ & $-$0&0597(29)  & $          $ & $         $ & $          $ & $          $ & $  1.00000 $ & $ -0.42932 $ & $ -0.03556 $ \\
 $a_{0,2}$ & $ 0$&31(15)    & $          $ & $         $ & $          $ & $          $ & $          $ & $  1.00000 $ & $ -0.06062 $ \\
 $a_{0,3}$ & $ 0$&06(91)    & $          $ & $         $ & $          $ & $          $ & $          $ & $          $ & $  1.00000 $ \\
  \hline
  \hline
  \end{tabular}
\end{table}

The truncation of the $z$ expansion introduces a possible systematic
error.  We take this into account by increasing the truncation order
until the central values and errors stabilize.  At this point, the
errors from the fit reflect the truncation error, and do not need to
be counted separately.  Table~\ref{tab:z_expansion_coeff1} shows that
the fit has stabilized by quadratic order.  We therefore take the
cubic fit, shown in the right panel of Fig.~\ref{fig:sysfit}, as our
preferred parameterization.  Table~\ref{tab:N3corr} gives the central
values, errors, and normalized correlation matrix for the series
coefficients $a_i$.  This information can be used to reproduce our
results for $f_+(w)$ and $f_0(w)$ over the full kinematic range, and in
particular, in combined lattice-and-experiment fits to obtain
$|V_{cb}|.$

\begin{figure}
    \centerline{
    \includegraphics[width=0.5\textwidth]{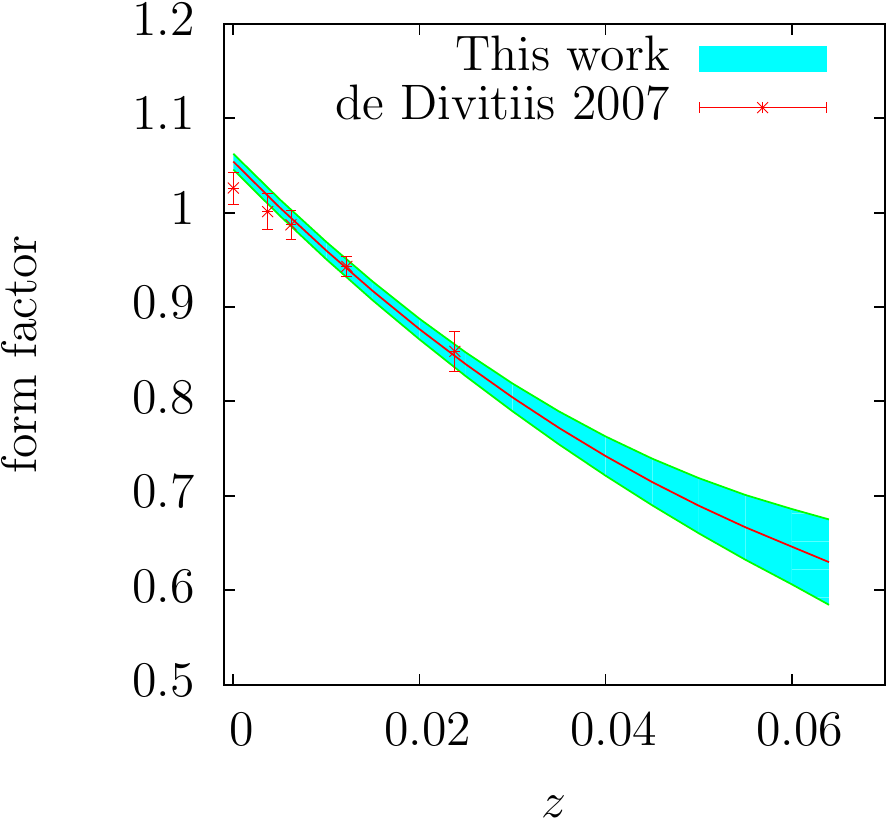}
    }
    \caption{Comparison of lattice-QCD results for the $B\to D\ell\nu$
      form factor $\mathcal{G}(z)$ at nonzero recoil from this work
      (curves with error bands) and Ref.~\cite{deDivitiis:2007ui}
      (points with error bars).  Errors on the data points from
      Ref.~\cite{deDivitiis:2007ui} include all uncertainties except
      for the unquantifiable error due to omitting sea-quark effects.}
    \label{fig:LQCDCompare}
\end{figure}

We compare our form-factor results with those of the most recent
lattice-QCD calculation of $B\to D \ell\nu$ at nonzero recoil in
Fig.~\ref{fig:LQCDCompare}.  Although this earlier calculation was
performed in quenched QCD, and thus is subject to an unquantifiable
systematic due to the omission of sea-quark effects, it uses
step-scaling~\cite{Guagnelli:2002jd} to control heavy-quark
discretization effects, plus multiple light-quark masses and lattice
spacings to control the mild chiral-continuum
extrapolation~\cite{deDivitiis:2007ui}.  Thus it is the best
calculation so far for $B\to D\ell\nu$ at nonzero recoil.  The two
calculations agree for all $w$ values, although the slope of $f_+(z)$
is somewhat steeper for the (2+1)-flavor result reported here.

\subsection{Determination of $|V_{cb}|$}

To obtain $|V_{cb}|$, we need lattice results for the form factors and
experimental values for $\bar \eta_{EW} |V_{cb}|f_+(w)$.  Because the
experimental value of the form factor at zero recoil suffers from
kinematic suppression, we prefer to fit the theoretical and
experimental data over the entire kinematic range.  For this work, we
use the 2009 $B$-tagged data from the BaBar
collaboration~\cite{Aubert:2009ac}, because it is the most precise to
date.\footnote{The Belle experiment presented preliminary measurements
  of $\bar \eta_{EW} |V_{cb}|f_+(w)$ at ICHEP
  2014~\cite{BelleICHEP14}.  Once these are finalized, our form-factor
  coefficients from Table~\ref{tab:N3corr} can be used to update
  $|V_{cb}|$ from a joint lattice-experiment fit with both the Belle
  and BaBar data (including experimental correlations).}
Reference~\cite{Aubert:2009ac} reports a systematic error of 3.3\% at
small $w$.  For present purposes, we take 3.3\% over the entire
kinematic range with 100\% correlation and combine this systematic in
quadrature with the reported (uncorrelated) statistical errors
\cite{Rotondo}.

Although the BaBar collaboration has applied some radiative
corrections to their published data, additional electroweak effects
still remain.  These include a Sirlin factor for the $W\gamma$ and
$WZ$ box diagrams \cite{Atwood:1989em} and a further Coulomb
correction for final-state interactions in $B^0$ decays.  The BaBar
collaboration reports that 37\% of the decays in their data sample
were $B^0$s, which results in a QED correction factor in the amplitude
of $1 + 0.37 \alpha/(2\pi)$.  We have assigned an uncertainty of $\pm
0.005$ to this correction to account for omitted electromagnetic
effects at intermediate distances.  When combined with the Sirlin
factor $\eta_{\rm EW} = 1.00662$ the net electroweak correction
becomes $\bar \eta_{\rm EW} = 1.011(5)$.  (We prefer to use
$\mathcal{G}(w)$ to denote the purely hadronic form factor, so in our
notation $\bar\eta_{EW}|V_{cb}|\mathcal{G}(w)$ corresponds to the
quantity often reported as $|V_{cb}|\mathcal{G}(w)$, and the ratio of
experimental to theoretical values must be divided by $\bar \eta_{\rm
  EW}$ to get $|V_{cb}|$.)

\begin{figure}
    \centerline{
    \includegraphics[width=0.475\textwidth]{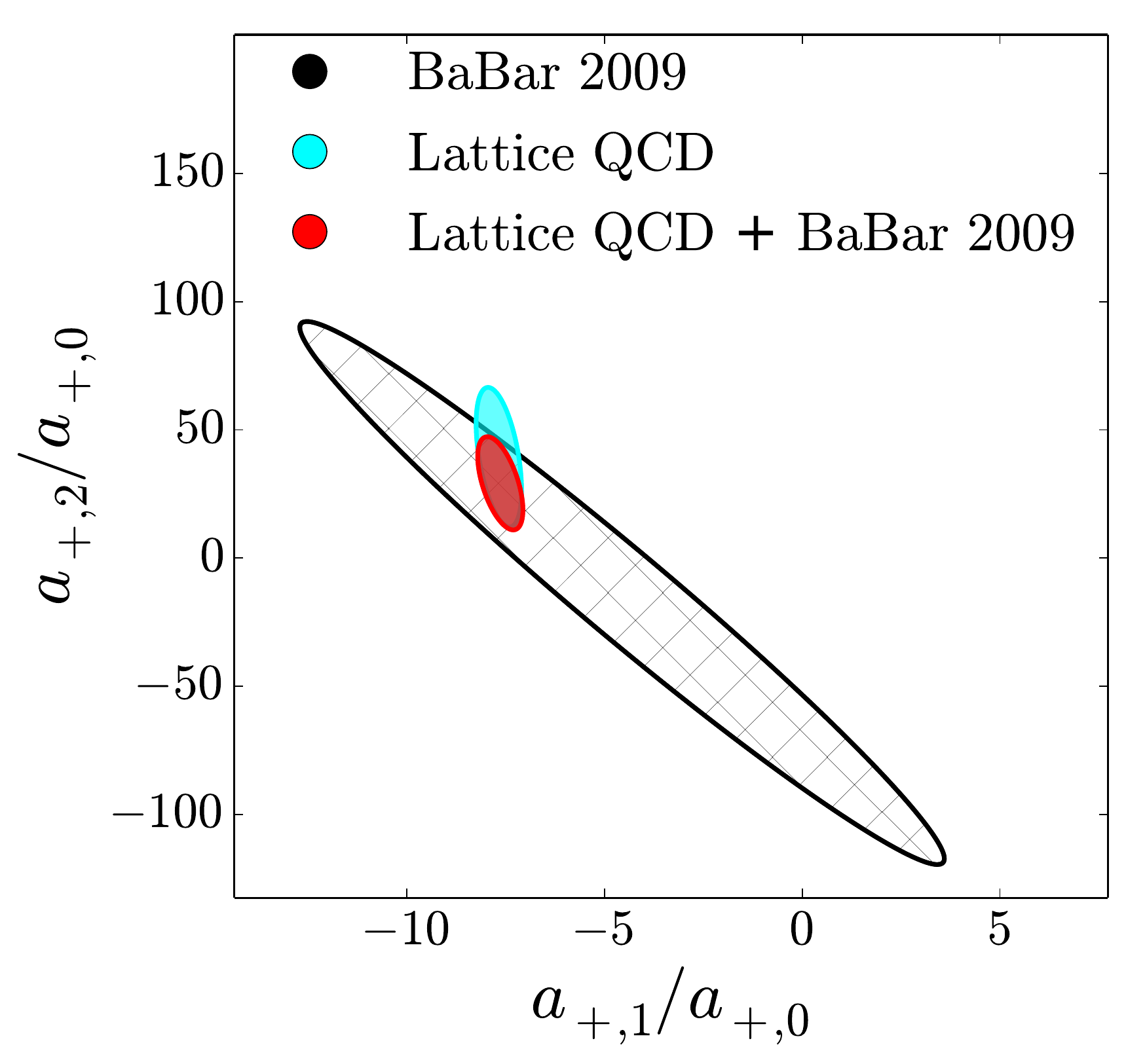} \hfill 
    \includegraphics[width=0.485\textwidth]{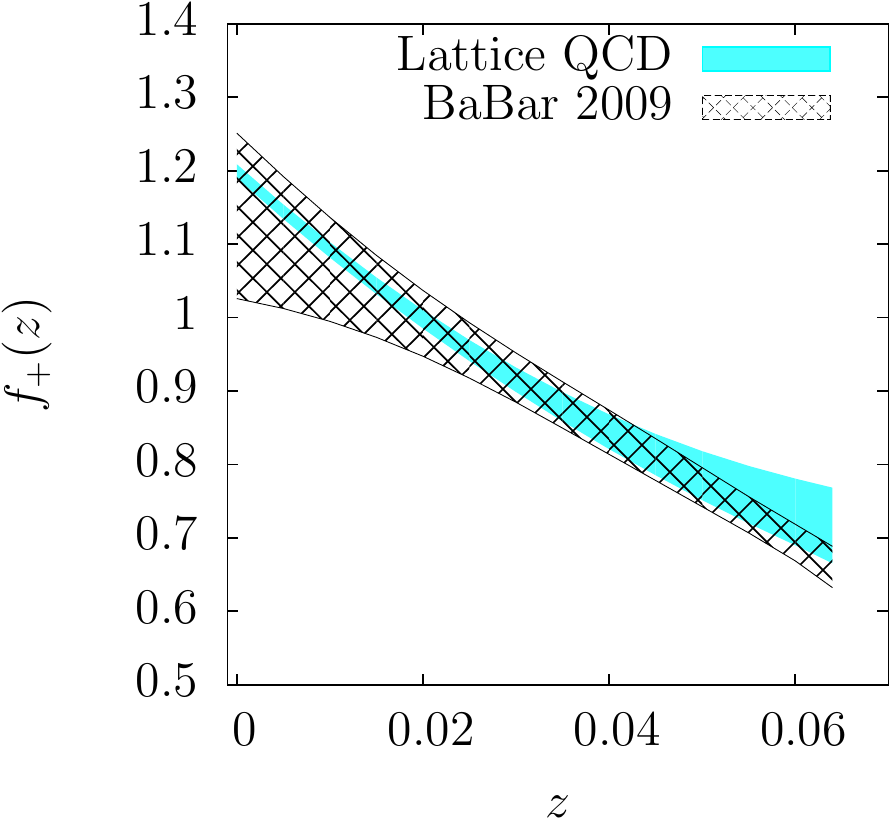}}
    \caption{Left: One sigma contour plots showing the
        correlation between the normalized slope $a_{+,1}/a_{+,0}$ and
          normalized curvature $a_{+,2}/a_{+,0}$ from $N = 3$
          $z$-expansion fits to either the BaBar experimental data
          alone, our lattice QCD results alone, and a joint fit to
          both. Right: vector form factor $f_+$ obtained from separate
          $z$-expansion fits of the 2009 BaBar experimental data
          (hatched band) and lattice form factors (solid band).}
    \label{fig:shape}
\end{figure}

Before performing a joint fit to the lattice and experimental data, we
compare the values of the shape parameters to check for consistency.
The left panel of Fig.~\ref{fig:shape} plots the 1-$\sigma$
constraints on the curvature $a_{+,2}/a_{+,0}$ versus slope
$a_{+,1}/a_{+,0}$ obtained from separate $N=3$ $z$-expansion fits of
the lattice data and the 2009 BaBar experimental data.  The results
are consistent, but the lattice data constrains the shape much better:
this is both because the lattice points are very precise at low
recoil, and because they are more correlated between $w$ values.
Given this consistency, we now proceed with the determination of
$|V_{cb}|$ from a combined fit of the two data sets.

\begin{figure}
    \includegraphics[width=0.5\textwidth]{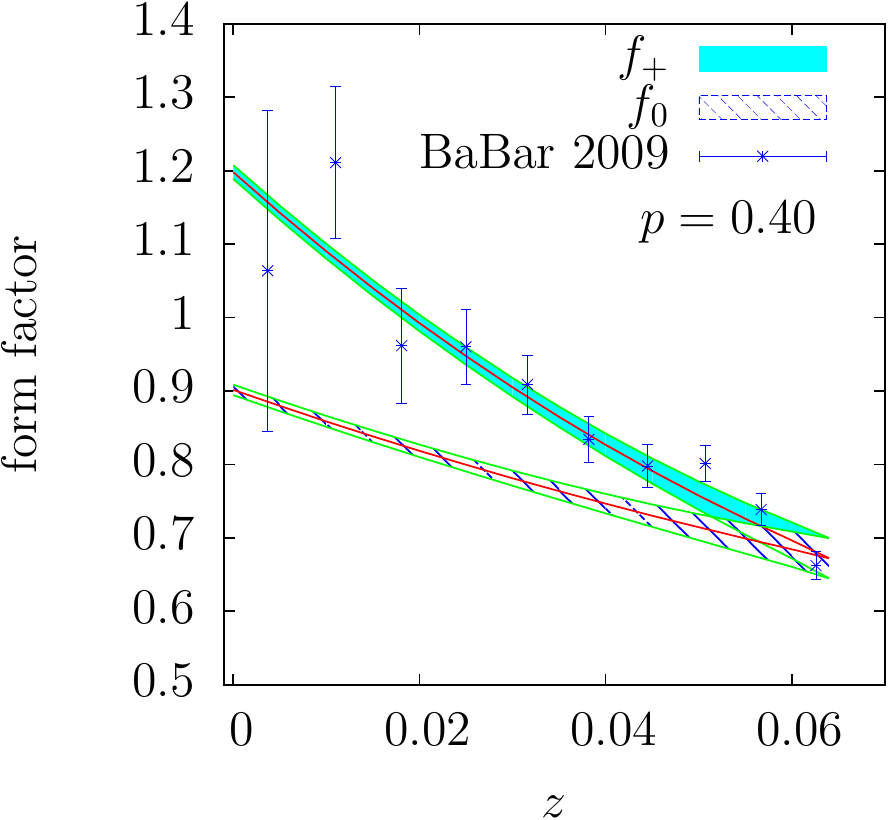}
    \caption{ Result of the preferred joint fit of the BaBar
      experimental data together with the lattice form factors. The
      plotted experimental points have been divided by our best-fit
      value of $\bar \eta_{\rm EW} |V_{cb}|$ and converted to $f_+$.}
    \label{fig:xf6}
\end{figure}

Table~\ref{tab:z_expansion_coeff2} shows the series coefficients and
goodness-of-fit obtained for combined fits of the lattice and
experimental data, imposing the kinematic constraint, for $N=2$--4.
Again, the fit, and in particular the error on $|V_{cb}|$, stabilizes
by quadratic order.  We choose $N=3$ for our preferred fit, and plot
the result in Fig.~\ref{fig:xf6}.

\begin{table}
    \caption{Best-fit values of the $z$-expansion parameters for
      different truncations $N$ from a joint fit to experimental data
      and lattice values. For completeness, the inferred value and
      error in $a_{0,0}$ is quoted. We also show the zero-recoil form
      factor $\mathcal{G}(1)$ and $|V_{cb}|$. }%
     \label{tab:z_expansion_coeff2}
    \begin{center}
         \begin{tabular}{cr@{.}lr@{.}lr@{.}l}
              \hline
              \hline
                 &  \multicolumn{2}{c}{$N=2$} &\multicolumn{2}{c}{$N=3$} &\multicolumn{2}{c}{$N=4$} \\
        \hline
    $a_{+,0}$          & 0&01260(10)  & 0&01261(10) & 0&01261(10)     \\
    $a_{+,1}$          & $-$0&096(3)  & $-$0&096(3) & $-$0&096(3)     \\
    $a_{+,2}$          & 0&37(8)      & 0&37(11)    & 0&37(11)        \\
    $a_{+,3}$          & \NA          & $-$0&05(90) & $-$0&05(90)     \\
    $a_{+,4}$          & \NA          & \NA         & $-$0&0(1.0)     \\\hline
    $a_{0,0}$          & 0&01140(9)   & 0&01140(9)  & 0&01140(9)      \\
    $a_{0,1}$          & $-$0&059(3)  & $-$0&059(3) & $-$0&059(3)     \\
    $a_{0,2}$          & 0&18(9)      & 0&19(10)    & 0&19(10)        \\
    $a_{0,3}$          & \NA          & $-$0&3(9)   & $-$0&3(9)       \\
    $a_{0,4}$          & \NA          & \NA         & $-$0&0(1.0)     \\\hline
    $\mathcal{G}(1)$   & 1&0527(82)   & 1&0528(82)  & 1&0528(82) \\
    $|V_{cb}|$         & 0&0396(17)   & 0&0396(17)  & 0&0396(17)      \\\hline
    $\chi^2/df$        & 8&4/10       & 8&3/10      & 8&3/10          \\
              \hline
              \hline
          \end{tabular}
    \end{center}
\end{table}

\subsection{Comment on the CLN parameterization}

The standard approach used by experimentalists to obtain $|V_{cb}|$ is
to use the Caprini, Lellouch, Neubert (CLN)
parameterization~\cite{Caprini:1997mu} to extrapolate the experimental
data to $w=1$.  Caprini, Lellouch, and Neubert use heavy-quark
symmetry to derive more stringent constraints on the coefficients of
the $z$-parameterization through ${\mathcal O}(z^3)$, resulting in a
function with only two free parameters, $f_+(0)$ and $\rho_1^2$:
\begin{equation}
  \frac{f_+(z)}{f_+(0)} = 1 - 8 \rho_1^2 z + (51 \rho_1^2 - 10) z^2 - (252 \rho_1^2 - 84) z^3 \,.
\label{eq:CLN}
\end{equation}
Use of the CLN parameterization in our analysis does not reduce the
quoted errors in $|V_{cb}|$ despite the introduction of additional
theoretical information.

The numerical values of the coefficients in Eq.~(\ref{eq:CLN}) have
theoretical uncertainties which can be estimated from the information
given in tables and plots from Ref.~\cite{Caprini:1997mu}.  To the
best of our knowledge, however, CLN fits to experimental data do not
incorporate the theoretical uncertainties discussed in
Ref.~\cite{Caprini:1997mu}, and may therefore be underestimating the
uncertainty in $|V_{cb}|$. We have attempted to quantify the
uncertainty from the use of the CLN form by incorporating the
theoretical uncertainties in the CLN parameters via Bayesian priors.
We did not find any difference in the error on $|V_{cb}|$ obtained
from fits with and without including these theoretical uncertainties
at the current level of precision. This is primarily because the $B\to
D \ell \nu$ data displays little evidence of curvature in $z$ within
the present errors, and does not constrain the coefficient of the
$z^3$ term.  Nevertheless, we do not quote the results of our CLN fits
in this work because we are more confident in the errors obtained from
the model-independent $z$-parameterization, Eq.~(\ref{eq:zexp}), which
can be used to obtain $|V_{cb}|$ even as the experimental and lattice
uncertainties become arbitrarily more precise.

%%%%%%%%%%%%%%%%%%%%%%%%%%%%%%%
\section{Discussion and outlook}
\label{sec:discussion}
%%%%%%%%%%%%%%%%%%%%%%%%%%%%%%%

We obtain
\begin{equation}
 |V_{cb}|=(39.6 \pm 1.7_{\rm QCD+exp} \pm 0.2_{\rm QED})\times 10^{-3}
\label{eq:Vcb}
\end{equation}
from our analysis of the exclusive decay $ B\rightarrow Dl\nu $ at
nonzero recoil, where the first error combines systematic and
statistical errors from both experiment and theory and the second
comes from the uncertainty in the correction for the final state
Coulomb interaction in the $B^0$ decays.  Because we provide the
series coefficients of a $z$ parameterization and their correlations,
the result for $|V_{cb}|$ in Eq.~(\ref{eq:Vcb}) can be updated
whenever new experimental information becomes available.

The combined error from lattice and experiment in $|V_{cb}|$ is about
4\%.  Because this error is obtained from a joint $z$-fit, the theory
and experimental errors cannot be strictly disentangled, but they can
be estimated as follows.  In the right panel of Fig.~\ref{fig:shape}
we plot the determinations of $f_+$ from separate $z$ fits to the
lattice form factors and to the experimental data.  Inspection of the
error bands shows that the combined error, which determines the
uncertainty on $|V_{cb}|$, is smallest at about $z \approx 0.025$ ($w
\approx 1.2$).  At this point, the experimental error is about 3.9\%
and the lattice error is about 1.4\%.  (Note that combining them in
quadrature yields a total that is close to the 4\% lattice+experiment
error on $|V_{cb}|$ from the joint fit.)  Thus the experimental error
currently limits the precision on $|V_{cb}|$ from this approach.  The
dominant uncertainty in the experimental data is the assumed 3.3\%
systematic error, which is used for all $w$ values in the joint fit.
Now that lattice-QCD results for the $B \to D \ell \nu$ form factors
are available at nonzero recoil, however, it is clearly worthwhile to
study and improve the systematic errors in the experimental data at
medium and large recoil.

It is interesting to compare the above nonzero-recoil result with the
result based on the standard method that uses only the zero-recoil
extrapolation of the experimental and theoretical form factors.  The
$z$ expansion fit to lattice-only data gives $\mathcal{G}(1) =
1.054(4)_{\rm stat}(8)_{\rm syst}$.  The BaBar collaboration quotes
$\bar\eta_{EW} |V_{cb}| \mathcal{G}(1) = 0.0430(19)_{\rm stat}(14)_{\rm
  syst}$ \cite{Aubert:2009ac} from its $B$-tagged data, which gives
$|V_{cb}| = (40.8 \pm 0.3_{\rm QCD} \pm 2.2_{\rm exp} \pm 0.2_{\rm
  QED})\times 10^{-3}$.  The result is consistent with the value from
nonzero recoil, but the error is larger, as expected.  Our zero-recoil
form factor is consistent with a previous, preliminary Fermilab/MILC
result of $\mathcal{G}(1) = 1.074(18)_{\rm stat}(16)_{\rm
  syst}$~\cite{Okamoto:2004xg}, but with significantly smaller
uncertainties due to the use of a much larger data set with several
lattice spacings and lighter pions. We also note that the systematic
error estimate for the earlier result did not include an estimate of
the heavy-quark discretization errors, one of the larger contributions
to the error in our new result.

\begin{figure}
    \centerline{\includegraphics[width=0.75\textwidth]{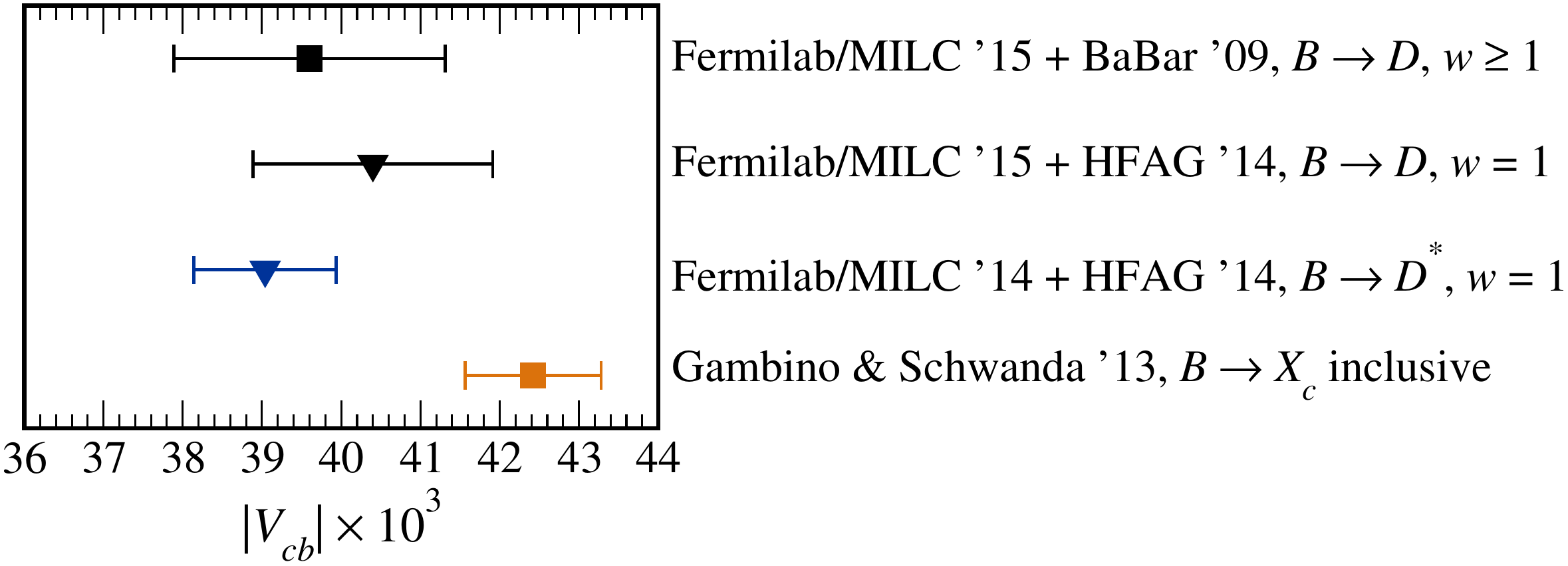}}
    \caption{Comparison of exclusive and inclusive determinations of $
      |V_{cb}|\times 10^3 $. Triangles denote an extrapolation to zero
      recoil, while squares use data over a wide kinematic range.  The
      color code is black, blue (dark gray), and orange (light gray)
      for $B\to D\ell\nu$, $B\to D^*\ell\nu$, and $B\to X_c\ell\nu$,
      respectively. }
    \label{fig:inex}
\end{figure}

We compare our result for $|V_{cb}|$ with other published
determinations from inclusive and exclusive decays in
Fig.~\ref{fig:inex}.  Our result is consistent with the determination
from our companion analysis of $ B\rightarrow D^* \ell\nu $ at zero
recoil, $|V_{cb}| = (39.04 \pm 0.53_{\rm QCD} \pm 0.49_{\rm exp} \pm
0.19_{\rm QED})\times 10^{-3}$ \cite{B2Dstar}.  The errors on
$|V_{cb}|$ from the current work are larger, however, because of the
larger errors in the experimental data. Our result is $1.5 \sigma$
lower than a recent inclusive (non-lattice) determination, $|V_{cb}| =
(42.4 \pm 0.9_{\rm thy+exp})\times 10^{-3}$ \cite{Gambino:2013rza},
which is also based on several experiments and employs data at nonzero
recoil.

We also plot the result for $|V_{cb}|$ in Fig.~\ref{fig:inex}
determined from only our zero-recoil lattice data, but using the best
experimental knowledge of the extrapolated quantity $\bar\eta_{EW}
|V_{cb}| {\cal G}(1)$.  The HFAG average value $\bar\eta_{EW} |V_{cb}|
{\cal G}(1)$ is $0.04264(72)_{\rm stat}(135)_{\rm syst}$
\cite{Amhis:2012bh}, which combines five experimental measurements
from ALEPH~\cite{Buskulic:1995hk}, Belle~\cite{Abe:2001yf},
BaBar~\cite{Aubert:2008yv,Aubert:2009ac}, and
CLEO~\cite{Bartelt:1998dq}.  From this value we obtain $|V_{cb}| =
(40.0 \pm 0.3_{\rm QCD} \pm 1.4_{\rm exp} \pm 0.2_{\rm QED}) \times 10^{-3}$.
This error is smaller than that from the analysis at nonzero recoil,
thanks to the additional experimental information, but only by about
10\%.  Thus combining lattice data at nonzero recoil with a single
experiment reduces the error on $|V_{cb}|$ by almost as much as adding
zero-recoil data from several experiments.  Clearly the error on
$|V_{cb}|$ from $B\to D\ell\nu$ at nonzero recoil can be further reduced
via a joint fit of the lattice form-factor data with additional
experimental measurements once correlations are available.

An interesting byproduct of our combined $z$-expansion fit to obtain
$|V_{cb}|$ is an improved determination of the $B\to D$ form factors
$f_+(q^2)$ and $f_0(q^2)$.  Because the lattice form factors are most
accurate at high $q^2$, while the experimental measurements are most
accurate at low $q^2$, they provide complimentary constraints on the
form-factor shape.  Table~\ref{tab:N3LQCDExptcorr} provides the
$z$-fit coefficients and correlation matrix from our preferred
combined lattice-experiment fit used to obtain our result for
$|V_{cb}|$ quoted in Eq.~(\ref{eq:Vcb}).  These represent our current
best knowledge of $f_+(q^2)$ and $f_0(q^2)$ for $B\to D$ semileptonic
decays, and can be used in other phenomenological applications.  Here
we use the results in Table~\ref{tab:N3LQCDExptcorr} to update our
calculation of the ratio ${\cal B}(B\to D \tau \nu )/{\cal B}(B \to D
\ell\nu ) $ in the Standard Model~\cite{Bailey:2012jg}.  We obtain
\begin{equation}
 R(D) = 0.299(11)\, ,
\label{eq:RD}
\end{equation}
which agrees with our previous determination $R(D) = 0.316(12)(7)$
in~\cite{Bailey:2012jg}, but is 2.0$\sigma$ lower than the BaBar
measurement $R(D) = 0.440(58)(42)$~\cite{Lees:2012xj}.  The error in
our new determination of $R(D)$ is about 20\% smaller than in
Ref.~\cite{Bailey:2012jg}, primarily due to the inclusion of the
experimental information on the shape of $f_+$ from the joint $z$-fit.

\begin{table}
  \caption{Central values, errors, and correlation matrix for the
    parameters of the joint cubic fit to the synthetic lattice data
    for $f_+$ and $f_0$ (including the kinematic constraint at
    $q^2=0$) and the experimental measurements of $\bar\eta_{EW}
    |V_{cb}| f_+(w)$.
  \label{tab:N3LQCDExptcorr}}
  \begin{tabular}{r|r@{.}l|rrrrrrr}
    \hline\hline
                 &&& \multicolumn{7}{c}{Correlation matrix} \\
 & \multicolumn{2}{c|}{value} & $a_{+,0}$    & $a_{+,1}$    & $a_{+,2}$    & $a_{+,3}$ &    $a_{0,1}$    & $a_{0,2}$    & $a_{0,3}$ \\
   \hline
 $a_{+,0}$  & $ 0$&01261(10) & $  1.00000 $ & $  0.24419 $ & $ -0.08658 $ & $  0.01207 $ & $  0.23370 $ & $  0.03838 $ & $ -0.05639 $  \\
 $a_{+,1}$  & $-$0&0963(33)  & $          $ & $  1.00000 $ & $ -0.57339 $ & $  0.25749 $ & $  0.80558 $ & $ -0.25493 $ & $ -0.15014 $  \\
 $a_{+,2}$  & $ 0$&37(11)    & $          $ & $          $ & $  1.00000 $ & $ -0.64492 $ & $ -0.44966 $ & $  0.66213 $ & $  0.05120 $  \\
 $a_{+,3}$  & $-$0&05(90)    & $          $ & $          $ & $          $ & $  1.00000 $ & $  0.11311 $ & $ -0.20100 $ & $  0.23714 $  \\
 $a_{0,1}$  & $-$0&0590(28)  & $          $ & $          $ & $          $ & $          $ & $  1.00000 $ & $ -0.44352 $ & $  0.02485 $  \\
 $a_{0,2}$  & $ 0$&19(10)    & $          $ & $          $ & $          $ & $          $ & $          $ & $  1.00000 $ & $ -0.46248 $  \\
 $a_{0,3}$  & $-0$&03(87)    & $          $ & $          $ & $          $ & $          $ & $          $ & $          $ & $  1.00000 $  \\
  \hline
  \hline
  \end{tabular}
\end{table}

The dominant errors in the lattice form factors come from statistics,
matching, and the chiral-continuum extrapolation, and can be reduced
through simulations at smaller lattice spacings and at physical quark
masses and from further study of the matching factors.  The MILC
Collaboration is currently generating (2+1+1)-flavor HISQ ensembles
with physical light quarks~\cite{Bazavov:2012xda}, which we anticipate
using for future calculations of $B\to D^{(*)}$ form factors.
Heavy-quark discretization errors are also important.  They can be
reduced with a more improved heavy-quark action such as that
proposed in Ref.~\cite{Oktay:2008ex}, and work on this is
underway~\cite{Bailey:2014zma,Bailey:2014jga}.

%%%%%%%%%%%%%%%%%%%%%%%%%%%%%%%
\section*{Acknowledgments} 
%%%%%%%%%%%%%%%%%%%%%%%%%%%%%%%

Computations for this work were carried out with resources provided by
the USQCD Collaboration, the National Energy Research Scientific
Computing Center and the Argonne Leadership Computing Facility, which
is funded by the Office of Science of the U.S.\ Department of Energy;
and with resources provided by the National Institute for
Computational Science and the Texas Advanced Computing Center, which
are funded through the National Science Foundation's Teragrid/XSEDE
Program.
This work was supported in part by the U.S.\ Department of Energy under grants
No.~DE-FG02-91ER40628 (C.B.,J.K.), 
No.~DE-FC02-06ER41446 (C.D., J.F., L.L.), 
No.~DE-SC0010120 (S.G.),
No.~DE-FG02-91ER40661 (S.G., R.Z.), 
No. DE-FG02-13ER42001 (D.D., A.X.K.),
No.~DE-FG02-ER41976 (D.T.); 
by the U.S.\ National Science Foundation under grants 
PHY10-67881 and PHY10-034278 (C.D.),
PHY14-17805~(J.L., D.D.),
PHY09-03571 (S.-W.Q.)
and PHY13-16748 (R.S.);
by the URA Visiting Scholars' program (C.M.B., D.D., A.X.K.);
by the MICINN (Spain) under grant FPA2010-16696 and Ram\'on y Cajal program (E.G.);
by the Junta de Andaluc\'ia (Spain) under Grants No.~FQM-101 and No.~FQM-6552 (E.G.);
by the European Commission (EC) under Grant No.~PCIG10-GA-2011-303781 (E.G.);
by the German Excellence Initiative and the European Union Seventh Framework Programme under grant agreement 
No.~291763 as well as the European Union's Marie Curie COFUND program (A.S.K.);
and by the Basic Science Research Program of the
National Research Foundation of Korea (NRF) funded by the Ministry of
Education (No. 2014027937) and the Creative Research Initiatives
Program (No. 2014001852) of the NRF grant funded by the Korean
government (MEST) (J.A.B.).
This manuscript has been co-authored by an employee of Brookhaven Science Associates, LLC, under Contract
No.\ DE-AC02-98CH10886 with the U.S.\ Department of Energy.
Fermilab is operated by Fermi Research Alliance, LLC, under Contract No.\ DE-AC02-07CH11359 with the United
States Department of Energy.

\clearpage
\appendix

%%%%%%%%%%%%%%%%%%%%%%%%%%%%%%%
\section{Heavy-quark mass correction}
\label{app:kappa} 
%%%%%%%%%%%%%%%%%%%%%%%%%%%%%%%

Heavy-quark masses ($\kappa$ values) are determined by requiring that
the kinetic masses of the $D_s$ and $B_s$ match their experimental
values.  The three-point and two-point functions in this study were
computed with $\kappa$ values from a preliminary tuning.  Final tuned
values differed slightly \cite{B2Dstar}, as shown in
Tables~\ref{tab:params2} and \ref{tab:rho}.

We therefore need to adjust the form factors and $w$ values accordingly.
This is done by repeating the computation of the ratios $R_+$,
$Q_+(\bm{p})$, and $\bm{R}_-(\bm{p})$ on the $a \approx 0.12$~fm,
$\hat m^\prime=0.2m_s^\prime$ ensemble for a few values of $\kappa_b$ and $\kappa_c$ in
the vicinity of the desired, tuned values.  These results permit
calculating the derivatives of the form factors with respect to the
quark masses.  We assume that these results, expressed in
dimensionless terms, can then be used to adjust form factors in our
other ensembles.

From Eqs.~(\ref{eq:renormR+})--(\ref{eq:renormR-}), we see that we
have the option of computing and applying these adjustments before or
after matching with the $\rho$ factors.  Because of the simplifying
steps taken in Appendix~\ref{app:hqerror} for
$\rhoV{4}(w)/\rhoV{4}(1)$ and $\rhoV{i}(w)/\rhoV{4}(w)$, we choose to
make the adjustments directly on unmatched quantities.  From
Eqs.~(\ref{eq:hplusQ}) and~(\ref{eq:hminusQ}), one sees that it is
convenient to study the mass dependence of
\begin{eqnarray}
    S_+ & = & \sqrt{R_+}Q_+ , \\
    S_- & = & \frac{\bm{R}_-\cdot\bm{x}_f}{\bm{x}_f^2} ,
\end{eqnarray}
and $\bm{x}_f^2S_-$.

Heavy-quark symmetry suggests that interpolations in inverse quark
masses will implement the quark-mass tuning most smoothly.  With the
Fermilab method~\cite{ElKhadra:1996mp}, the quark mass is identified
with the kinetic mass:
\begin{equation}
    \frac{1}{m_2a} = \frac{2}{m_0a(2+m_0a)} + \frac{1}{m_0a+1} ,
\end{equation}
where we compute the bare quark mass $m_0a$ from the tadpole-improved,
tree-level formula
\begin{equation}
    m_0a = \frac{1}{u_0} \left(\frac{1}{2\kappa} - \frac{1}{2\kappa_\text{cr}}\right) .
\end{equation}
Here, $u_0$ is the tadpole parameter, and $\kappa_\text{cr}$ is the
value of $\kappa$ such that the lightest pseudoscalar meson mass
vanishes.  Thus, below we compute slopes of $S_+$, $S_-$, and
$\bm{x}_f^2S_-$ with respect to $\xi_c=1/(m_{2c}r_1)$ and
$\xi_b=1/(m_{2b}r_1)$.
\begin{figure}
    \includegraphics[width=0.32\textwidth]{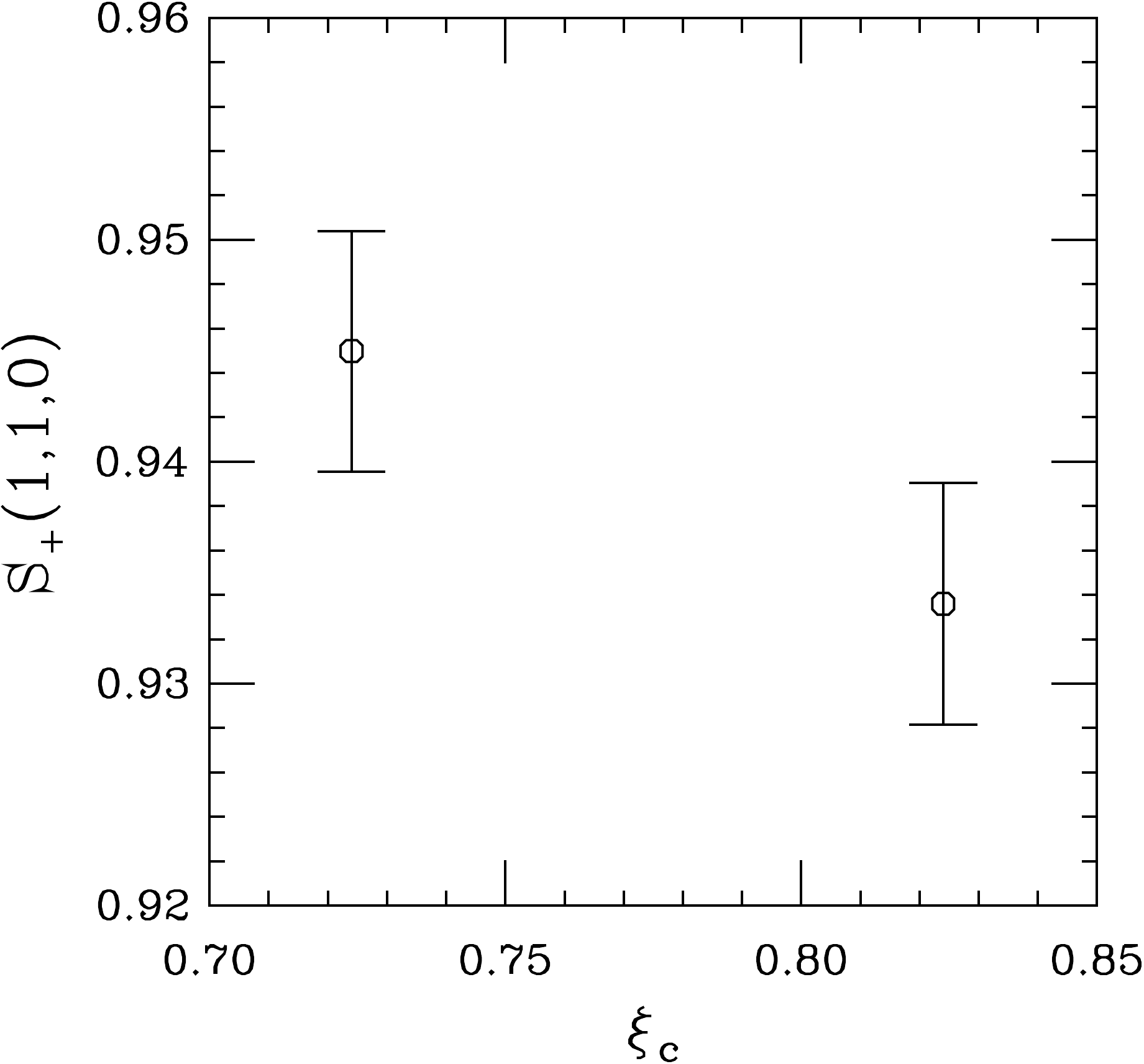} \hfill
    \includegraphics[width=0.32\textwidth]{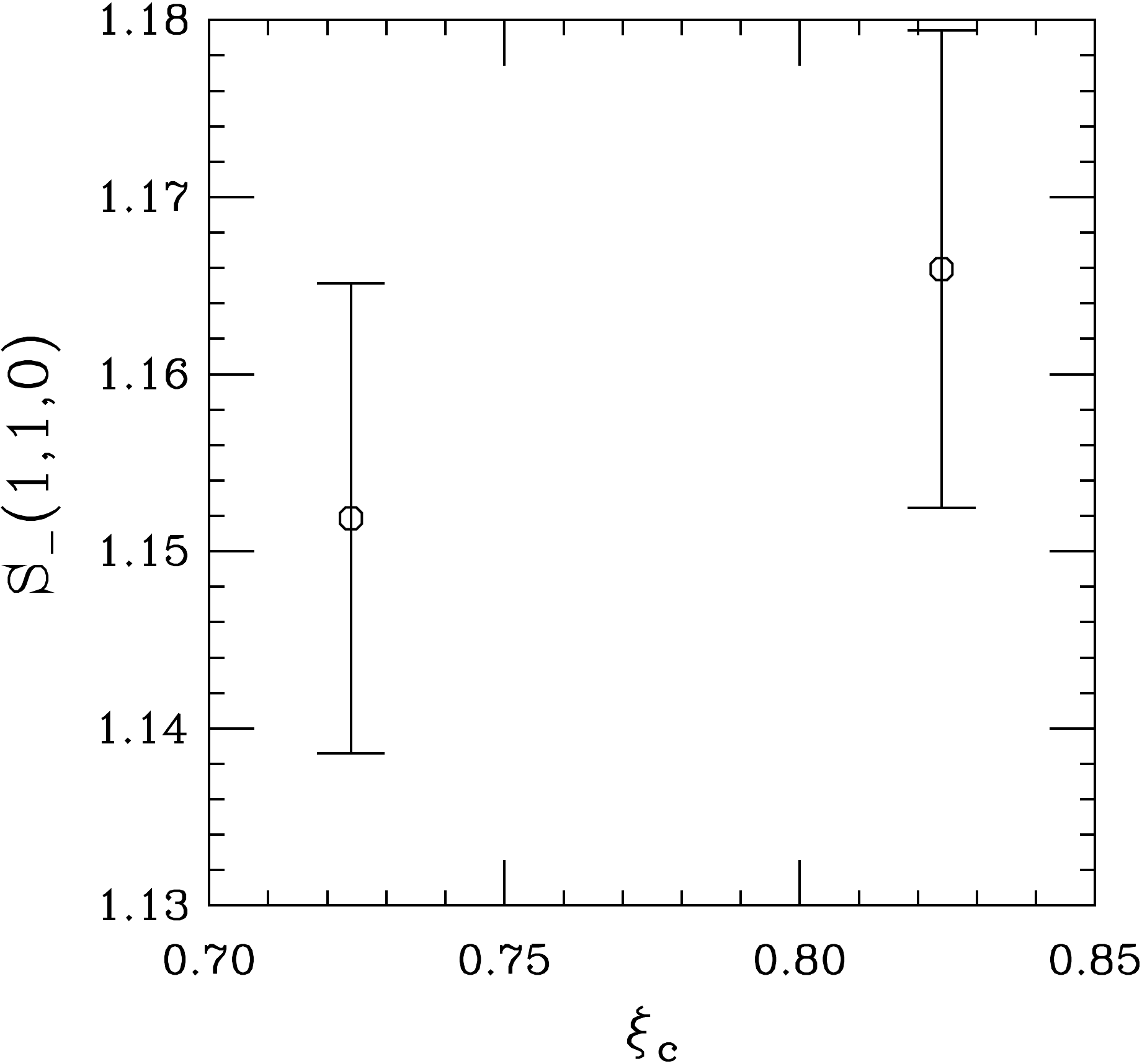} \hfill
    \includegraphics[width=0.32\textwidth]{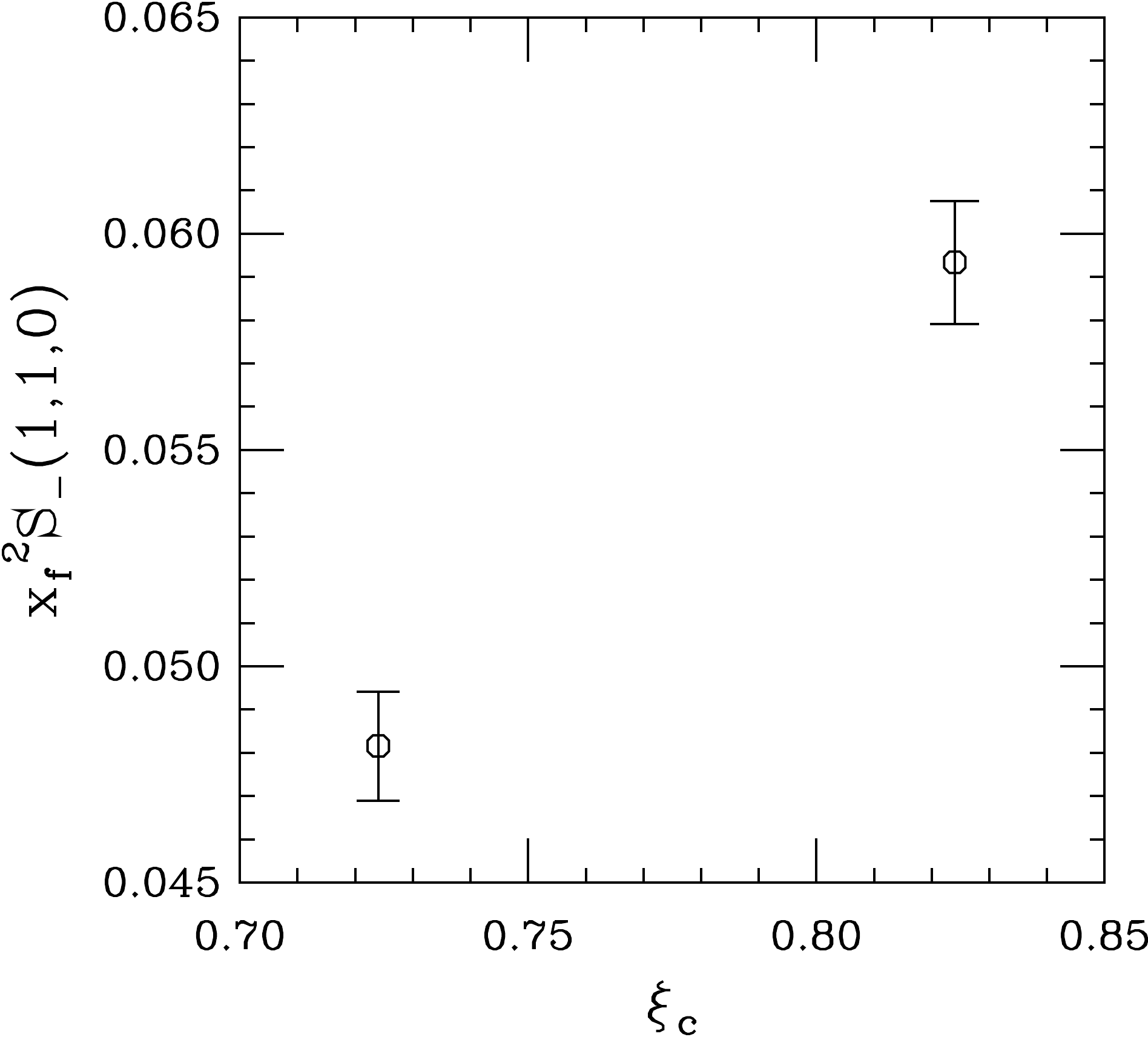} 
    \caption{
        Heavy quark mass dependence on the $a \approx 0.12$ fm, $\hat m^\prime=0.2m_s^\prime$ ensemble at momentum 
        $2\pi(1,1,0)/L$.
        Left to right: $S_+$, $S_-$, and $\bm{x}_f^2S_-$, respectively, {\it vs.} inverse charm-quark kinetic 
        mass~$\xi_c=(m_{2c}r_1)^{-1}$.}
    \label{fig:charm-dependence}
\end{figure}
\begin{figure}
    \includegraphics[width=0.32\textwidth]{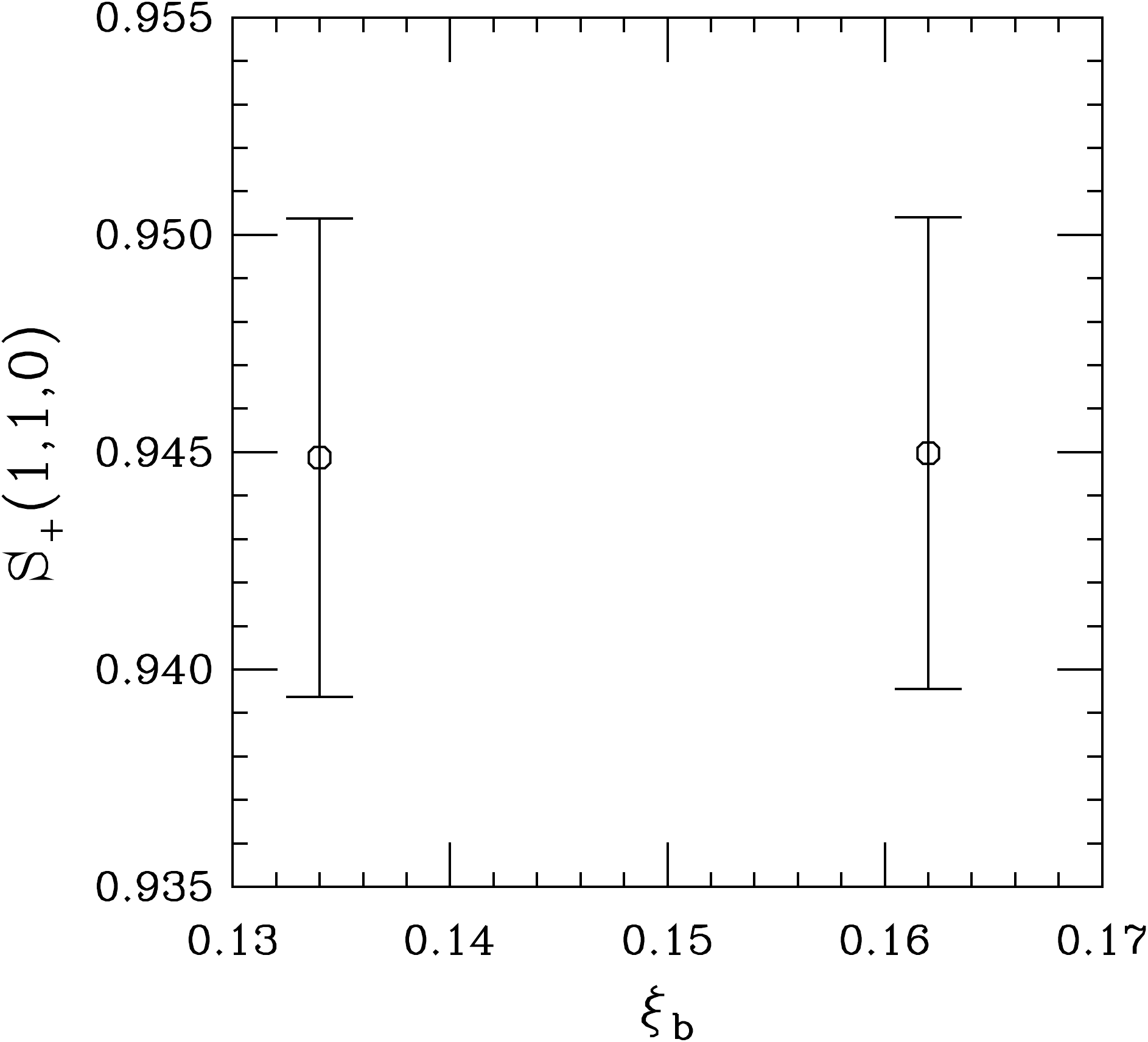} \quad
    \includegraphics[width=0.32\textwidth]{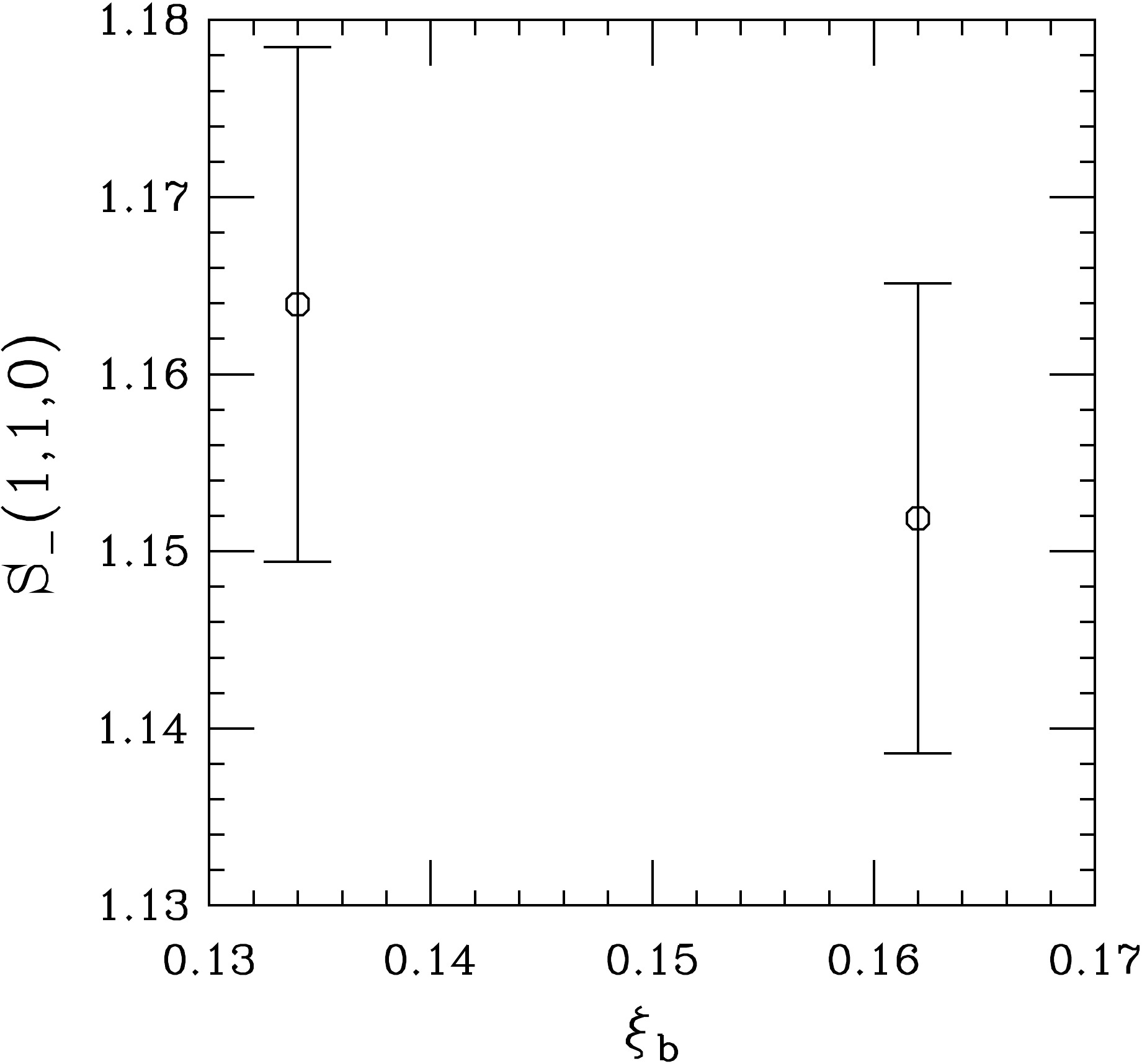}
    \caption{
        Heavy quark mass dependence on the $a \approx 0.12$ fm, $\hat m^\prime=0.2m_s^\prime$ ensemble at momentum 
        $2\pi(1,1,0)/L$.
        $S_+$~(left) and $S_-$ (right) vs.\ inverse bottom-quark kinetic mass~$\xi_b=(m_{2b}r_1)^{-1}$.}
    \label{fig:bottom-dependence}
\end{figure}
The results of the computations with varying quark masses are shown in Figs.~\ref{fig:charm-dependence}
and~\ref{fig:bottom-dependence}.

Because the corrections in the charm and bottom masses are small, it
suffices to work to first order in the inverse mass shift.
Heavy-quark symmetry also suggests that the leading mass dependence of
$S_+(w=1)=\sqrt{R_+}$ is quadratic, of the form~$(\xi_c-\xi_b)^2$.
Therefore, the leading shift in $\xi_c$ is suppressed by $\xi_b$, and
the leading shift in $\xi_b$ is suppressed by~$\xi_c$.  Below, we
neglect the former effect but keep the latter, since it is suppressed
only by~$\xi_c$.  Furthermore, by construction $\bm{x}_f^2S_-\to0$ as
$w\to1$ for all quark masses, and, therefore, the derivative with
respect to $\xi_c$ also vanishes at $w=1$.  On the other hand, neither
$S_-$ nor its derivatives vanish at $w=1$.

\begin{figure}
    \includegraphics[width=0.32\textwidth]{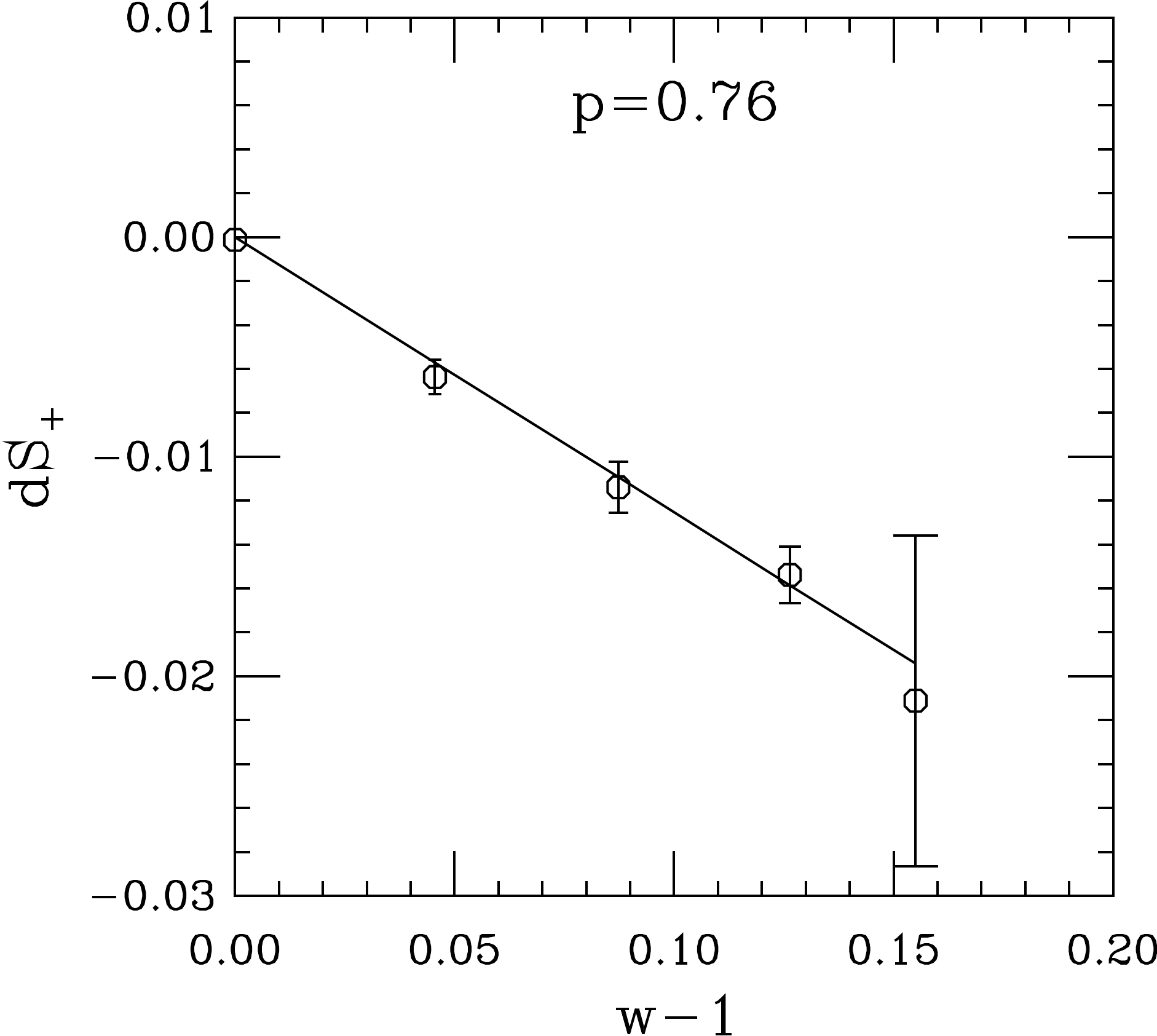} \hfill
    \includegraphics[width=0.32\textwidth]{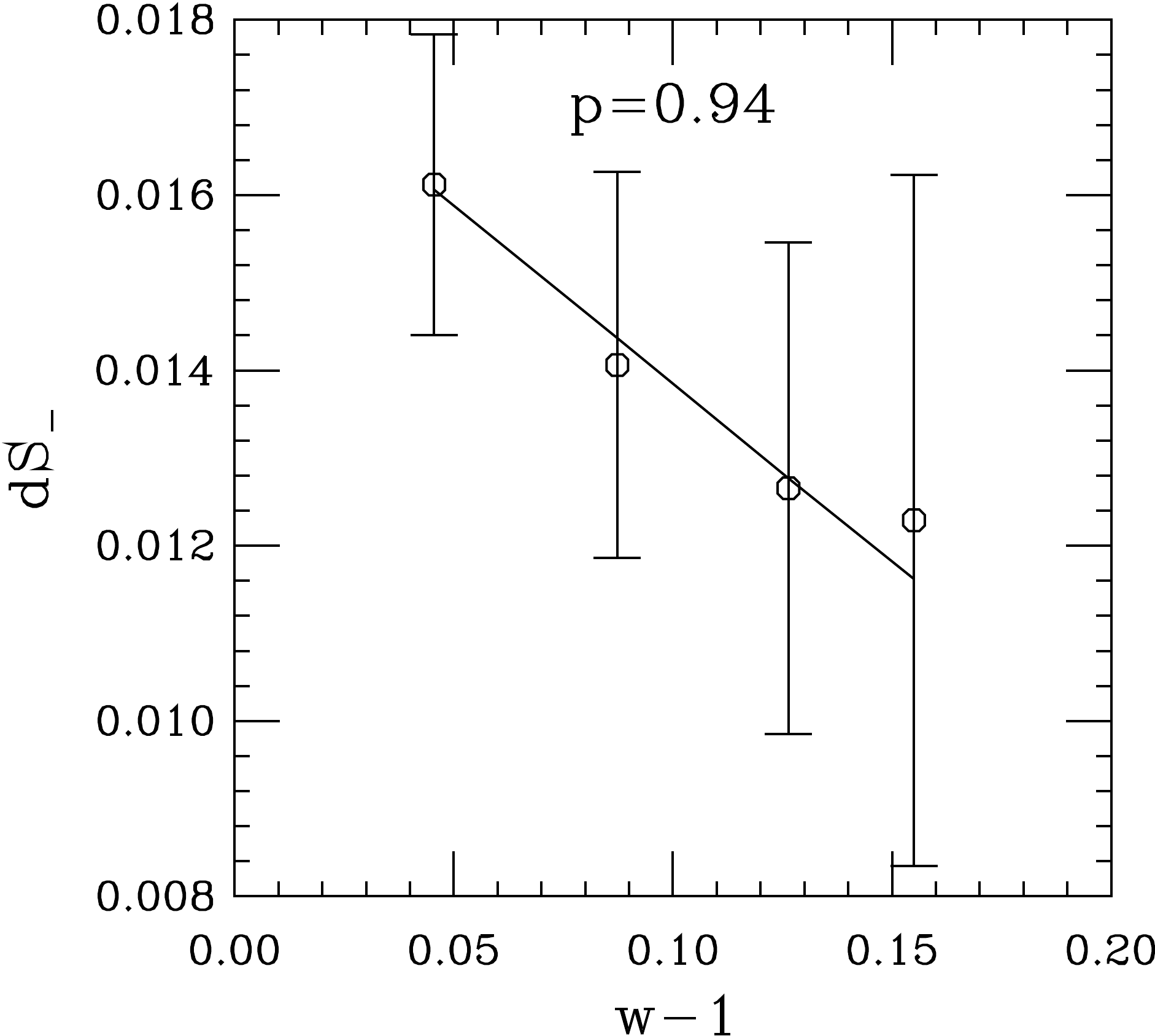} \hfill
    \includegraphics[width=0.32\textwidth]{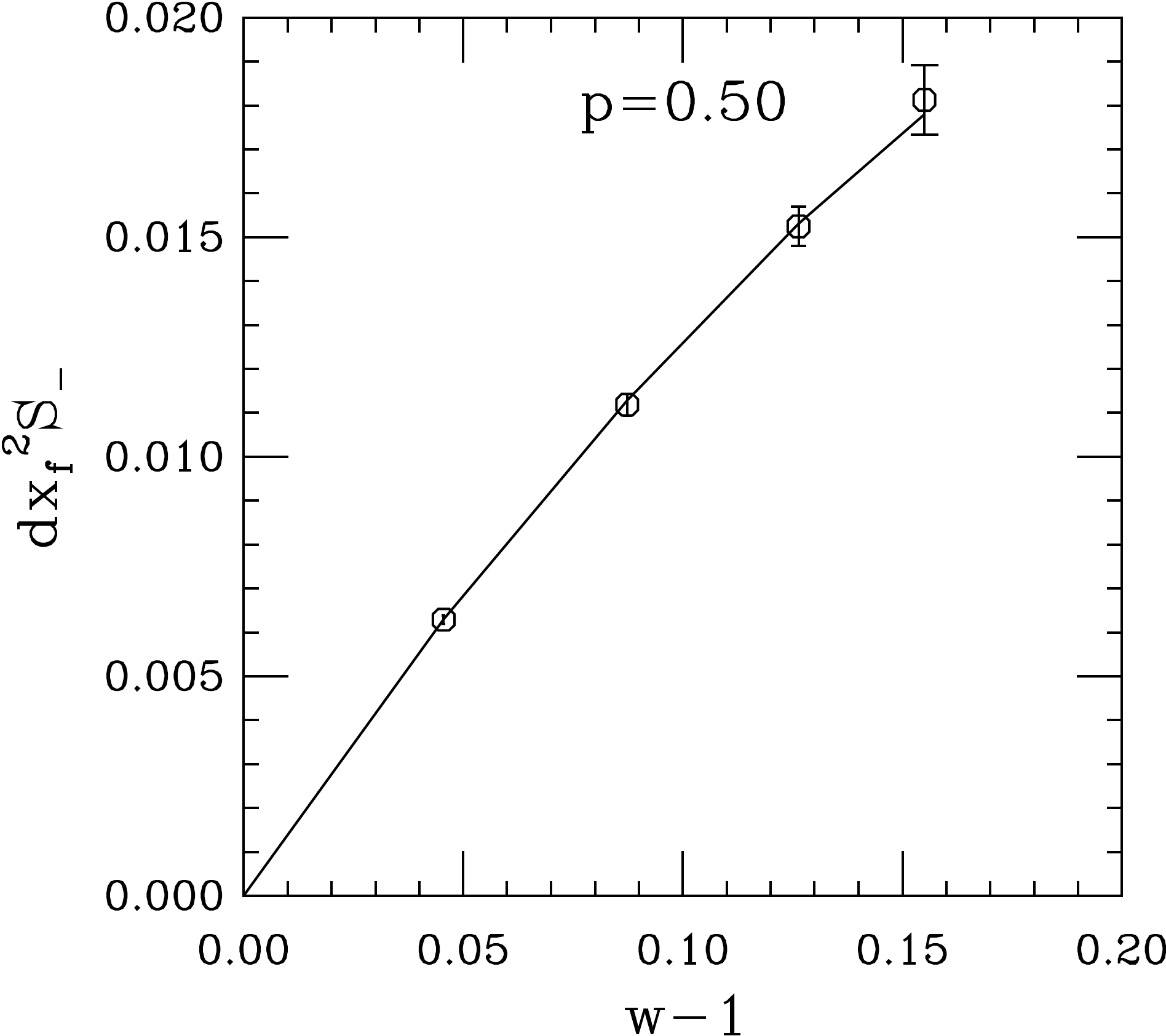}
    \caption{ Effect of heavy-quark mass shifts on the $a \approx
      0.12$~fm, $\hat m^\prime = 0.2 m_s^\prime$ ensemble as the charm-quark mass
      parameter is increased from $\kappa_c = 0.1254$ to~0.1280.  Left
      to right: $dS_+$, $dS_-$, and $d(\bm{x}_f^2S_-)$, respectively,
      vs.~$w-1$.}
    \label{fig:charm-shift}
\end{figure}

\begin{figure}
    \includegraphics[width=0.32\textwidth]{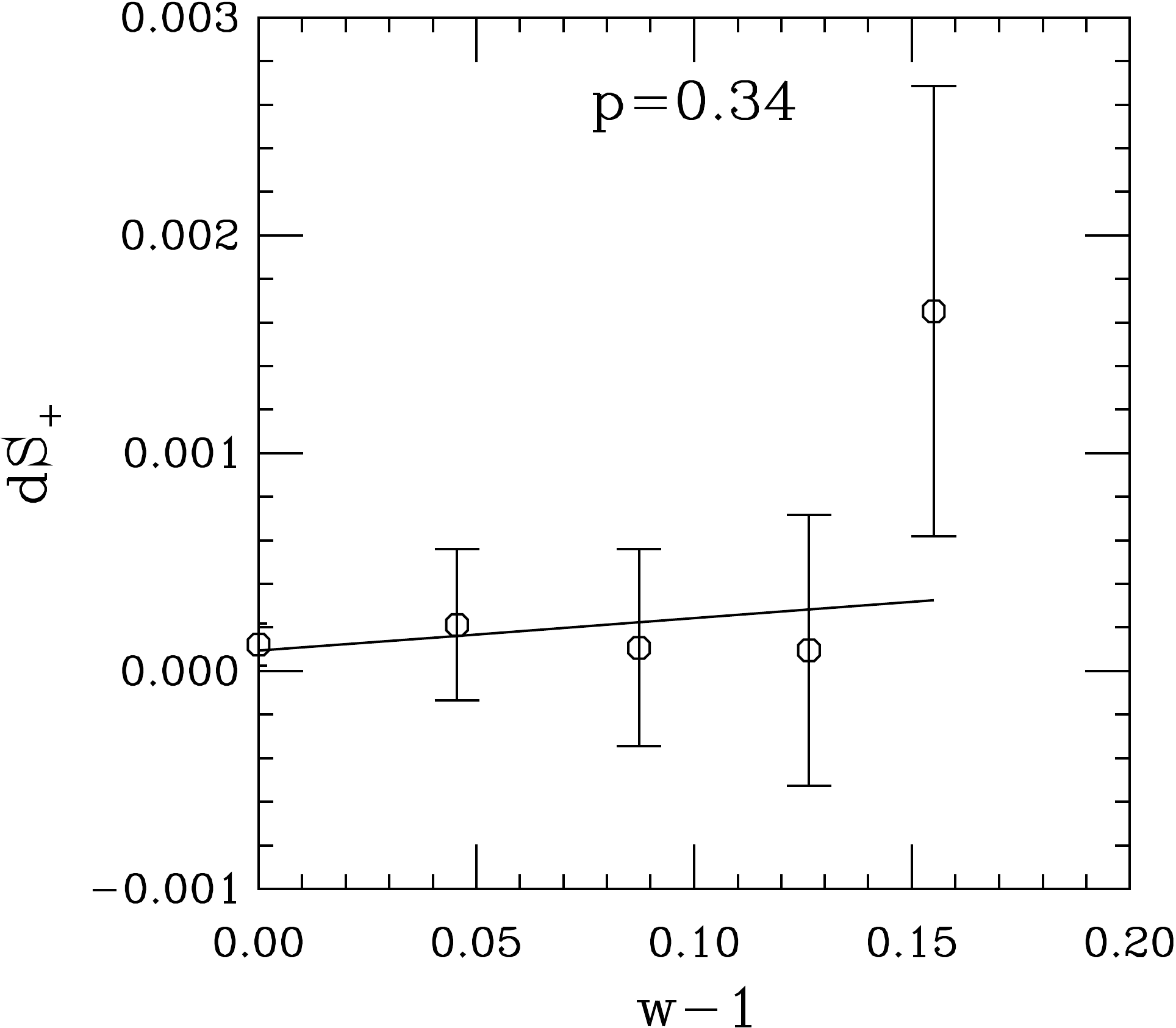} \quad\quad
    \includegraphics[width=0.32\textwidth]{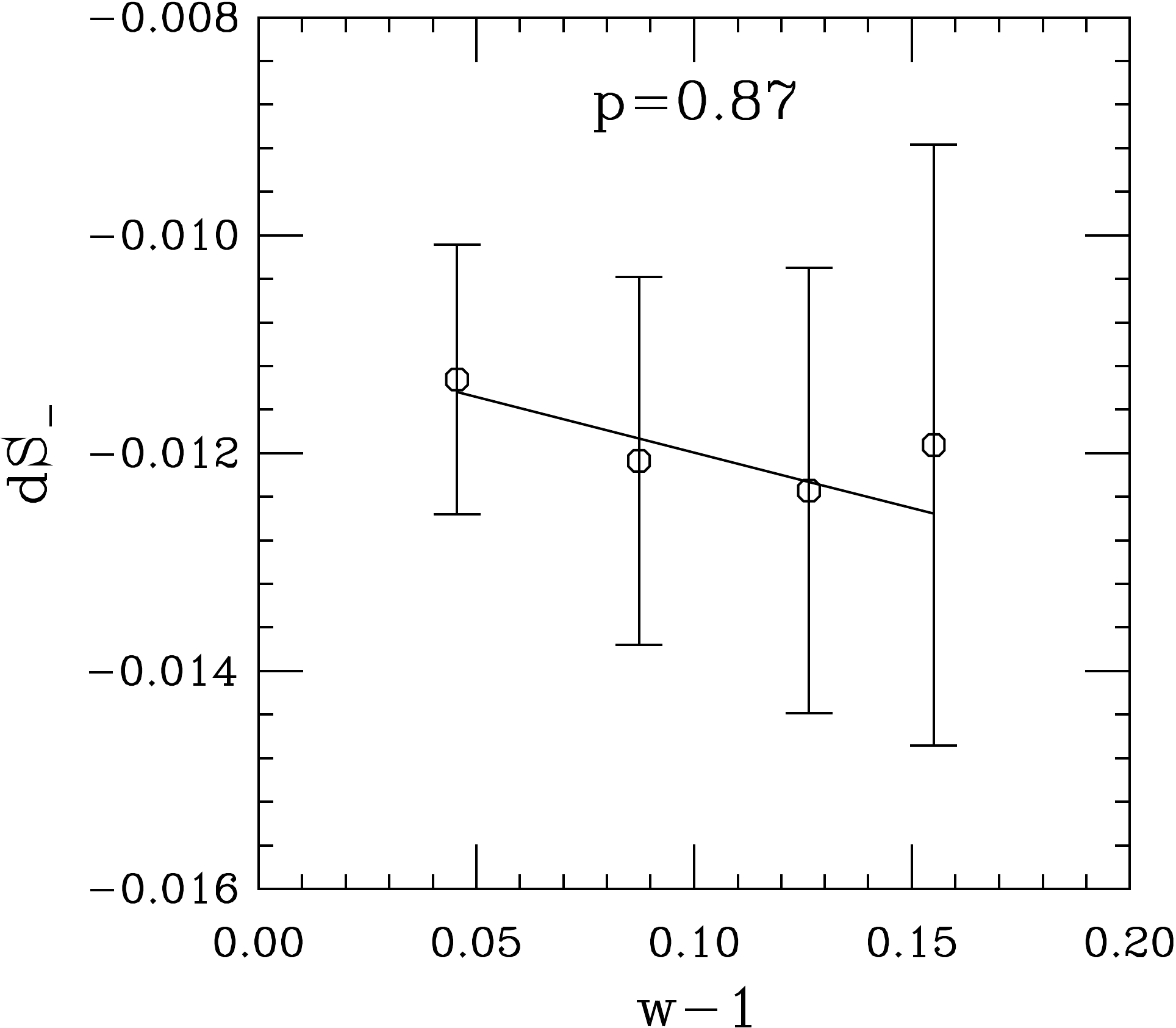}
    \caption{ Effect of heavy-quark mass shifts on the $a \approx
      0.12$~fm, $\hat m^\prime = 0.2 m_s^\prime$ ensemble as the bottom-quark mass
      parameter is increased from $\kappa_b=0.0860$ to~0.0901
      vs.~$w-1$.  Left to right: $dS_+$ and $dS_-$, respectively,
      vs.~$w-1$.}
    \label{fig:bottom-shift}
\end{figure}

\pagebreak
Because of the narrow range of $w$, $1\le w<1.16$, for our data, one
should expect a linear approximation in~$w$ to suffice for the
quark-mass adjustments.  Indeed, only $\bm{x}_f^2 S_-$ requires a
quadratic, as shown in Figs.~\ref{fig:charm-shift}
and~\ref{fig:bottom-shift}.
Therefore, we introduce
\begin{eqnarray}
    \frac{dS_+}{d\xi_c} & = & r_{+,1,c}(w-1)\, , \\
    \frac{dS_-}{d\xi_c} & = & r_{-,0,c} + r_{-,1,c}(w-1)\, , \\
    \frac{d(\bm{x}_f^2S_-)}{d\xi_c} & = & r_{x,1,c}(w-1) + r_{x,2,c}(w-1)^2\, , \\
    \frac{dw}{d\xi_c} & = & r_{w,1,c}(w-1)\, , \\
    \frac{dS_+}{d\xi_b} & = & r_{+,0,b} + r_{+,1,b}(w-1)\, , \\
    \frac{dS_-}{d\xi_b} & = & r_{-,0,b} + r_{-,1,b}(w-1)\, . 
\end{eqnarray}
(The notation for the slope parameters $r_{f,n,q}$ encodes a form factor label $f$,
a polynomial coefficient index $n$, and a quark mass label $q$.)
Fits to our data then yield
\begin{eqnarray}
    r_{+,1,c} & = & -0.72(5)\, , \\
    r_{-,0,c} & = & 0.102(11)\, , \\
    r_{-,1,c} & = & -0.23(14)\, , \\
    r_{x,1,c} & = & 0.851(14)\, , \\
    r_{x,2,c} & = & -1.22(16)\, , \\
    r_{+,0,b} & = & 0.0042(41)\, , \\
    r_{+,1,b} & = & 0.07(21)\, , \\
    r_{-,0,b} & = & -0.49(5)\, , \\
    r_{-,1,b} & = & -0.46(69)\, ,
\end{eqnarray}
As discussed above, we expect $r_{+,0,b}$ to be of order $\xi_c$, or approximately 0.83.
In fact, it is much smaller.

\begin{figure}
    \includegraphics[width=0.32\textwidth]{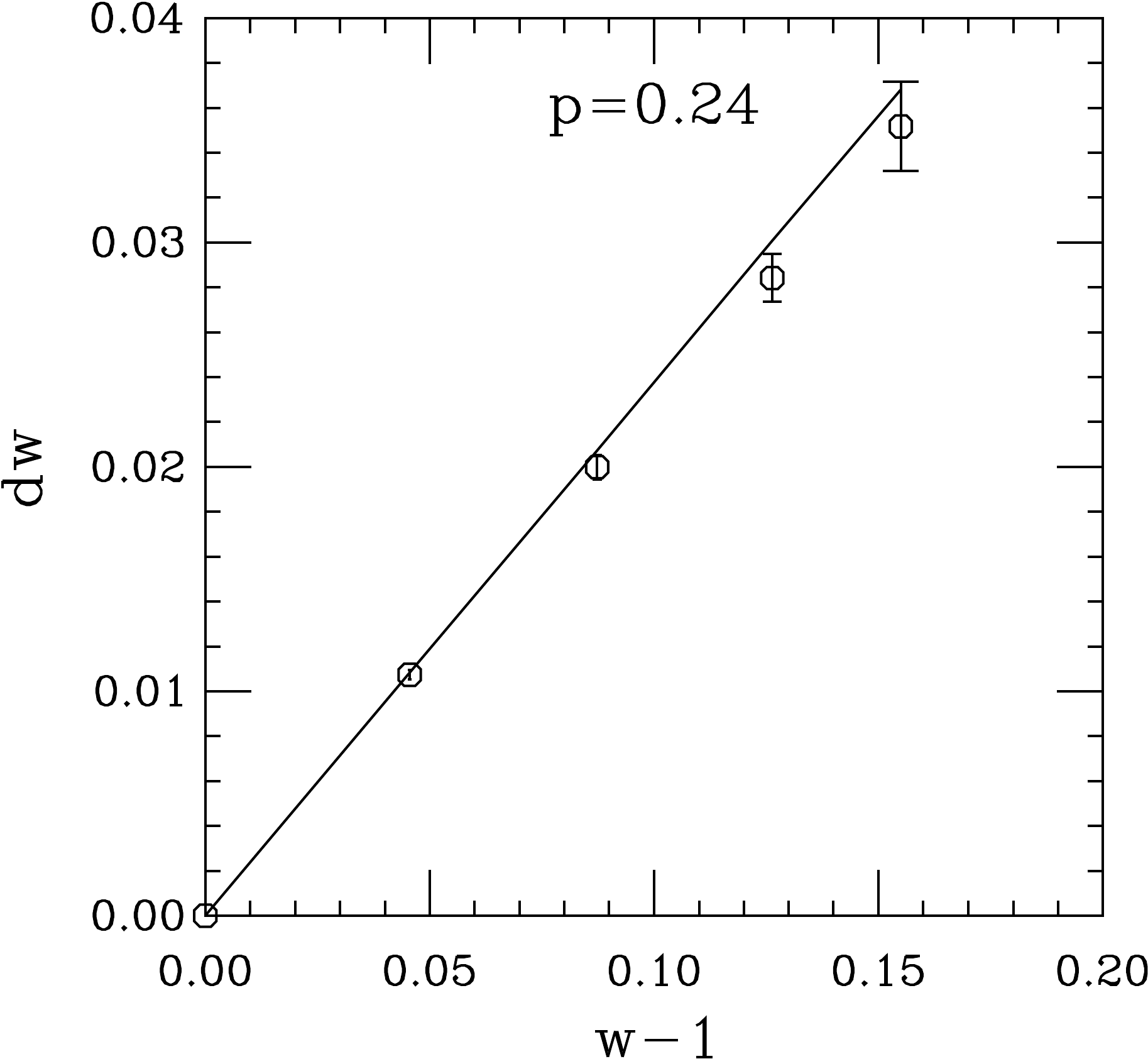} \quad\quad
    \caption{Effect of heavy-quark mass shifts on the $a \approx
      0.12$~fm, $\hat m^\prime = 0.2 m_s^\prime$ ensemble.  Shift in recoil variable
      $w_i$ vs.\ $w-1$ as the charm-quark mass parameter is increased
      from $\kappa_c = 0.1254$ to~0.1280.}
    \label{fig:dw}
\end{figure}

We compute the correlation functions at discrete values of the recoil
momentum of the $D$ meson, resulting in discrete values $w_i$, which
are determined from Eqs.~(\ref{eq:xf}) and~(\ref{eq:w}).  The recoil
variable $w_i$ is determined dynamically from diagonal vector current
matrix elements involving the $D$ meson, so it varies with the charm
quark mass, but not the bottom quark mass.  We take the convention
that when we shift both quark masses, we shift $w_i$ to $w_i'$ and we
shift $S_+(w_i)$ to $S_+'(w_i')$, and similarly for $S_-$ and
$\bm{x}_f^2S_-$.  As can be seen from Fig.~\ref{fig:dw}, the data
support a linear approximation for the shift in $w_i$ also.

The effect of the kappa adjustment on the values of $w$, $h_+$, and
$h_-$ is illustrated in Table~\ref{tab:tableformfactor} for the $a \approx 0.12$ fm,
$\hat m^\prime = 0.14 m_s^\prime$ ensemble where the tuning adjustment decreases both
$\kappa_c$ and $\kappa_b$ from their simulation values.
    
\begin{table}
    \caption{
        Unadjusted and adjusted values of $w$, $h_+$, and $h_-$ for the $a \approx 0.12$~fm, 
        $\hat{m}' = 0.14m'_s$ ensemble. For this illustration only statistical errors are shown.}
    \label{tab:tableformfactor}
    \begin{ruledtabular}
    \begin{tabular}{lllllll}
            momentum  & raw $w$ & shifted $w$ & raw $h_+$ & tuned $h_+$ & raw $h_-$     & tuned $h_-$    \\
            \hline
            000       & 1       & 1       & 1.0391(53)   & 1.0390(53)  &  --         &  --            \\ 
            100       & 1.0465  & 1.0426  & 0.9812(61)   & 0.9849(62)  & 0.0041(111) & 0.0008(111) \\ 
            110       & 1.0896  & 1.0822  & 0.9388(70)   & 0.9457(70)  & 0.0072(134) & 0.0031(135) \\ 
            111       & 1.1299  & 1.1191  & 0.8978(97)   & 0.9073(97)  & 0.0139(165) & 0.0092(166) \\ 
            200       & 1.1553  & 1.1424  & 0.8789(120)  & 0.8898(121) & 0.0336(205) & 0.0286(206)
        \end{tabular}
    \end{ruledtabular}
\end{table}

%%%%%%%%%%%%%%%%%%%%%%%%%%%%%%%
\clearpage
\section{Heavy-quark discretization effects}
\label{app:hqerror}
%%%%%%%%%%%%%%%%%%%%%%%%%%%%%%%

We use the heavy-quark effective theory (HQET) to derive the form of
heavy-quark discretization
effects~\cite{Kronfeld:2000ck,Harada:2001fj}.  In this appendix, we
apply this formalism to derive the matching procedure from lattice
gauge theory to continuum QCD,
cf.\ Eqs.~(\ref{eq:renormR+})--(\ref{eq:renormR-}).  We also use it to
derive power-law discretization effects, both at nonzero recoil
($w>1$) and at zero recoil ($w=1$) where heavy-quark symmetry is more
constraining.  In the last subsection of the appendix, we also present
numerical estimates for the discretization errors.

\subsection{Formalism}

We describe the underlying lattice gauge theory (LGT) with an
effective Lagrangian, asserting the relation
\begin{equation}
    \mathcal{L}_\text{LGT} \doteq \bar{h}(iv\cdot{D} - m_1)h +
        \frac{\bar{h}D_\perp^2h}{2m_2} +
        \frac{\bar{h}s\cdot Bh}{2m_B} + 
        \frac{\bar{h}[D_\perp^\alpha,iE_\alpha]h}{8m_D^2} +
        \frac{\bar{h}s_{\alpha\beta}\{D_\perp^\alpha,iE^\beta\}h}{4m_E^2} + \cdots,
    \label{eq:L-HQET}
\end{equation}
where $\doteq$ can be read ``has the same matrix elements as.''  Here,
$v$ is a four vector specifying the rest frame of the heavy-light
meson, such that $v^2=-1$; the heavy-quark field $h$ satisfies
$\vslash h=ih$; and $s_{\alpha\beta}=-i\sigma_{\alpha\beta}/2$.  Then,
$D_\perp^\mu=D^\mu+v^\mu\,v\cdot\!D$ is the covariant derivative
orthogonal to $v$, $B^{\alpha\beta}=(\delta^\alpha_\mu+v^\alpha
v_\mu)F^{\mu\nu}(\delta^\beta_\nu+v^\beta v_\nu)$ is the
chromomagnetic field (in the $v$ frame), and $E^\beta=-v_\alpha
F^{\alpha\beta}$ is the chromoelectric field (in the $v$ frame).  The
HQET description for continuum QCD has the same structure
\begin{equation}
    \mathcal{L}_{\text{QCD}} \doteq \bar{h}(iv\cdot{D} - m)h +
        \frac{\bar{h}D_\perp^2h}{2m} +
        \frac{z_B\bar{h}s\cdot Bh}{2m} + 
        \frac{z_D\bar{h}[D_\perp^\alpha,iE_\alpha]h}{8m^2} +
        \frac{z_E\bar{h}s_{\alpha\beta}\{D_\perp^\alpha,iE^\beta\}h}{4m^2} + \cdots.
    \label{eq:L-HQET-QCD}
\end{equation}
In this framework, matching and improvement boil down to choosing the
parameters of the lattice Lagrangian, such that the
Eq.~(\ref{eq:L-HQET}) reproduces Eq.~(\ref{eq:L-HQET-QCD}) term by
term.

The rest mass $m_1$ does not influence matrix elements or mass splittings.
In the Fermilab method, therefore, one tunes $\kappa$ so that
\begin{equation}
    \frac{1}{2m_2}=\frac{1}{2m},
\end{equation}
and $c_{\text{SW}}$ so that
\begin{equation}
    \frac{1}{2m_B}=\frac{z_B}{2m} = \frac{1 + \mathcal{O}(\alpha_s)}{2m},
    \label{eq:cSWtune}
\end{equation}
where the second equality follows because
$z_B=1+\mathcal{O}(\alpha_s)$.  In this work, we tune $\kappa$ via the
heavy-strange meson mass; for details of our procedures, see
Appendix~C of Ref.~\cite{B2Dstar}.  Furthermore, we choose
$c_{\text{SW}}$ at the tadpole-improved tree level, which makes the
coefficient of the $\mathcal{O}(\alpha_s)$ error in
Eq.~(\ref{eq:cSWtune}) small~\cite{Nobes:2005dz}.

The Fermilab vector current, Eq.~(\ref{eq:Vlat}), has an HQET
description too.  Through dimension four~\cite{Harada:2001fj}
\begin{equation}
	V^\mu \doteq \bar{C}_{V_\parallel}^\text{LGT} v^\mu \bar{c}_{v'}b_v  +
		\bar{C}_{V_\perp}^\text{LGT} \bar{c}_{v'}i\gamma^\mu_\perp b_v +
		\bar{C}_{V_{v'}}^\text{LGT} v^{\prime\mu}_\perp \bar{c}_{v'}b_v -
		\sum_{a=1}^{14} \bar{B}_{Va}^\text{LGT} \bar{\mathcal{Q}}^\mu_{Va} + \cdots.
	\label{eq:V-HQET-4}
\end{equation}
The continuum-QCD current $\mathcal{V}^\mu$ can be described in the
same way albeit with different $\bar{C}$ and $\bar{B}$ coefficients,
denoted in this paper by omitting the label ``LGT.''  Then
$\ZV{\mu}V^\mu\doteq\mathcal{V}^\mu$ if the $Z$ factors are chosen to
be \cite{Harada:2001fj}
\begin{align}
  \ZV{4} \equiv \, \bar{Z}_{V_\parallel} & = \bar{C}_{V_\parallel}/\bar{C}_{V_\parallel}^\text{LGT}, \\
  \ZV{i} \equiv \, \bar{Z}_{V_\perp}     & = \bar{C}_{V_\perp}/\bar{C}_{V_\perp}^\text{LGT}.
\end{align}
In practice, of course, such matching is only approximate.  For
example, the one-loop calculation of $\rhoV{4}$, explained in
Sec.~\ref{sec:PT}, leads to a matching error of order $\alpha_s^2$.

With the Fermilab currents most of the fourteen dimension-four
coefficients $\bar{B}_{Va}^\text{LGT}$ vanish at the tree level; the
same holds for continuum QCD and the $\bar{B}_{Va}$.  The exceptions
multiply the operators
\begin{align}
	\bar{\mathcal{Q}}^\mu_{V1} & =  - v^\mu   \bar{c}_{v'}\Dslash_{\perp} b_v, 
    \label{eq:Q1} \\
	\bar{\mathcal{Q}}^\mu_{V2} & =  \bar{c}_{v'}i\gamma^\mu_{\perp} \Dslash_{\perp} b_v,
    \label{eq:Q2} \\
	\bar{\mathcal{Q}}^\mu_{V4} & =  -{v'}^\mu \bar{c}_{v'}\lDslsh_{\perp'}b_v,
    \label{eq:Q4} \\
	\bar{\mathcal{Q}}^\mu_{V5} & =  \bar{c}_{v'}\lDslsh_{\perp'}i\gamma^\mu_{\perp'}b_v.
    \label{eq:Q5} 
\end{align}
At the tree level, their coefficients are
\begin{align}
    \bar{Z}_{V_\parallel} \bar{B}_{V1}^\text{LGT} & =  
    \bar{Z}_{V_\perp}     \bar{B}_{V2}^\text{LGT} \equiv \frac{1}{2m_{3b}}, 
    \label{eq:Bm3b} \\
    \bar{Z}_{V_\parallel} \bar{B}_{V4}^\text{LGT} & =  
    \bar{Z}_{V_\perp}     \bar{B}_{V5}^\text{LGT} \equiv \frac{1}{2m_{3c}}. 
    \label{eq:Bm3c} 
\end{align}
The improvement terms in the current, namely $d_1$, are chosen so that 
\begin{equation}
    \frac{1}{2m_3} = \frac{1}{2m_2} + \mathcal{O}(\alpha_sa),
    \label{eq:Bm3}
\end{equation}
for operators with label $a\in\{1,2,4,5\}$.  The other
$\bar{B}_{Va}^\text{(LGT)}$ are of order $\alpha_s$ from the outset.

\subsection{Matching factors}

Equation~(\ref{eq:renormR+}) is well known from earlier
work~\cite{Hashimoto:2001nb,Harada:2001fj}.  To establish
Eqs.~(\ref{eq:renormQ+}) and~(\ref{eq:renormR-}), let us start by
defining $h_\pm^\text{LGT}(w)$ for the lattice current $V^\mu$ in
analogy with the decomposition in Eq.~(\ref{eq:h+-}).  These form
factors $h_\pm^\text{LGT}(w)$ are \emph{not} the right-hand sides of
Eqs.~(\ref{eq:hplusQ}) and~(\ref{eq:hminusQ}).  The task here is to
show how the ratios cancel some of the cutoff effects in
$h_\pm^\text{LGT}(w)$.  Sometimes it is convenient to choose arbitrary
$v$ and $v'$ when working out consequences of the HQET.  The
kinematics of our lattice-QCD correlators correspond to $v=(i,\bm{0})$
and $v'=(iw,\bm{v}')$.

The simplest case is the definition of the velocity via $D(\bm{0})\to
D(\bm{p})$ matrix elements:
\begin{equation}
    \bm{x}_f(\bm{p}) = \bm{v}'\frac{h_+^\text{LGT}(\bm{p})-h_-^\text{LGT}(\bm{p})}%
        {(w+1)h_+^\text{LGT}(\bm{p}) - (w-1)h_-^\text{LGT}(\bm{p})}
         = \frac{\bm{v}'}{w+1},
    \label{eq:x=v}
\end{equation}
because $h_-^\text{LGT}=0$ for a flavor-conserving transition.  This
property follows from time-reversal invariance of the chosen current
and arises independent of any matching considerations.  The expression
for $w$ in Eq.~(\ref{eq:w}) then follows immediately from
$w^2=1+{\bm{v}'}^2$ (when $\bm{v}=\bm{0}$).

Similarly, the other ratios are
\begin{align}
    Q_+(\bm{p}) & =  \frac{(w+1)h_+^\text{LGT}(w) - (w-1)h_-^\text{LGT}(w)}{2h_+^\text{LGT}(1)}, 
    \label{eq:Q-LGT} \\
    \bm{R}_-(\bm{p}) & =  \bm{v}'
        \frac{h_+^\text{LGT}(w) - h_-^\text{LGT}(w)}{(w+1)h_+^\text{LGT}(w) - (w-1)h_-^\text{LGT}(w)},
    \label{eq:R-LGT}
\end{align}
with $w=w(\bm{p})$.  These form factors, of course, are for the
flavor-changing process.

Using the trace formalism explained in Ref.~\cite{Kronfeld:2000ck}, it
is straightforward to obtain the following expressions for
$h_\pm^\text{LGT}(w)$:
\begin{align}
    h_+^\text{LGT}(w) & =             \bar{C}_+^\text{LGT}(w) \Xi(w) + \frac{w-1}{2}\left\{
        \bar{B}_+^\text{LGT}(w) \left[2\xi_3(w) - \bar{\Lambda}\xi(w)\right] -
        \bar{B}_+^{\prime\text{LGT}}(w) \bar{\Lambda}\xi(w) \right\},
    \label{eq:hplusLGT} \\
    h_-^\text{LGT}(w) & =  \half(w+1)\bar{C}_-^\text{LGT}(w) \Xi(w) +
        \bar{B}_-^\text{LGT}(w) \left[2\xi_3(w) - \bar{\Lambda}\xi(w)\right] -
        \bar{B}_-^{\prime\text{LGT}}(w) \bar{\Lambda}\xi(w), \hspace*{3.5em}
    \label{eq:hminusLGT}
\end{align}
neglecting higher-dimension terms.
The leading-dimension, short-distance coefficients are
\begin{align}
    \bar{C}_+^\text{LGT}(w) & =  \bar{C}_{V_\parallel}^\text{LGT}(w) + \half(w-1) \bar{C}_-^\text{LGT}(w),
    \label{eq:CplusLGT} \\
    \bar{C}_-^\text{LGT}(w) & =  \bar{C}_{V_\parallel}^\text{LGT}(w) - \bar{C}_{V_\perp}^\text{LGT}(w) - 
        (w+1)\bar{C}_{V_{v'}}^\text{LGT}(w) .
    \label{eq:CminusLGT}
\end{align}
The $\bar{B}_\pm^{(\prime)\text{LGT}}$ each contain several of the fourteen $\bar{B}_{Va}^\text{LGT}$ in
Eq.~(\ref{eq:V-HQET-4}), and the detailed expressions are not illuminating.
The Isgur-Wise function $\xi(w)$ and its generalizations $\xi_3(w)$ and
\begin{equation}
    \Xi(w) = \xi(w) + \Sigma_2 A_1(w) + \Sigma_B\left[3A_3(w) + 2(w-1)A_2(w)\right]
    \label{eq:Xi=xi}
\end{equation}
parameterize the long-distance physics.
In the context of lattice gauge theory, their discretization effects arise only from the light degrees of
freedom.
In Eq.~(\ref{eq:Xi=xi}), $\xi(1)=1$ and $A_1(1)=A_3(1)=0$ by flavor conservation in HQET.
In order to have compact formulas, the function $\Xi$ contains some short-distance information, namely the
mass combinations
\begin{align}
    \Sigma_X = \frac{1}{2m_{Xc}} + \frac{1}{2m_{Xb}}, \quad X\in\{2,B,3\},
    \label{eq:SigmaX}
\end{align}
which depend on the short-distances $a$ and $m_Q^{-1}$.

When using HQET to describe the heavy-quark limit of continuum QCD,
the algebra is identical.  The difference lies in the short-distance
coefficients: in the notation used here,
$\bar{C}_{V_\parallel}^\text{LGT}$ etc.\ simply lose the superscript
``LGT''.  Further, discretization effects of the light degrees of
freedom disappear from the HQET quantities $\bar{\Lambda}$, $\xi(w)$,
$\xi_3(w)$, and $A_i(w)$.

To derive the matching factors, we focus on the leading-dimension
term.  Then one finds
\begin{align}
    Q_+(\bm{p}) & =  \frac{w+1}{2}\frac{\bar{C}_{V_\parallel}^\text{LGT}(w)}
        {\bar{C}_{V_\parallel}^\text{LGT}(1)}\Xi(w), \\
    \bm{R}_-(\bm{p}) & =  \frac{\bm{v}'}{w+1}
        \frac{\bar{C}_{V_\perp}^\text{LGT}(w)+(w+1)\bar{C}_{v'}^\text{LGT}(w)}
        {\bar{C}_{V_\parallel}^\text{LGT}(w)} .
\end{align}
Thus, to match these quantities to continuum QCD, one must multiply
$Q_+$ and $\bm{R}_-$ by
\begin{align}
   \frac{\rhoV{4}(w)}{\rhoV{4}(1)}  \equiv \; \frac{\rho_{V_\parallel}(w)}{\rho_{V_\parallel}(1)} & = 
        \frac{\bar{C}_{V_\parallel}(w)}{\bar{C}^\text{LGT}_{V_\parallel}(w)}
        \frac{\bar{C}^\text{LGT}_{V_\parallel}(1)}{\bar{C}_{V_\parallel}(1)},
    \label{eq:Z4Q+} \\
  \frac{\rhoV{i}(w)}{\rhoV{4}(w)}  \equiv  \; \frac{\rho_{V_{v'}}(w)}{\rho_{V_\parallel}(w)} & = 
        \frac{\bar{C}_{V_\perp}(w)+(w+1)\bar{C}_{v'}(w)}%
            {\bar{C}_{V_\perp}^\text{LGT}(w)+(w+1)\bar{C}_{v'}^\text{LGT}(w)}
        \frac{\bar{C}^\text{LGT}_{V_\parallel}(w)}{\bar{C}_{V_\parallel}(w)}, 
    \label{eq:Z4R-}
\end{align}
respectively, to obtain $\mathcal{Q}_+$ and $\bm{\mathcal{R}}_-$ in
Eqs.~(\ref{eq:renormQ+}) and~(\ref{eq:renormR-}).

One-loop calculations of the $w$ dependence of these coefficients are
not available, however.  (The algebra with $\bm{p}\neq\bm{0}$ is much
more voluminous.)  We shall proceed with a further approximation for
each of the two factors multiplying $Q_+$ and~$R_-$.  By construction
in Eq.~(\ref{eq:Z4Q+}),
\begin{equation}
    \frac{\rho_{V_\parallel}(w)}{\rho_{V_\parallel}(1)} = 
   1 + \mathcal{O}\left(\alpha_s(w-1)\right).
\end{equation}
Because the $w$ dependence arises only from the vertex diagram---the
tadpoles on the legs cancel---the computed coefficient should, like
those in Table~\ref{tab:rho}, be small.  Furthermore we note that
$w-1<0.16$ and that the $w$ dependence disappears when
$m_ca\to0$. Hence we neglect this one-loop contribution and take
$\rhoV{4}(w)/\rhoV{4}(1) = 1$. For the $\mathcal{O}(\alpha_s)$ error
we use the following form:
\begin{equation}
   \frac{\rhoV{4}(w)}{\rhoV{4}(1)} = 1 \pm 
        \alpha_V(2/a) {\rhoV{4}^{[1]}}_{\text{max}} (w-1) m_{2c}a,
    \label{eq:rhov4error}
\end{equation}
where the values for $\alpha_V(2/a)$ are listed in Table~\ref{tab:rho}, and 
\begin{equation}
{\rhoV{4}^{[1]}}_\text{max} = 0.1 \label{eq:B30}
\end{equation}
is an upper bound on the size of the observed one-loop corrections to
\rhoV{4}(1). In the mass region of interest, $\rhoV{4}^{[1]} <
     {\rhoV{4}^{[1]}}_\text{max}$.

Equation~(\ref{eq:rhov4error}) gives an estimate of the error in the
ratio $\rhoV{4}(w)/\rhoV{4}(1)$.  The zero-recoil $\rhoV{4}(1)$ is
calculated at one-loop order in lattice perturbation theory and
tabulated in Table~\ref{tab:rho}. We estimate the
$\mathcal{O}(\alpha_s^2)$ truncation error, in the spirit of
Ref.~\cite{B2Dstar}, by taking the coefficient as twice the largest
first-order coefficient, $2 {\rhoV{4}^{[1]}}_\text{max} = 0.2$. Hence,
the error due to omitted higher order corrections is estimated as
\begin{equation}
    \pm 2 {\rhoV{4}^{[1]}}_\text{max} \alpha^2_V(2/a).
    \label{eq:rhov4error1}
\end{equation}
The two errors are combined in quadrature to obtain the total
systematic error in $\rhoV{4}(w)$:
\begin{equation}
  \pm \rhoV{4}(1) \sqrt{ [ {\rhoV{4}^{[1]}}_\text{max} \alpha_V(2/a) (w-1) m_{2c} a]^2 +
      [ 2 {\rhoV{4}^{[1]}}_\text{max} \alpha^2_V(2/a)/ \rhoV{4}(1)]^2 } \,.
    \label{eq:rhov4error-quad}
\end{equation}

For the factor in Eq.~(\ref{eq:Z4R-}) for $R_-$, note that most of our
ensembles have $m_ca<0.4$ and recall that as $m_ca\to0$ with $m_ba$
fixed, the short-distance coefficients of the HQET with two
heavy-quark fields tend to those with one heavy-quark field (for
bottom) and a Dirac field (for charm).  As shown in
Ref.~\cite{Harada:2001fj},
\begin{align}
    \lim_{m_ca\to 0} \bar{Z}_{V_\parallel}(w) & =  Z_{V_\parallel}, \\
    \lim_{m_ca\to 0} \bar{Z}_{V_\perp}(w) & =  Z_{V_\perp}, \\
    \lim_{m_ca\to 0} \bar{Z}_{V_\perp}(w) \bar{C}^\text{LGT}_{V_{v'}}(w) & =  \bar{C}_{V_{v'}}(w); 
\end{align}
the unbarred coefficients have no $w$ dependence~\cite{Harada:2001fi}.
In practice, the error in these equations is of order $\alpha_s(a)m_ca$.
We shall neglect this contribution and use
\begin{equation}
  \frac{\rhoV{i}(w)}{\rhoV{4}(w)}  \equiv \; \frac{\rho_{V_{v'}}(w)}{\rho_{V_\parallel}(w)} = \frac{Z_{V_\perp}}{Z_{V_\parallel}} .
\end{equation}
The one-loop calculation of the right-hand side can be done at zero
recoil and is, thus, much simpler.  The one-loop result is given in
the right-most column of Table~\ref{tab:rho}.  To account for the
error due to the neglected $\mathcal{O}( \alpha_s \, m_c a)$
contribution, as in Eq.~(\ref{eq:B30}) we consider the size of the
one-loop coefficient for the range of $b$-quark masses used in this
calculation, finding $\rho^{[1]} \leq 0.352$. With
$\rho^{[1]}_\text{max} = 0.352$ we take the error as
\begin{equation}
    \pm\alpha_V(2/a) \rho^{[1]}_\text{max} m_{2c}a \;. 
    \label{eq:rhoVierror}
\end{equation}

\subsection{Discretization errors at nonzero recoil \boldmath ($w>1$)}

Power-law discretization effects arise from the higher-dimension terms
in Eqs.~(\ref{eq:CplusLGT}) and~(\ref{eq:CminusLGT}).  The
discretization errors can be found by comparing the HQET description
of lattice gauge theory to that of continuum QCD, as follows: %
substitute Eqs.~(\ref{eq:hplusLGT}) and~(\ref{eq:hminusLGT}) into
Eqs.~(\ref{eq:Q-LGT}) and~(\ref{eq:R-LGT}), multiply by the matching
factors as in Eqs.~(\ref{eq:renormQ+}) and~(\ref{eq:renormR-}), and
form the combinations in Eqs.~(\ref{eq:hplusQ})
and~(\ref{eq:hminusQ}).  The resulting HQET descriptions of the form
factors are
\begin{align}
    h_+(w) & =             \bar{C}_+(w) \Xi(w) + \frac{w-1}{2}\left\{
        \bar{B}_+^\text{mis}(w) \left[2\xi_3(w) + \bar{\Lambda}\xi(w)\right] +
        \bar{B}_-^{\prime\text{mis}}(w) \bar{\Lambda}\xi(w) \right\},
    \label{eq:hplusR} \\
    h_-(w) & =  \half(w+1)\bar{C}_-(w) \Xi(w) +
        \bar{B}_-^\text{mis}(w) \left[2\xi_3(w) - \bar{\Lambda}\xi(w)\right] -
        \bar{B}_-^{\prime\text{mis}}(w) \bar{\Lambda}\xi(w), \hspace*{3.5em}
    \label{eq:hminusR}
\end{align}
where%
\footnote{The continuum QCD analogs of Eqs.~(\ref{eq:Bbar+mis}) and~(\ref{eq:Bbar-mis}) can be obtained by 
erasing the superscript ``LGT'' and simplifying with Eqs.~(\ref{eq:CplusLGT}) and~(\ref{eq:CminusLGT}).
The result becomes, as expected, trivial.} 
\begin{align}
    \bar{B}_+^{(\prime)\text{mis}}(w) & =     
    \frac{\bar{B}_+^{(\prime)\text{LGT}}(w)}{\bar{C}_{V_\perp}^\text{LGT}(w)+(w+1)\bar{C}_{v'}^\text{LGT}(w)} 
        \left( \bar{C}_+(w) - \frac{w+1}{2}
        \frac{\bar{C}_-^\text{LGT}(w) \bar{C}_{V_\parallel}(w)}{\bar{C}_{V_\parallel}^\text{LGT}(w)} \right),
    \nonumber \\ & -  
    \frac{\bar{B}_-^{(\prime)\text{LGT}}(w)}{\bar{C}_{V_\perp}^\text{LGT}(w)+(w+1)\bar{C}_{v'}^\text{LGT}(w)} 
        \left( \bar{C}_-(w) -
        \frac{\bar{C}_-^\text{LGT}(w) \bar{C}_{V_\parallel}(w)}{\bar{C}_{V_\parallel}^\text{LGT}(w)} \right),
    \label{eq:Bbar+mis} \\
    \bar{B}_-^{(\prime)\text{mis}}(w) & =  
    \frac{\bar{B}_-^{(\prime)\text{LGT}}(w)}{\bar{C}_{V_\perp}^\text{LGT}(w)+(w+1)\bar{C}_{v'}^\text{LGT}(w)} 
        \left( \bar{C}_+(w) - w\bar{C}_-(w) + \frac{w-1}{2}
    \frac{\bar{C}_-^\text{LGT}(w) \bar{C}_{V_\parallel}(w)}{\bar{C}_{V_\parallel}^\text{LGT}(w)} \right),
    \nonumber \\ & -  \frac{w^2-1}{4}
    \frac{\bar{B}_+^{(\prime)\text{LGT}}(w)}{\bar{C}_{V_\perp}^\text{LGT}(w)+(w+1)\bar{C}_{v'}^\text{LGT}(w)} 
        \left( \bar{C}_-(w) -
        \frac{\bar{C}_-^\text{LGT}(w) \bar{C}_{V_\parallel}(w)}{\bar{C}_{V_\parallel}^\text{LGT}(w)} \right).
    \label{eq:Bbar-mis}
\end{align}
As long as the matching of the dimension-three currents is carried out to order~$\alpha_s^\ell$, the parts 
of Eqs.~(\ref{eq:Bbar+mis}) and~(\ref{eq:Bbar-mis}) entailing the $\bar{C}$ coefficients collapses such that
\begin{equation}
    \bar{B}^{(\prime)\text{mis}}_\pm = \bar{B}^{(\prime)}_\pm + \mathcal{O}(\alpha_s^{\min(k,\ell)+1}),
\end{equation}
where the dimension-four currents have been matched through order~$\alpha_s^k$.
In particular at the tree level ($k=0$),
\begin{align}
    \bar{B}_\pm^{\text{mis}} & =  \frac{1}{2m_{c3}}\pm\frac{1}{2m_{b3}}, 
    \label{eq:Bbar+-tree} \\
    \bar{B}_\pm^{\prime\,\text{mis}} & =  0,
    \label{eq:Bbar'+-tree}
\end{align}
while in continuum QCD, $\bar{B}_\pm=1/2m_c\pm1/2m_b$ and $\bar{B}'_\pm=0$.
Thus, we have tree-level matching in the dimension-four currents, with errors from this source of the form
\begin{equation}
    \texttt{error}_{3,\pm} = \left[f_3(m_{0c}a) \pm f_3(m_{0b}a)\right]\bar{\Lambda}a.
    \label{eq:error3}
\end{equation}
Here $af_3(m_0a)=1/2m_3-1/2m_2$, and the factor of~$\bar{\Lambda}$ is
a power-counting estimate of the HQET matrix element; $\bar{\Lambda}$
is the scale of nonperturbative QCD as it pertains to heavy-light
mesons, roughly the difference between the heavy-light-meson and
heavy-quark masses.

Another discretization error arises from the function $\Xi(w)$ in
lattice gauge theory and continuum QCD.  In LGT, the kinetic and
chromomagnetic masses appear.  In this way, one finds that the
mismatch in $\Sigma_B$ in $\Xi(w)$ yields an error
\begin{equation}
    \texttt{error}_B = \left[f_B(m_{0c}a) + f_B(m_{0b}a)\right](w-1)\bar{\Lambda}a,
    \label{eq:errorB}
\end{equation}
taking the functions $A_i$ to be of order $\bar{\Lambda}$ and building
in the fact that the contribution vanishes as $w\to1$.  Similarly to
above, $af_B(m_0a)=1/2m_B-1/2m_2$, which, for our choice of
$c_\text{SW}$, is of order~$\alpha_s$.

Combining the two kinds of errors ($\oplus$ means to add in quadrature),
\begin{align}
    h_+(\ref{eq:hplusQ})-h_+(\text{cont.}) & =  \texttt{error}_B \oplus \half(w-1) \texttt{error}_{3,+},
    \label{eq:error+} \\
    h_-(\ref{eq:hminusQ})-h_-(\text{cont.}) & =  \texttt{error}_{3,-}.
    \label{eq:error-}
\end{align}
Because $\bar{C}_-$ vanishes at the tree level, the contribution to
the error in $h_-$ from $\bar{C}_-\texttt{error}_B$ is suppressed by
an addition factor of $\alpha_s$ and, thus, omitted here.  Note that
$\texttt{error}_{3,+}$ in $h_+(w)$ is multiplied by $(w-1)$, whereas
$\texttt{error}_{3,-}$ in $h_-(w)$ is not; cf.\ Eqs.~(\ref{eq:hplusR})
and~(\ref{eq:hminusR}).  Our choices for the functions $f_B(m_0a)$ and
$f_3(m_0a)$ are discussed below; cf.\ Eqs.~(\ref{eq:fB})
and~(\ref{eq:f3}).

\subsection{Discretization errors at zero recoil \boldmath ($w=1$)}

Because the next-to-leading-dimension discretization effects are
suppressed by $\alpha_s$, the next-to-next-to-leading-dimension
effects may be of the same size.  This is especially true at zero
recoil, where the next-to-leading contributions to $h_+$ vanish.  To
capture the leading discretization errors of $h_+(1)$, therefore, one
needs the dimension-five temporal vector current (with
$v'=v$)~\cite{Kronfeld:2000ck}:
\begin{align}
    \ZVcb{4}V^4 = -\ZVcb{4} v\cdot V \doteq \bar{C}_{V^{cb}_\parallel} \bar{c}_vb_v & +  
        z^{(1,1)}_{V^{cb}1} \frac{\bar{c}_v\loarrow{D}_\perp\cdot D_\perp b_v}{2m_{3c}\;2m_{3b}} +
        z^{(1,1)}_{V^{cb}s} \frac{\bar{c}_v\loarrow{D}_\perp^\alpha s_{\alpha\beta}D_\perp^\beta b_v}
            {2m_{3c}\;2m_{3b}} 
    \label{eq:V-HQET-5} \\ & + 
        \eta^{(0,2)}_{V^{cb}D_\perp^2} \frac{\bar{c}_vD_\perp^2b_v}{8m_{D_\perp^2b}^2} +
        \eta^{(0,2)}_{V^{cb}sB} \frac{\bar{c}_vs\cdot Bb_v}{8m_{sBb}^2} +
        \eta^{(0,2)}_{V^{cb}\alpha E} \frac{\bar{c}_vi\Eslash b_v}{4m_{\alpha Eb}^2} 
    \nonumber \\ & + 
        \eta^{(2,0)}_{V^{cb}D_\perp^2} \frac{\bar{c}_v\loarrow{D}_\perp^2b_v}{8m_{D_\perp^2c}^2} +
        \eta^{(2,0)}_{V^{cb}sB} \frac{\bar{c}_vs\cdot Bb_v}{8m_{sBc}^2} +
        \eta^{(2,0)}_{V^{cb}\alpha E} \frac{\bar{c}_vi\Eslash b_v}{4m_{\alpha Ec}^2},
    \nonumber
\end{align}
and similarly for $-v\cdot\mathcal{V}$.  For the currents defined in
Sec.~\ref{sec:formalism}, as well as for the continuum currents, the
$\eta$-coefficients and $z$-coefficients in Eq.~(\ref{eq:V-HQET-5})
all take the form $1+\mathcal{O}(\alpha_s)$.  The $\eta$-like
coefficients and associated masses in Eq.~(\ref{eq:V-HQET-5}) drop out
of the analysis.

From Eqs.~(7.23)--(7.29) of Ref.~\cite{Kronfeld:2000ck}, the HQET
expansions through $\mathcal{O}(\bar{\Lambda}^2)$ of the matrix
elements are
\begin{equation}
    \sqrt{\mathcal{R}_+} =
            \eta_V W^{(0)}_{00} + \overline{W}^{(2)}_{00},
    \label{eq:BVBW}
\end{equation}
where $\eta_V$ is an HQET-to-QCD matching factor that starts with 1 in
perturbative QCD, and
\begin{align}
    W^{(0)}_{00} & =  1 - \half\Delta_2^2D - 3\Delta_2\Delta_BE - \half\Delta_B^2(R_1+3R_2),
    \label{eq:W000} \\
    \overline{W}^{(2)}_{00} & =  -\half\Delta_3^2
        \left[z_{V1}^{(1,1)}\mu_\pi^2 - z_{Vs}^{(1,1)}\mu_G^2\right],
    \label{eq:W200}
\end{align}
where $D$, $E$, $R_1$, $R_2$, $\mu_\pi^2$, and $\mu_G^2$ are HQET
matrix elements of order~$\bar{\Lambda}^2$.%
\footnote{Ref.~\cite{Kronfeld:2000ck} used a notation setting $\mu_\pi^2=-\lambda_1$ and
$\mu_G^2=3\lambda_2$.} Also,
\begin{equation}
    \Delta_I = \frac{1}{2m_{Ic}} - \frac{1}{2m_{Ib}},\quad I=2,B,3, 
\end{equation}
are combinations of the mass coefficients in Eqs.~(\ref{eq:L-HQET})
and~(\ref{eq:Bm3}).  Beyond the leading~1, the terms in $W^{(0)}_{00}$
come from double insertions of the kinetic and chromomagnetic
interactions.  $\overline{W}^{(2)}_{00}$ stems from the dimension-five
currents in Eq.~(\ref{eq:V-HQET-5}).

Taking the difference between these expressions and the analogous ones
for continuum QCD, one sees that the error in $W^{(0)}_{00}$ stems
from
\begin{equation}
    \frac{1}{2m_{Bh}} - \frac{z_B}{2m_{2h}} = af_B(m_{0h}a).
\end{equation}
The coefficients $z_{J\bullet}^{(1,1)}=1+\mathcal{O}(\alpha_s)$; also
$1/m_{3h}\to 1/m_h+\mathcal{O}(\alpha_sa)$ [compare
  Eqs.~(\ref{eq:Bm3b}) and~(\ref{eq:Bm3c})].  Thus, the error entering
$\overline{W}^{(2)}_{00}$ stems from
\begin{equation}
    \Delta_3^2z^{(1,1)}_{V\bullet} - \Delta_2^2z^{(1,1)}_{V\bullet} = 2a[f_3(m_{0c}a)-f_3(m_{0b}a)]\Delta_2,
\end{equation}
with $f_3$ of order~$\alpha_s$ for our choices.
Thus, errors in $\rhoV{4}\sqrt{R_+}$ stem from the mismatches
\begin{align} 
    W^{(0)}_{00}(\text{LGT}) - W^{(0)}_{00}(\text{cont.}) & =  
        - a\Delta_2\left[f_B(m_{0c}a)-f_B(m_{0b}a)\right](R_1+3R_2+3E),
    \label{diff:W000} \\
    \overline{W}^{(2)}_{00}(\text{LGT}) - \overline{W}^{(2)}_{00}(\text{cont.}) & =  
        - a\Delta_2\left[f_3(m_{0c}a)-f_3(m_{0b}a)\right](\mu_\pi^2-\mu_G^2).
    \label{diff:W200}
\end{align}
In estimating heavy-quark discretization errors, we use these results
at $w=1$, where the more generic effects in Eqs.~(\ref{eq:error+})
and~(\ref{eq:error-}) are much smaller.

\subsection{Numerical estimates}

For the mismatch functions $f_B$ and $f_3$ in Eqs.~(\ref{eq:errorB})
and~(\ref{eq:error3}), and in Eqs.~(\ref{diff:W000})
and~(\ref{diff:W200}), we use the functional
forms~\cite{Bazavov:2011aa}
\begin{align}
	f_B(m_0a) & =  \frac{\alpha_s}{2(1+m_0a)}, \label{eq:fB} \\
	f_3(m_0a) & =  \frac{\alpha_s}{2(2+m_0a)}. \label{eq:f3}
\end{align}
To estimate the HQET matrix elements, we take $\bar{\Lambda}=450~\text{MeV}$,%
\footnote{Here, 450~MeV is not an estimate of $M_B-m_b$, but simply a practical number for power-counting 
estimates.}
\begin{align}
    \mu_G^2 = \case{3}{4}(M_{B^*}^2 - M_B^2) = 0.364~\text{GeV}^2 & =  (603~\text{MeV})^2,
    \label{eq:muG2} \\
    \mu_\pi^2(1~\textrm{GeV}) = 0.424\pm0.042~\text{GeV}^2 & =  (651\pm32~\text{MeV})^2.
    \label{eq:mupi2}
\end{align}
We do not have estimates for $D$, $E$, $R_1$, and $R_2$ as good as
Eqs.~(\ref{eq:muG2}) and~(\ref{eq:mupi2}), but in Ref.~\cite{B2Dstar}
we found that we could explain the discretization effects at zero
recoil in $B\to D^*$ with $|R_1+3R_2+3E|\lesssim(450~\text{MeV})^2$.

We take the typical $\alpha_V(q^*)$ to be $0.262$ on the
$a\approx0.09$~fm lattices, and we use one-loop
running to obtain $\alpha_V(q^*)$ at the other lattice spacings.
\begin{table}[tp]
    \centering
    \caption[tbl:diffs]{Absolute difference of $h_\pm(w)$ from mismatches
        in the heavy-quark Lagrangian and current.
        We take $\bar{\Lambda}=450$~MeV,
        $\mu_\pi^2=0.424~\textrm{GeV}^2$, and
        $\mu_G^2=0.364~\textrm{GeV}^2$.
        We further estimate the quantity $|R_1+3R_2+3E|$ with
        $\bar{\Lambda}^2$.
        The totals are obtained from Eqs.~(\ref{eq:error+}),
        (\ref{eq:error-}), and (\ref{diff:W000})~and~(\ref{diff:W200})
        for $h_+(w)$, $h_-$, and $h_+(1)$, respectively.
        The column for $h_+(w)$ must be multiplied by $(w-1)$.
        The difference is estimated using the $a=0.09$~fm
        lattice as a baseline.}
    \label{tbl:diffs}
    \begin{tabular}{ccccrrr}
    \hline\hline
    $a$~(fm) & $\alpha_V(q^*)$ & $~~m_{0b}a~~$ & $~~m_{0c}a~~$ & $h_+(w)~$ & $h_-(w), \forall w$ & $h_+(1)~$  \\
    \hline
     0.120   &       0.300     & 2.462 & 0.532 & $-0.0095$ & $-0.0030~$ & $-0.0011$ \\ 
     0.090   &       0.261     & 1.664 & 0.362 & $0.0000$ & $0.0000~$ & $0.0000$ \\ 
     0.060   &       0.220     & 1.123 & 0.240 & $0.0109$ & $0.0021~$ & $0.0011$ \\ 
     0.045   &       0.198     & 0.808 & 0.176 & $0.0160$ & $0.0029~$ & $0.0016$ \\ 
    \hline\hline
    \end{tabular}
\end{table}

\begin{table}[tp]
    \centering
    \caption[tbl:errors]{Absolute error on $h_\pm(w)$ from mismatches in the heavy-quark Lagrangian and 
        current.
		We take $\bar{\Lambda}=450$~MeV, $\mu_\pi^2=0.424~\textrm{GeV}^2$, and 
        $\mu_G^2=0.364~\textrm{GeV}^2$.
        We further estimate the quantity $|R_1+3R_2+3E|$ with $\bar{\Lambda}^2$.
		The columns for $h_+(w)$ correspond to the chromomagnetic mismatch [``$B$'', Eq.~(\ref{eq:errorB})], 
        the current mismatch [``$3$'', Eq.~(\ref{eq:error3})], and their quadrature sum [``$\oplus$'', 
        Eq.~(\ref{eq:error+})]; these columns must be multiplied by $(w-1)$.
        The column for $h_-(w)$ comes from the mismatch in Eq.~(\ref{eq:error-}).
        The columns for $h_+(1)$ correspond to the second-order mismatch of the Lagrangian 
        [``$W_{00}^{(0)}$'', Eq.~(\ref{diff:W000})] and the second-order mismatch of the current 
        [``$\overline{W}_{00}^{(2)}$'', Eq.~(\ref{diff:W200})], and their quadrature sum (``$\oplus$'').}
    \label{tbl:errors}
    \begin{tabular}{cccc@{\quad}ccc@{\quad}c@{\quad}ccc}
    \hline\hline
    $a$~(fm) & $\alpha_V(q^*)$ & $~~m_{0b}a~~$ & $~~m_{0c}a~~$ & \multicolumn{3}{c}{$h_+(w)$} & $h_-(w)$ & \multicolumn{3}{c}{$h_+(1)$}  \\
                &                    &                 &
				& $B$ & 3 & $\oplus$ & $\forall w$ & $W_{00}^{(0)}$
				& $\overline{W}_{00}^{(2)}$ & $\oplus$ \\
    \hline
     0.120   &       0.300     & 2.462 & 0.532 & 0.0382 & 0.0125 & 0.0402 & 0.0069 & 0.0033 & 0.0005 & 0.0033 \\ 
     0.090   &       0.261     & 1.664 & 0.362 & 0.0293 & 0.0092 & 0.0307 & 0.0040 & 0.0023 & 0.0003 & 0.0023 \\ 
     0.060   &       0.220     & 1.123 & 0.240 & 0.0190 & 0.0057 & 0.0198 & 0.0019 & 0.0012 & 0.0001 & 0.0012 \\ 
     0.045   &       0.198     & 0.808 & 0.176 & 0.0141 & 0.0041 & 0.0147 & 0.0010 & 0.0007 & 0.0001 & 0.0007 \\ 
    \hline\hline
    \end{tabular}
\end{table}

In Table~\ref{tbl:diffs}, we show results from using these inputs to
compute the differences that Eqs.~(\ref{diff:W000})
and~(\ref{diff:W200}) predict, using the lattice with
$a\approx0.09$~fm as the baseline.  The estimates of the differences
are compatible with the lattice-spacing dependence that can been seen
for $w=1$ in Fig.~\ref{fig:hplusvsa}, and can be inferred for $w>1$
from Fig.~\ref{fig:hplushminus}.  For example, the error in $h_+(w)$
grows slowly with $w$, both from Table~\ref{tbl:diffs} (adding in
quadrature the right-most column with $(w-1)$ times the fifth column)
and Fig.~\ref{fig:hplushminus}.  Because the \emph{differences} from
lattice to lattice are well described by the theory, we can proceed to
use the same ideas to estimate the difference from each lattice to the
continuum.  The results of these calculations are shown in
Table~\ref{tbl:errors}.  For our final error estimates on the vector
and scalar form factors, we take the absolute errors on $h_+$ and
$h_-$ in Table~\ref{tbl:errors} at $a\approx0.06$~fm, and combine them
in quadrature following Eqs.~(\ref{eq:f+fromh}) and~(\ref{eq:f0fromh})
that relate $f_+$ and $f_0$ to $h_+$ and $h_-$.  The resulting
expressions for the absolute errors as a function of recoil are
\begin{align}
    \texttt{error}_+ &= \left[0.0198(w -1)\frac{1+r}{2\sqrt{r}}\right] \oplus
        \left[0.0019\frac{1-r}{2\sqrt{r}}\right] \oplus 0.0012\frac{1+r}{2\sqrt{r}}, \\
    \texttt{error}_{0\,} &= \left[0.0198(w^2 -1)\frac{\sqrt{r}}{1+r}\right] \oplus
        \left[0.0019(w-1)\frac{\sqrt{r}}{1-r}\right] \oplus 0.0012(w+1)\frac{\sqrt{r}}{1+r}.
\end{align}
These lead to estimates that range from 0.1--0.4\% for both $f_+$ and
$f_0$ in our range of simulated lattice $w$ values.

%%%%%%%%%%%%%%%%%%%%%%%%%%%%%%%

\bibliographystyle{apsrev4-1}
\bibliography{B2Dprd}

\end{document}